\documentclass[11pt,twocolumn]{iopart}
 
\usepackage{mathrsfs}
\usepackage{color}
\usepackage{amsfonts}
\usepackage{amsthm,amssymb}
\usepackage{graphicx} 

\usepackage{etoolbox}

\usepackage{lipsum}% just for the example

\makeatletter
\def\@mkboth#1#2{}
\newlength\appendixwidth
\preto\appendix{\addtocontents{toc}{\protect\patchl@section}}
\newcommand{\patchl@section}{%
  \settowidth{\appendixwidth}{\textbf{Appendix }}%
  \addtolength{\appendixwidth}{1.5em}%
  \patchcmd{\l@section}{1.5em}{\appendixwidth}{}{\ddt}%
}
\makeatother

\begin{document}

\title[McVittie spacetimes in non-flat FLRW backgrounds]{Local properties and global structure of McVittie spacetimes with non-flat FLRW backgrounds}
\author{Brien C. Nolan}
\address{Centre for Astrophysics and Relativity, School of Mathematical Sciences, Dublin City University, Glasnevin, Dublin 9, Ireland.}
\eads{\mailto{brien.nolan@dcu.ie}}

\begin{abstract}
McVittie spacetimes embed the vacuum Schwarzschild(-(anti) de Sitter) spacetime in an isotropic, Friedmann-Lema\^{i}tre-Robertson-Walker (FLRW) background universe. The global structure of such spacetimes is well understood when the FLRW background is spatially flat. In this paper, we study the global structure of McVittie spacetimes with spatially non-flat FLRW backgrounds. 
We derive some basic results on the metric, curvature and matter content of these spacetimes and provide a representation of the metric that makes the study of their global properties possible. In the closed case, we find that at each instant of time, the spacetime is confined to a region bounded by a (positive) minimum and a maximum area radius, and is bounded either to the future or to the past by a scalar curvature singularity. This allowed region only exists when the background scale factor is above a certain minimum, and so is bounded away from the Big Bang singularity, as in the flat case. In the open case, the situation is different, and we focus mainly on this case. In $K<0$ McVittie spacetimes, radial null geodesics originate in finite affine time in the past at a boundary formed by the union of the Big Bang singularity of the FLRW background and a hypersurface (of varying causal character) which is non-singular in the sense of scalar curvature. Furthermore, in the case of eternally expanding open universes with $\Lambda\geq0$, we prove that black holes are ubiquitous: ingoing radial null geodesics extend in finite affine time to a hypersurface that forms the boundary of the region from which photons can escape to future null infinity. We determine the structure of the conformal diagrams that can arise in the open case. Finally, we revisit the black hole interpretation of McVittie spacetimes in the spatially flat case, and show that this interpretation holds also in the case of a vanishing cosmological constant, contrary to a previous claim of ours. \end{abstract}
%\pacs{04.20.Dw, 04.20.Ex}

\maketitle

%\tableofcontents

%\renewtitle[McVittie spacetimes in non-flat FLRW backgrounds]{Local properties and global structure of McVittie spacetimes with non-flat FLRW backgrounds}

\newtheorem{comments}{Comment}
\newtheorem{theorem}{Theorem}[section]
\newtheorem{proposition}{Proposition}[section]
\newtheorem{lemma}{Lemma}[section]
\newtheorem{corollary}{Corollary}[section]
\newtheorem{definition}{Definition}[section]
\newtheorem{observation}{Observation}[section]

\newcommand{\be}{\begin{equation}}
\newcommand{\ee}{\end{equation}}
\newcommand{\bes}{\begin{eqnarray*}}
\newcommand{\ees}{\end{eqnarray*}}
\newcommand{\beq}{\begin{eqnarray}}
\newcommand{\eeq}{\end{eqnarray}}
\newcommand{\tao}{\tau_0}
\newcommand{\tac}{\tau_{c1}}
\newcommand{\tacc}{\tau_{c2}}
\newcommand{\real}{\mathbb{R}}
\newcommand{\rat}{\mathbb{Q}}
\newcommand{\dig}{\mathbb{D}}
\newcommand{\integer}{\mathbb{Z}}
\newcommand{\ds}[1]{\displaystyle{#1}}
\newcommand{\pd}[2]{\frac{\partial #1}{\partial #2}}
\newcommand{\pdd}[2]{\frac{\partial^2#1}{\partial{#2}^2}}
\newcommand{\poo}{\partial\Omega_0}
\newcommand{\pom}{\partial\Omega_{2m}}
\newcommand{\vx}{\vec{x}}
\newcommand{\co}{{\cal{O}}}
\newcommand{\vep}{V_\epsilon}
\newcommand{\aep}{A_\epsilon}
\newcommand{\bep}{B_\epsilon}
\newcommand{\wep}{W_\epsilon}
\newcommand{\ui}{u_{\rm{isco}}}
\newcommand{\sigmcv}{\Sigma_{McV}}
\newcommand{\mmcv}{{\cal{M}}_{McV}}
\newcommand{\sigrw}{\Sigma_{RW}}
\newcommand{\mrw}{{\cal{M}}_{RW}}
\newcommand{\mm}{{\cal{M}}}
\newcommand{\sig}{\sigma}
\newcommand{\sigr}{\sigma_r}
\newcommand{\sigh}{{\hat{\sigma}}_1}
\newcommand{\tsig}{t_\sig}
\newcommand{\rsig}{r_\sig}
\newcommand{\tsigd}{\dot{\tsig}}
\newcommand{\tasigd}{\dot{\tasig}}
\newcommand{\rsigd}{\dot{\rsig}}
\newcommand{\tasig}{\tau_\sig}
\newcommand{\et}{\vec{e}_1}
\newcommand{\choice}{{1,2}}
\newcommand{\delmcv}{\Delta_{McV}}
\newcommand{\delrw}{\Delta_{RW}}
\def\eqq{\stackrel{\sig}{=}}
\newcommand{\cj}{{\cal{J}}}
\newcommand{\hor}{{\cal{H}}}
\newcommand{\horp}{{\cal{H}}_+}
\newcommand{\horm}{{\cal{H}}_-}
\newcommand{\alr}{{\cal{A}}}
\newcommand{\balr}{{\partial\Omega_{\{\kappa=0\}}}}
\newcommand{\alrt}{\partial{\cal{A}}_{T}}
\newcommand{\alrs}{\partial{\cal{A}}_{S}}
\newcommand{\kap}{\kappa}
\newcommand{\rst}{r_*}
\newcommand{\rh}{\hat{r}}
\newcommand{\fixme}[1]{\textcolor{red}{[\textbf{#1}}]}
\renewcommand{\qed}{\hfill$\blacksquare$}
\newcommand{\qedapp}{\hfill$\square$}
\newcommand{\ep}{\epsilon}
\newcommand{\startproof}{\noindent\textbf{Proof: }}
\newcommand{\bark}{\bar{K}}
\newcommand{\omk}{\Omega_{\{\kappa=0\}}}
\newcommand{\rl}{r_\lambda}
\newcommand{\kl}{{\cal{K}}_\lambda}
\newcommand{\mcv}{$\textrm{McV}{\!}_4\,$}

%\DeclareMathOperator\arcsinh{arcsinh}

%\newtheorem{exercise}{Exercise}
%\newcounter{problem}
%\newcounter{exercise}
%%\newcounter{definition}
%\newcounter{example}
%%\newcounter{lemma}
%\newcounter{hp}
%\newcounter{question}
%\newcounter{comment}

%\newcounter{comment}
%\newenvironment{comment}[1][\thedefinition]{%
      %\stepcounter{comment}%
      %\pagebreak[2]\medskip\par\noindent%
      %\textbf{Comment \arabic{comment}}%
      %\,\,\nopagebreak% 
			%}\smallskip\par%

%\sffamily

%\newcounter{prop}
%
%\newenvironment{prop}[1][\theprop]{%
%      \stepcounter{prop}%
%      \pagebreak[2]\bigskip\par\noindent%
%      \textbf{Proposition \arabic{prop}}%
%      \medskip\nopagebreak\par\noindent}%
%
%

%========================================INTRODUCTION===============================================================================
%========================================INTRODUCTION===============================================================================
%========================================INTRODUCTION===============================================================================
%========================================INTRODUCTION===============================================================================

%\section{Introduction}
%Describe McVittie and key results; note that 1933 McV solutions don't have bound mass; note existence and uniqueness result from 1998 and elliptic integral. Key global features of the K=0 case. 
%
%\section{The background} As below.
%
%\section{Metric, matter and curvature} As below. Then lead on to three key global features. 
%
%\section{Boundary of the allowed region}
%

\section{Introduction and summary}

In this paper, we revisit the question of finding and understanding  solutions of the Einstein equations that represent cosmological black holes. This question dates back to 1933, when McVittie \cite{mcvittie1933mass} found solutions of the Einstein field equations that yield, respectively, (i) the vacuum Schwarzschild metric (or the Schwarzschild-de Sitter metric when a cosmological constant is included) and (ii) the FLRW metric when appropriate limits are taken. More precisely, McVittie's metric solves the Einstein equations for a perfect  fluid, and contains a parameter $M$ and a free function $a(t)$ such that: (i) when $a(t)$ is chosen so that the spacetime is vacuum, then the line element is that of Schwarzschild (de Sitter) spacetime with mass parameter $M$ and (ii) when we set $M=0$, the metric is that of a spatially flat FLRW spacetime with scale factor $a(t)$. Furthermore, at fixed time, various geometric and physical quantities are asymptotic, at large spatial distances, to their corresponding FLRW values. (The FLRW metric which emerges in this limit is referred to as the FLRW background.) McVittie also presented solutions which sought to embed the Schwarzschild spacetime in non-flat FLRW spacetimes, in the sense described above. However, the metrics presented by McVittie do not satisfy (i) - these spacetimes do not possess a vacuum limit - and display other characteristics which run contrary to the desired scenario of an isolated mass embedded in an FLRW background. 

The issue of whether or not McVittie spacetimes do indeed represent black holes embedded in isotropic universes is of course a question about the global structure of these spacetimes. The key results on this question were first established in \cite{nolan1999point}, and the black hole interpretation was identified in \cite{kaloper2010mcvittie}. These results relate to the behaviour of radial null geodesics: particle and photon orbits were studied in \cite{nolan2014particle}, where the existence of bound particle and photon orbits in large classes of McVittie spacetimes was proven. Thus the global understanding of spatially flat McVittie metrics is well advanced (see also \cite{lake2011more,da2013expansion}). As first identified in \cite{nolan1999point} a feature of such metrics, in the case when the FLRW background is expanding, is the appearance of a mildly singular (see \cite{nolan1999bpoint}) spatial hypersurface forming the past boundary of the spacetime, located at $r=2M$ where $r$ is the area-radius of the (spherically symmetric) spacetime, and $M$ is the mass parameter identified above. This locus reappears as part of a future boundary in the case of a non-negative cosmological constant. 

One way to seek to understand these features is to study the corresponding scenario in the spatially non-flat cases. One must first determine the metric of such spacetimes. In the case of a point mass (as represented by the Schwarzschild(-de Sitter) spacetime) embedded in an FLRW universe with negatively curved spatial slices, the existence and uniqueness of the relevant metric was established in \cite{nolan1998point}. The approach used encountered two difficulties: first, it did not generalise to the case of postive spatial curvature and second, the resulting spacetime metric was written, in comoving coordinates, in a form that required the use of elliptic integrals \textit{implicit in one of the metric functions}. Furthermore, in an earlier study, the use of co-moving coordinates hindered the correct identification of spacetime boundaries \cite{sussman1988spherically}.
%In particular, integrating the field equations results in 
%\be
%-\frac14\ln\left(\frac{x}{x+4}\right) + B(t)
%= \int \frac{du}{(4u^2-16Mu^3+a^{-2}(t))^{1/2}},
%\label{elliptic}
%\ee
%where $x=4\sinh^2(R/2)$, $(t,R)$ are respectively a cosmic time and comoving radial coordinate, and $u=u(t,R)$ is a metric function. The function $B$ is set equal to zero by using an asymptotic condition, and $a(t)$ is the scale factor of the FLRW background. See p.\ 4 of \cite{nolan1998point}. 
This form of the solution makes the study of the global structure of the spacetime considerably more difficult than that of the spatially flat case, where there is an explicit representation of the metric in terms of elementary functions of the space-time coordinates, and where these coordinates have a clear physical and/or geometric interpretation. 

The principal aims of this paper are twofold: First, we show how the spatially non-flat McVittie metric may be written in a form where there is an \textit{explicit} representation of the metric in terms of special functions of the space-time coordinates, and where these coordinates have a clear physical and/or geometric interpretation. Second, we use this form of the metric to determine the global structure of non-flat McVittie spacetimes. As in \cite{nolan1998point}, we adopt an axiomatic approach, and argue that the conditions we impose on the metric are natural from the point of view of our purpose: to derive solutions of the Einstein equations representing a point mass or spherical black hole embedded in an FLRW universe. These conditions yield the spatially flat McVittie metric of \cite{mcvittie1933mass} (as studied extensively in the papers cited above); the negatively curved McVittie metric whose existence and uniqueness was established in \cite{nolan1998point}, and generalise to yield a new interpretation of a metric corresponding to the case of positive spatial curvature of the FLRW background. 

%Before summarising the content of this paper, we make the following note about the FLRW background of these McVittie spacetimes.

As emphasised in \cite{da2013expansion}, the choice of scale factor $a(t)$ of the FLRW background strongly influences the global structure of a spatially flat McVittie spacetime. This will also be the case for non-flat McVittie spacetimes. Our starting point is the whole family of FLRW spacetimes. With each member of this family we associate a maximal open interval $I$ on which the scale factor $a$ is $C^2$ and satisfies $a(t)>0$ for all $t\in I$. We take the cosmic time function $t$ to increase into the future, and we define the Hubble function $H=\theta/3$, where $\theta$ is the expansion of the fluid flow lines. We use $K$ to represent the curvature index of the FLRW spacetimes. We will introduce further restrictions on the class of FLRW backgrounds in which we are interested: see Section \ref{background} below. 

Section 2 comprises, in a sense, the first half of the paper. We define spacetimes representing \textit{a point mass in an isotropic universe} (Definition \ref{def1}) as spherically symmetric, shear-free, perfect fluid solutions of the Einstein equations with Weyl curvature satisfying 
\be \Psi_2 = -\frac{M}{r^3}, \label{weyl-form1} \ee
where $M$ is a constant and $r$ is the area radius of the spherically symmetric spacetime, and which satisfy two other conditions relating to the FLRW background. We then prove a theorem (Theorem \ref{thm1}) that provides the line element of such spacetimes as well as other details. This serves to identify the class of spacetimes under consideration, and to clarify their relationship to the Schwarzschild family of spacetimes, and to the FLRW family of spacetimes. 

The remainder of the paper is devoted to the study of the global structure of the spacetimes described in Section \ref{metric}, focussing on the cases where the FLRW curvature index $K\neq 0$. In Section 3, we make some necessary comments on the class of FLRW background spacetimes under consideration. In Section \ref{metric}, we encounter an invariantly defined function $\kappa$ (\ref{kap-new-def}) which must be positive throughout the spacetime. In Section 4, we develop some technical details relating to the zero-set of $\kappa$ - the boundary of the allowed region. Section 5 initiates the discussion proper of the global structure of McVittie spacetimes with non-flat FLRW backgrounds. We establish results relating to the (apparent) horizon and to the radial null geodesics (RNGs) of the spacetimes. Sections 6-8 deal with the case where the FLRW curvature index $K$ is negative. In Section 6, we study the past evolution of RNGs, and thereby establish the nature of the past boundary of the spacetimes. In Section 7, we study the future evolution of RNGs, and show that the spacetimes being studied possess black hole horizons. The conformal diagrams of the spacetimes are presented in Section 8. In Section 9, we discuss the case $K>0$, and in Section 10, we revisit the case $K=0$, correcting a previous statement of ours relating to the black hole nature of these spacetimes. Section 11 contains some concluding comments.

We use the curvature conventions of \cite{wald1984general}, and use units in which $G=c=1$. The symbol $\blacksquare$ is used to indicate the end of a proof, or the absence of a proof where it was not felt necessary. A number of proofs have been relegated to the appendix in order to make the paper more readable. This is flagged with the symbol $\square$ at the end of the statement of the relevant result. We use a prime (e.g.\ $A'(t)$) to denote a derivative with respect to argument, except for geodesics, where an overdot (e.g.\ $\dot{r}$) is used to denote derivative with respect to parameter (proper time, affine parameter).

%%%%%%%%%%%%%%%%%%%%%%%%%%%%%%
%%%%%%%%%%%%%%%%%%%%%%%%%%%%%%
%%%%%%%%%%%%%%%%%%%%%%%%%%%%%%

\section{Metric, matter and curvature}\label{metric}

We begin the discussion with the following definition: 

\begin{definition}\label{def1}
A spacetime $({\mm},g)$ represents a \textbf{point mass in an isotropic universe} if it satisfies the following four conditions:
\begin{itemize}
\item[(C1)] the metric is a spherically symmetric solution of the Einstein equations (with a cosmological constant) coupled to a shear-free perfect fluid that inherits the spherical symmetry of the spacetime;
\item[(C2)] the Newman-Penrose Weyl curvature invariant $\Psi_2$ has the form 
\be \Psi_2 = -\frac{M}{r^3},\label{weyl-form} \ee
where $M$ is a non-negative constant and $r$ is the area radius of the spherically symmetric spacetime;
\item[(C3)] there is a number $K\in\real$ and a function $a\in C^2(I,\real_+)$ (where $I$ is an interval) such that the invariantly defined functions $A(t), f(t)$ of (\ref{chi-new-def}) and (\ref{kap-new-def}) satisfy 
\be A(t)=-\frac{K}{a^2(t)},\quad e^{2f(t)} = H^2(t), \label{C3} \ee
where $t$ is a cosmic time function defined by the fluid flow and $H(t)$ is the Hubble function of the FLRW universe with curvature index $K$ and scale factor $a(t)$; 
\item[(C4)] 
\begin{itemize}
\item[(i)] for $K<0$, 
\be \forall t\in I,\quad \lim_{r\to+\infty} P(t,r) = P_{FLRW}(t) \label{C4i} \ee
where $P_{FLRW}$ is the pressure of the FLRW universe with curvature index $K$ and scale factor $a(t)$;
\item[(ii)] for $K>0$, the fluid pressure is homogeneous in the limit $M\to0$:
\be \forall t\in I,\quad \lim_{M\to 0} P(t,r)= P_0(t). \label{C4ii} \ee
\end{itemize}
\end{itemize}
\end{definition}

The analysis of the remainder of this section corresponds to a proof of the following theorem, which summarises the main results of the first half of this paper. 

\begin{theorem}\label{thm1}
Given constants $M\geq0$ and $K\in\mathbb{R}$ and a function $a\in C^2(I,\real_+)$, there is a spacetime $({\mm},g)$ satisfying the conditions $(C1)-(C4)$. This spacetime has the following properties:  
\begin{enumerate} 
\item the line element may be written in the form 
\be ds^2 = -\sig^2 \left(H^{-2}-r^2\kappa^{-1}\right)dt^2 -2\sig r\kappa^{-1}dtdr +\kappa^{-1}dr^2+r^2d\omega^2, \label{mcv-lel} \ee
where $\kappa$ is defined in (\ref{kap-new-def}), $\sig$ is defined by (\ref{sig-k-zero}), (\ref{sig-k-neg}) and (\ref{sig-k-pos}) respectively in the cases $K=0, K<0$ and $K>0$, $\mm=\Omega\times\mathbb{S}^2$ where 
\be \Omega=\{(t,r): \kappa(t,r)>0, t\in I, r\geq0\}\label{omega-def1} \ee
and $d\omega^2$ is the standard line element on the unit sphere; 
\item the function $\sigma$ and hence the spacetime are uniquely determined by conditions (C1)-(C4) in the cases $K\leq 0$; for $K>0$, $\sigma$ is determined up to an arbitrary function of $t$ that depends on the parameter $M$;
\item in the limit $M=0$, the line element is that of an FLRW spacetime with scale factor $a(t)$ and curvature index $K$;
\item the energy density of the spacetime is spatially homogeneous and is given by 
\be 8\pi\mu + \Lambda = 3(H^2+Ka^{-2}), \label{mu-def}\ee
while the  pressure is given by 
\be 8\pi P - \Lambda = -\sig^{-1}\pd{}{t}(H^2+Ka^{-2})-3(H^2+Ka^{-2}), \label{P-def}\ee
and in the limit $M=0$, these give respectively the energy density and pressure of the FLRW universe with scale factor $a$ and curvature index $K$;
\item the fluid expansion is given by $\theta = 3H(t)$;
\item choosing $a(t)$ so that $\mu=P=0$ yields the line element of (i) Schwarzschild spacetime (if $\Lambda=0$ and $K\leq 0$), (ii) Schwarzschild-de Sitter spacetime (if $\Lambda>0$, without restriction on $K$) and (iii) Schwarzschild-anti de Sitter spacetime (if $\Lambda<0$ and $K<0$). 
\end{enumerate}
\end{theorem}

In what follows, we will refer to $M$ as the \textit{(Schwarzschild) mass parameter}, and we will refer to $K, a(t)$ and related quantities as \textit{background} terms. The fluid expansion of the spacetime described by Definition \ref{def1} is identical to that of the FLRW background. Thus references below to the spacetime as expanding or collapsing apply simultaneously to the point mass spacetime and to the corresponding FLRW background. As we will see below, the conclusions of Theorem \ref{thm1} for $K=0$ follow from (C1)-(C3) of Definition \ref{def1}; hence there is no need for a third option in (C4) to cover this case. We now proceed with the proof of Theorem \ref{thm1}.
% \end{comments}
% Now I can refer to part (i) of Definition \ref{def1}.

Shear-free, spherically symmetric perfect fluid spacetimes are described in Section 16.2.2 of \cite{stephani2009exact}. In co-moving coordinates $(\tau,\rho)$, the line element takes the form 
\be ds^2 = -\left(\pd{\lambda}{\tau}\right)^2e^{-2f(\tau)}d\tau^2 +e^{2\lambda}(d\rho^2 + \rho^2 d\omega^2), \label{lel2}\ee
and the fluid flow vector is 
\be \vec{u} =\left(\pd{\lambda}{\tau}\right)^{-1}e^{f(\tau)}\pd{}{\tau}. \label{uvec} \ee
%This form of the line element is invariant under a rescaling of the comoving coordinate $\rho$ and, in addition, under the coordinate transformations 
%\be \rho\to \bar{\rho}=\rho^{-1},\qquad \tau\to\bar{\tau}(\tau), \label{coord-trans}\ee 
%where $\tau\to\bar{\tau}(\tau)$ is an arbitrary $C^1$ function. Under this coordinate transformation, the metric retains the form of (\ref{lel2}):
%\be ds^2 = -\left(\pd{\bar{\lambda}}{\bar{\tau}}\right)^2e^{-2\bar{f}(\bar{\tau})}d\bar{\tau}^2 +e^{2\bar{\lambda}}(d\bar{\rho}^2 + \bar{\rho}^2 d\omega^2). \label{lel3}\ee
%with 
%\be \bar{f}(\bar{\tau})=f(\tau),\qquad e^{\bar{\lambda}}=\rho^2e^\lambda. \label{coord-metric}\ee
The term $f(\tau)$ is an invariant of the spacetime, and is related to the fluid expansion by 
\be \theta = 3e^{f(\tau)}. \label{exp} \ee
The Einstein equations yield (among other conditions, which are explored below) the pressure-isotropy condition
\begin{eqnarray} e^\lambda\left(\pdd{\lambda}{\rho}-\left(\pd{\lambda}{\rho}\right)^2-\frac{1}{\rho}\pd{\lambda}{\rho}\right)= - \phi(\rho),\label{eeq2} 
\end{eqnarray}
where $\phi=\phi(r)$ is an arbitrary function of integration. In spherical symmetry, the Newman-Penrose Weyl scalar $\Psi_2$ is a scalar invariant of the spacetime. It can be shown that in the present case, 
\be \phi(\rho) = 3e^{3\lambda}\Psi_2. \label{phi-psi} \ee
Thus (\ref{weyl-form}) yields $\phi(\rho)=-3M/\rho^3$. We note that defining $x=\rho^2$ and $F(x)=\phi(\rho)/4\rho^2$, this yields
\be F(x) = -(2bx)^{-5/2}, \quad b = \frac12\left(\frac{4}{3M}\right)^{2/5}. \label{Kust} \ee
This functional form of $F$ is a necessary and sufficient condition for the energy density to be homogeneous: $\mu=\mu(\tau)$  (but we note that the interpretation of the constant $b$ is crucial to our purpose.) It follows that we are in the class of spacetimes considered by Kustaanheimo \cite{kust1947}; see Table 16.3 of \cite{stephani2009exact}. Defining $u=(2bx)^{-1/2}e^{-\lambda}$ then leads to the first integral (see Eq.\ (16.42) of \cite{stephani2009exact})
\be \int \frac{du}{\sqrt{\frac23u^3+b^2u^2+3b(e^{2f(\tau)}-\frac{8\pi}{3}\mu(\tau))}}= \frac{\log\rho}{2b} + \beta(\tau), \label{1642} \ee
for some function of integration $\beta(\tau)$ (another function of integration arises in the integrand; this must necessarily equate to $e^{2f}-8\pi\mu/3$ as indicated). The integral on the left hand side here may be written as an elliptic integral - but as is evident, is implicit in the metric function $\lambda$. Sussman \cite{sussman1987spherically} has shown how properties of elliptic functions may be used to invert the functional dependence, and so obtain the solution \textit{explicitly} in terms of elliptic functions of certain co-moving coordinates. We take a complementary approach which we find to be better suited to the understanding of the global structure of the spacetime  (see our comments in Section \ref{sub:elliptic} below). The spacetimes being considered correspond to members of Sussman's NKQ (neutral Kustaanheimo-Qvist) class of metrics \cite{sussman1987spherically}. They are distinct from McVittie's 1933 solutions \cite{mcvittie1933mass}, except for the $K=0$ case.

We observe that the PDE (\ref{eeq2}) can be written in the form
\be \pd{}{\rho}\left\{\rho^2\left(\pd{v}{\rho}\right)^2\right\} = \pd{}{\rho}\left\{v^2-2Mv^3\right\},\label{eeq2a}\ee
where $v=r^{-1}=\rho^{-1}e^{-\lambda}$.
Integrating and rewriting as an equation for $r=r(\tau,\rho)$ yields
\be \rho^2\left(\pd{r}{\rho}\right)^2 = r^2(1-\frac{2M}{r}+A(\tau)r^2),\label{eeq2b}\ee
where $A=A(\tau)$ is an arbitrary function of integration. The significance of the function on the right hand side of this equation will become apparent. We note for future reference that it must be non-negative. We now make the coordinate transformation
\be (\tau,\rho) \to (t,r) = (\tau,\rho e^{\lambda(\tau,\rho)}). \label{coord-transf}\ee
Using (\ref{lel2}) and (\ref{eeq2b}), we find that in these coordinates the line element takes the form 
\be ds^2 = -\sigma^2 \left(e^{-2f(t)}-r^2\kappa^{-1}\right)dt^2 -2\sigma r\kappa^{-1}dtdr +\kappa^{-1}dr^2+r^2d\omega^2, \label{lel4}\ee
where
\be \kappa(t,r) = 1-\frac{2M}{r}+A(t)r^2, \label{chi-def}\ee
and
\be \sigma(t,r) = r^{-1}\pd{r(\tau,\rho)}{\tau}=\pd{\lambda}{\tau}.\label{sigma-def}\ee

The usage $A(t)=A(\tau)$ is consistent on account of the structure of the coordinate transformation (\ref{coord-transf}). This functional dependence of the scalar $A$ may be expressed in the coordinate invariant form $(\delta^\alpha_\beta+u^\alpha u_\beta)\nabla_\alpha A=0$.

The remaining Einstein equations give relations for the energy density  $\mu$ and the isotropic pressure $P$ and an equation for the metric function $\sigma$:
\begin{eqnarray}
8\pi\mu + \Lambda = 3(e^{2f(t)}-A(t)), \label{rho-def} \\
8\pi P - \Lambda = \sigma^{-1}\pd{}{t}(A(t)-e^{2f(t)})+3(A(t)-e^{2f(t)}), \label{press-def}\\
(1-\frac{2M}{r}+A(t)r^2)\pd{\sigma}{r}-(\frac{M}{r^2}+A(t)r)\sigma=\frac12A'(t)r.\label{sig-eqn}
\end{eqnarray}

Apart from isometries of the 2-sphere, the only remaining coordinate freedom in the line element (\ref{lel4}) corresponds to a rescaling of the time coordinate $t$:
\be t\to \bar{t}(t). \label{coord-t}\ee
%The form of the line element is preserved:
%\be ds^2 = -\bar{\sigma}^2 \left(e^{-2\bar{f}(\bar{t})}-r^2\bar{\kappa}^{-1}\right)d\bar{t}^2 -2\bar{\sigma} r\bar{\kappa}^{-1}d\bar{t}dr +\bar{\kappa}^{-1}dr^2+r^2d\omega^2, \label{lel5}\ee
%where
%\begin{eqnarray} \bar{\kappa}(\bar{t},r) &=& 1-\frac{2M}{r}+\bar{A}(\bar{t})r^2, \label{chi-bar-def}\\
%\bar{A}(\bar{t}) &=& A(t),\label{A-bar}\\
%\bar{f}(\bar{t}) &=& f(t),\label{f-bar}\\
%\bar{\sigma}(\bar{t},r) &=& \frac{\sigma(t,r)}{\bar{t}'(t)}. \label{sig-bar}
%\end{eqnarray}
%
%Since the line element is otherwise invariantly defined, 
Consideration of this coordinate freedom shows that the functions $f(t), A(t)$ are scalar invariants of the spacetime, and that the 1-form $\sigma(t,r)dt$ is also an invariant. The invariant nature of the functions $f,A$ is reinforced by their appearance in the scalars
\be
\chi(t,r) := g^{\alpha\beta}\nabla_\alpha r\nabla_\beta r = 1-\frac{2M}{r}+(A(t)-e^{2f(t)})r^2, \label{chi-new-def} 
\ee
and $\kappa(t,r)$, which has the invariant definition
\be 
\kappa(t,r) := h^{\alpha\beta}\nabla_\alpha r\nabla_\beta r = 1-\frac{2M}{r}+A(t)r^2, \label{kap-new-def}
\ee
where
\be h^{\alpha\beta}=g^{\alpha\beta}+u^\alpha u^\beta \label{spatial-metric} \ee
is the (inverse) spatial metric relative to the fluid flow vector $u^\alpha$.
Now observe that the FLRW universe with curvature index $K$, scale factor $a$ and Hubble function $H$ has 
\begin{eqnarray}
\chi_{FLRW} &=& 1-(H^2+\frac{K}{a^2})r^2,\label{chi-flrw}\\
\kappa_{FLRW} &=& 1-\frac{K}{a^2}r^2,\label{kap-flrw}
\end{eqnarray}
and so comparing with (\ref{chi-new-def}) and (\ref{kap-new-def}) with $M=0$ yields the relations of (C3). We note that while a choice of $K$ and $a(t)$ uniquely determines $A$ and $f$ in (\ref{C3}), the converse is not true: for any $\lambda\in\real\setminus\{0\}$, the pairs $(K,a(t))$ and $(\lambda^2K,\lambda a(t))$ both determine the same $A,f$. This rescaling freedom is equivalent to the rescaling freedom in the comoving radial coordinate $\rho$ mentioned above. As is commonly done in FLRW spacetimes, we remove this freedom by specifying, without loss of generality, that $K\in\{-1,0,+1\}$. 

Since the derivative operator in (\ref{kap-new-def}) is \textit{spatial}, we must have the inequality, 
\be \kappa = h^{\alpha\beta}\nabla_\alpha r\nabla_\beta r> 0.\ee 
Thus there is a fundamental restriction to the \textit{allowed region} which we define to be the region 
\be \Omega := \{(t,r): 1-\frac{2M}{r}-K\frac{r^2}{a^2} >0, t\in I, r > 0 \}. \label{allowed-def}\ee
In the case $K=0$, this is the familiar restriction of the McVittie line element to the region $r>2M$. The implications of this restriction are more complicated in the cases when $K\neq 0$, and will be discussed in more detail below. For now, we restrict ourselves to applying standard results on the roots of cubic equations:

\begin{lemma}{\label{lem-allowed}}
In the case $K=-1$, the allowed region has the form 
\be \Omega_{(-1)} = \{(t,r): r>r_{1,(-1)}(t), t\in I\}, \label{allowed-k-neg}\ee
where
\be r_{1,(-1)}(t) = \frac{2}{\sqrt{3}}a(t)\sinh\left(\frac{1}{3}\sinh^{-1}\left(\frac{3\sqrt{3}{M}}{a(t)}\right)\right),\label{r2-plus}\ee
and in the case $K=+1$, the allowed region has the form
\be \Omega_{(+1)} = \{(t,r): r_{1,(+1)}(t)<r<r_{2,(+1)}(t), t\in I\}, \label{allowed-k-pos}\ee
where
\begin{eqnarray}
r_{1,(+1)}&=& \frac{2a}{\sqrt{3}}\cos\left(\rho_0-\frac{2\pi}{3}\right) \in (0,\frac{a}{\sqrt{3}}),\label{r1-pos} \\
r_{2,(+1)}&=& \frac{2a}{\sqrt{3}}\cos\rho_0 \in (\frac{a}{\sqrt{3}},a),\label{r2-pos} \\
\rho_0 &=&\frac{1}{3}\left(\pi-\arccos\left(\frac{3\sqrt{3}M}{a}\right)\right)\in [\frac{\pi}{6},\frac{\pi}{3}].\label{rho0-def}
\end{eqnarray}
\hfill{$\blacksquare$}
\end{lemma}

\begin{lemma}\label{lem:a-m-cond}
For $K=+1$, the allowed region is non-empty at time $t$ if and only if 
\be a(t) >3\sqrt{3}M. \label{a-m-cond}\ee
\hfill{$\blacksquare$}
\end{lemma}

%\fixme{We will assume henceforth that the allowed region is non-empty. Not needed?}

%We note that in the zero-mass limit, the quantities in Lemma \ref{lem-allowed} have the values appropriate to the respective FLRW backgrounds:
%\be \lim_{M\to 0} r_{1,(-1)} = \lim_{M\to 0} r_{1,(+1)} = 0, \lim_{M\to 0} r_{2,(+1)} = a(t). \label{r-bdy-limits} \ee
%This suggests that the boundary of the allowed region might not correspond to a singularity of the spacetime (as in the $K=0$ case, and as suggested by the form of the line element (\ref{mcv-lel})), but may correspond to a coordinate singularity, similar to what one sees in the $K>0$ FLRW spacetime written in area-radius coordinates.

At this stage, the remaining unknown in the line element (\ref{mcv-lel}) is the function $\sig$, which satisfies the PDE (\ref{sig-eqn}). We can (formally) write down the general solution of this linear equation: 
\be \sig= s(t)\kap^{1/2}+\frac12A'(t)\kap^{1/2}\int r\kap^{-3/2} dr. \label{sig-sol}\ee
(We abuse the notation slightly: the function of integration $s(t)$ is implicit in the indefinite integral.) Due to the different nature of the allowed region in the cases $K=0, K<0$ and $K>0$, we treat each case separately.

%We note that having imposed (C3), we can rewrite the expressions (\ref{rho-def}) and (\ref{press-def}) for the energy density and pressure as (\ref{mu-def}) and (\ref{P-def}) respectively. 

\subsection{Uniqueness considerations.}

\subsubsection{The line element for $K=0$.}
This is the only one of the three cases in which the integral of (\ref{sig-sol}) can be evaluated explicitly in terms of elementary functions. From the definitions above, we have $A\equiv 0$ and $\kap=1-2M/r$ in this case, and so we find 
\be \sig = s(t)(1-\frac{2M}{r})^{1/2}. \label{sig-sol-zero} \ee
At this point we can use the coordinate freedom $t\to \bar{t}(t)$ to introduce a new time coordinate satisfying 
\be H(\bar{t})d\bar{t} = s(t)dt. \ee
Using invariance of the 1-form $\sigma dt$ then yields (without loss of generality)
\be {\sig}({t},r) = H({t})(1-\frac{2M}{r})^{1/2}, \label{sig-k-zero}\ee
and so we recover the familiar $K=0$ McVittie line element of \cite{mcvittie1933mass} - \cite{nolan1998point}:
\be ds^2 = -(1-\frac{2M}{r}-H^2r^2)dt^2-2Hr(1-\frac{2M}{r})^{-1/2}dtdr+(1-\frac{2M}{r})^{-1}dr^2+r^2d\omega^2.\ee
This line element follows uniquely under conditions (C1)-(C3) with $K=0$.

%\begin{comments} The fluid flow vector of the spacetimes described by Definition \ref{def1} is given by 
%\be \vec{u} = \sigma^{-1}H\frac{\partial}{\partial t} + rH\frac{\partial}{\partial r}. \label{4-vel} \ee
%Thus $u^\alpha\nabla_\alpha t=\sigma^{-1}H$, and so in the case $K=0$, we have $u^\alpha\nabla_\alpha t=s^{-1}H(1-2M/r)^{-1/2}$. Hence the coordinate choice that yields $s(t)=H(t)$ corresponds to choosing $t$ to be proper time along the fluid flow lines of the FLRW background - i.e.\ $\vec{u}(t)|_{M=0}=1$. With this choice, the Hubble function takes the familiar form $H(t)=a'(t)/a(t)$. 
%\end{comments}

\subsubsection{The line element for $K<0$.}

Our main aim here is to see how the condition (C4) allows us to identify uniquely a line element satisfying the other conditions of Definition \ref{def1}. Two options arise naturally to achieve a unique identification: an asymptotic condition at large $r$ (which would be feasible as the allowed region extends to arbitrarily large values of $r$ in the case $K<0$), or a condition involving the `zero mass' limit $M\to 0$. It is of interest to compare these options. Recall that we can set $K=-1$ without loss of generality. 

In this case, we can write the general solution of (\ref{sig-eqn}) as 
\be \sig = s_M(t)\kap^{1/2} -\frac12A'(t)\kap^{1/2}\int_r^{+\infty} \bar{r}\kap^{-3/2}(t,\bar{r})d\bar{r},\quad r\geq r_{1,(-1)},\label{sig-sol-neg}\ee
where the subscript emphasises that the function of integration may depend upon the parameter $M$.  We can now use the freedom in the choice of the time coordinate to specify that $t$ is the proper time along the fluid flow lines in the background spacetime, and so we may write $H(t)=a'(t)/a(t)$ and $A'(t)=2Ka^{-2}H$.

Since
\be \kap = 1-\frac{2M}{r}+\frac{r^2}{a^2},\ee
we see that 
\be r_{1,(-1)}<2M \quad \hbox{ for all } M>0.\ee

%\fixme{Non sequiter - move. We note that if $a(t)\to+\infty$ as $t\to+\infty$ (which is a case of interest - see Section 4 below), then 
%\be \lim_{t\to+\infty} r_{1,(-1)}(t) =2M. \label{r2-limit} \ee}

For $r>2M$, it is straightforward to show that 
\be \int_r^{+\infty} \bar{r}\kap^{-3/2}(t,\bar{r})d\bar{r}<\frac{a^3}{r},\label{Sig-upr}\ee
whereas just  using positivity of $M$ in the integrand yields
\be \int_r^{+\infty} \bar{r}\kap^{-3/2}(t,\bar{r})d\bar{r}>\frac{a^3}{\sqrt{a^2+r^2}}. \label{Sig-lr}\ee
Thus, for each fixed $t$,
\be \lim_{r\to +\infty} \kap^{1/2}\int_r^{+\infty} \bar{r}\kap^{-3/2}(t,\bar{r})d\bar{r} = a^2(t). \label{int-limit}\ee
This yields (for each fixed $t\in I$)
\be \sig(t,r) \sim H(t) +\frac{s_M(t)}{a(t)}r,\quad r\to+\infty. \label{sig-limit}\ee
It follows from (\ref{P-def}) that for each fixed $t$ with $s(t)\neq 0$ we have
\be \lim_{r\to\infty} 8\pi P(t,r)-\Lambda = -3(H^2+Ka^{-2}).\ee
This does not align with the behaviour we are seeking to model: what we look for is that for each fixed $t$, 
\be \lim_{r\to\infty} 8\pi P(t,r)-\Lambda = 8\pi P_{FLRW} -\Lambda,\ee
where $P_{FLRW}$ is the pressure of the FLRW background with scale factor $a$ and curvature index $K$. This would necessitate
\begin{eqnarray}
\lim_{r\to\infty} 8\pi P(t,r)-\Lambda &=& -H^{-1}\partial_t(H^2+Ka^{-2})-3(H^2+Ka^{-2}). \label{P-limit}
\end{eqnarray}
Comparing (\ref{sig-limit}) and (\ref{P-def}), we see that the required condition on $s$ is $s_M(t)\equiv 0$. Thus the function $\sig$ of the line element (\ref{mcv-lel}) is given, in this case, by
\be \sig(t,r) = \frac{H}{a^2}\left(1-\frac{2M}{r}+\frac{r^2}{a^2}\right)^{1/2}\int_r^\infty \bar{r}\left(1-\frac{2M}{\bar{r}}+\frac{\bar{r}^2}{a^2}\right)^{-3/2} d\bar{r}. \label{sig-k-neg}\ee

While this gives a satisfactory way of identifying a unique solution in the case $K<0$, we cannot apply this argument in the case $K>0$, where at each time $t$, the area radius $r$ has a finite maximum value. Thus it is of interest to seek an alternative route to uniqueness, which nevertheless maintains a physical interpretation. An option that naturally suggests itself is to consider the behaviour of the pressure in the limit $M\to 0$. This requires the following result, which can be established by applying the dominated convergence theorem to interchange the limit and integral (see e.g. \cite{apostol1974mathematical}). 
%We note that a little care is needed to take account of the fact that the integrand is not defined for all $r>0$ and all $M>0$; this problem can be circumvented by starting with an initial value of $M$ that is sufficiently small. 

\begin{lemma}\label{M-lim-neg} For $K<0$ and for all $r>0$, 
\be \lim_{M\to 0} \int_r^{+\infty} \bar{r}\kap^{-3/2}(t,\bar{r})d\bar{r} = a^2(1+\frac{r^2}{a^2})^{-1/2},
\label{integral-limit-neg} \ee
and hence
\be \lim_{M\to 0} \sig(t,r) = (1+\frac{r^2}{a^2})^{1/2}s_0(t)+H, \label{sig-lim-neg} \ee
where $s_0(t)=\lim_{M\to 0}s_M(t)$. 
\hfill{$\blacksquare$}
\end{lemma}

We can then write down this result:

\begin{lemma}\label{lem:equiv} For $K<0$, (i) and (ii) below are equivalent, (iii)-(v) and equivalent and (i) implies (iii).
\begin{itemize}
\item[(i)] $\lim_{r\to+\infty} P(t,r) = P_{FLRW}$;
\item[(ii)] $s_M(t)=0$ for all $M\geq0$;
\item[(iii)] $s_0(t)=0$;
\item[(iv)] $\lim_{M\to 0} P(t,r) = P_0(t)$ (i.e. the pressure is homogenous in the limit);
\item[(v)] $\lim_{M\to 0} P(t,r) = P_{FLRW}$.
\end{itemize}
\hfill{$\blacksquare$}
\end{lemma}

%\begin{comments}
%We see that in the case $K<0$, condition (C4-i) fully specifies the function of integration $s(t)$, and yields a unique solution of (\ref{sig-eqn}) and hence a unique line element. However, the `zero-mass limit' condition of item (iv) in Lemma \ref{lem:equiv} is not sufficient to obtain a unique solution. We expand on this point below.
%\end{comments}

\subsubsection{The line element for $K>0$.} Taking $K=+1$ without loss of generality, we have
\be \kap = 1-\frac{2M}{r}-\frac{r^2}{a^2}, \label{kap-k-pos} \ee
and the allowed region is given by (\ref{allowed-k-pos}). As in the previous section, our aim here is to see how the condition (C4) picks out solutions of (\ref{sig-eqn}), and thus yields a unique class of line elements. In writing down the general solution of this PDE for $K<0$, we used the fact that $r$ can be arbitrarily large to use the form (\ref{sig-sol-neg}), which then allowed us to interpret $s$ and apply the appropriate boundary condition. In the present case, we seek to do something similar. This requires that we identify a `preferred' value of $r$ analogous to the way that $r\to+\infty$ is `preferred' for $K<0$. We observe that
\be \kappa|_{r=3M}=\frac13-\frac{9M^2}{a^2}. \label{kap-3m}\ee
This quantity is positive precisely when the allowed region is non-empty. In other words, for all relevant values of $t$, $r_{1,(+1)}(t)<3M<r_{2,(+1)}$ (see also the discussion in Section \ref{allowed boundary} from Lemma \ref{lem:sig-zero} onwards, and Section \ref{positive curvature}). It follows that the integrand below is real for all $r$ with $ r_{1,(+1)}<r<r_{2,(+1)}$ and so 
%We choose the value of $r\in(r_{1,(+1)},r_{2,(+1)})$ at which $\kap$ attains its maximum. This unique maximum is easily found to be 
%\be r_*(t) = (Ma^2)^{1/3}. \label{rst-def} \ee
we can write the general solution of (\ref{sig-eqn}) as 
\be \sig(t,r) = s_M(t)\kap^{1/2}+a'a^{-3}\kap^{1/2}\int_{3M}^r \bar{r}\kap^{-3/2}(t,\bar{r})d\bar{r}, \quad r_{1,(+1)}<r<r_{2,(+1)}.
\label{sig-sol-pos} \ee
As in the $K<0$ case, we have specified the time coordinate to be proper time along the FLRW background fluid flow lines, and so $H=a'/a$.

Corresponding to Lemma \ref{M-lim-neg}, we have the following result:
%The proof is essentially the same: for a given $r\in(0,a)$, we choose a sufficiently small initial value of $M$ to ensure that the integrand is defined and bounded throughout the domain of integration. Additional care is needed to take account of the fact that the lower limit of integration depends on $M$. 

\begin{lemma}\label{M-lim-pos} For $K>0$ and for all $r\in(0,a)$, 
\be \lim_{M\to 0} \int_{3M}^r \bar{r}\kap^{-3/2}(t,\bar{r})d\bar{r} = \int_0^r \bar{r}(1-\frac{\bar{r}^2}{a^2})^{-3/2} d\bar{r} \label{integral-limit-pos} \ee
and hence
\be \lim_{M\to 0} \sigma(t,r) = (s_0-H)(1-\frac{r^2}{a^2})^{1/2}+H. \label{sig-lim-pos} \ee
\hfill{$\blacksquare$}
\end{lemma}

Then we see from (\ref{P-def}) that the condition (C4-ii) yields the boundary condition $s_0=H$. Thus for $K>0$, we have 
\be \sig(t,r) = s_M(t)\kap^{1/2}+a^{-2}H\kap^{1/2}\int_{3M}^r \bar{r}\kap^{-3/2}(t,\bar{r})d\bar{r}, \quad r_{1,(+1)}<r<r_{2,(+1)}
\label{sig-k-pos} \ee
with $s_0(t)=H(t)$. However, $s_M(t)$ may include $O(M)$ terms. These must be of a form that ensures that $s_M(t)$ is an invariant function of the proper time coordinate $t$. This suggests that it should be constructed from combinations of $Ka^{-2}$ and $H$ and their derivatives - but there remains considerable freedom in this choice. It is difficult to identify a meaningful condition in the $K>0$ case that would yield a unique spacetime, and we will not make any proposal in this regard. However, even with the freedom remaining in the choice of the function $s_M(t)$, we can identify universal global features of this family of spacetimes, as we will see below.

\subsection{Proof of Theorem \ref{thm1}}

We note at this point that the proof of parts (i) - (v) of Theorem \ref{thm1} is complete (see (\ref{exp}) for part (v)). The form of the line element is established in (\ref{lel4}), with the condition (C3) giving the form of $f(t)$. Lemma \ref{M-lim-neg} and Lemma \ref{M-lim-pos} (along with a corresponding trivial result in the case $K=0$) give the limit of $\sig$ when $M\to 0$, and thus prove parts (iii) and (iv) of the theorem. 

\subsection{The matter-free limit}

Thus it remains to prove part (vi) of Theorem \ref{thm1}. This is straightforward. From (\ref{mu-def}) and (\ref{P-def}), we see that the necessary and sufficient condition for the spacetime to be matter free (i.e.\ $\mu=P=0$) is that 
\be H^2+Ka^{-2} = \frac{\Lambda}{3}. \label{matter-free} \ee
Recalling that $H=a'/a$, we see that this ODE has solutions for $a(t)$ (i) when $\Lambda = 0$ and $K\leq 0$; (ii) when $\Lambda >0$ for any value of $K$ and (iii) when $\Lambda <0$ and $K<0$. The conclusion that the resulting line element is that of Schwarzschild ((-anti)-de Sitter) spacetime follows from the `Birkhoff-with-$\Lambda$' theorem (see e.g.\ \cite{schleich2010simple}).  
%\newpage

\subsection{Representation by elliptic integrals}\label{sub:elliptic}
As we see from (\ref{mcv-lel}), (\ref{sig-k-zero}), (\ref{sig-k-neg}) and (\ref{sig-k-pos}), the line elements that arise from Definition \ref{def1} can be written either in terms of elementary functions (when $K=0$) or in terms of the integrals arising in (\ref{sig-k-neg}) and (\ref{sig-k-pos}) when $K=\pm1$. We point out here that these latter integrals can be written in terms of elliptic integrals. In the case $K<0$, the relevant transformations are given in Lemma \ref{sig-at-bdy} - see e.g.(\ref{sig-form1}), (\ref{Sig-int-def}). The integral appearing here can be expressed in terms of elliptic integrals of the first and second kind (see e.g.\ chapter 3 of \cite{gradshteyn2014table}). We emphasise that this situation is qualitatively different to that of \cite{nolan1998point} summarised above, where the $K<0$ solution was written as an \textit{implicit} function of an elliptic integral (cf. (\ref{1642})). As noted above, Sussman \cite{sussman1987spherically} has shown how to write the metric \textit{explicitly} in terms of elliptic functions. Furthermore, he applied this in a subsequent analysis of the global structure of shear-free, spherically symmetric perfect fluid spacetimes \cite{sussman1988spherically}. However, it seems that the use of co-moving coordinates is not ideal for this purpose, as some mis-identifications of boundary surfaces arose (\textit{cf.} \cite{nolan1999point}, p.1232). In the present paper, the solution is also written explicitly as an elliptic integral, but in non-comoving coordinates. As we will see below, this allows us to probe the global structure of the spacetime in a way that is not possible with the implicit representation of \cite{nolan1998point}. Before proceeding with the analysis of the global structure, we must give further details on the FLRW background spacetimes. 

%%%%%%%%%%%%%%%%%%%%%%%%%%%%%%
%%%%%%%%%%%%%%%%%%%%%%%%%%%%%%
%%%%%%%%%%%%%%%%%%%%%%%%%%%%%%

\section{More on the background}\label{background} 

In Definition \ref{def1}, we gave conditions on a spherically symmetric spacetime intended to capture the idea of a point mass embedded in an otherwise isotropic universe. In order to specialise to those spacetimes of most interest to cosmology, we place some constraints on the background scale factor. Consistent with \cite{nolan2014particle}, we make the following definition. 

\begin{definition}\label{def2} A spacetime $({\mm},g)$ representing a point mass in an expanding universe, in the sense of Definition \ref{def1}, is said to be \textbf{an initially expanding McVittie spacetime with a Big Bang background} if the following conditions on the scale factor $a$ hold: there exists $t_f>0$ such that
\begin{itemize}
\item[(i)] $a\in C^2((0,t_f),\mathbb{R}_+)$;
\item[(ii)] $\lim_{t\to 0^+} a(t)=0$;
\item[(iii)] the dominant and strong energy conditions hold in the background: $\mu_{FLRW}(t)>0$, $\mu_{FLRW}(t)+P_{FLRW}(t)>0$ and $\mu_{FLRW}(t)+3P_{FLRW}(t)>0$ for all  $t\in (0,t_f)$;
\item[(iv)] if $t_f<+\infty$, then $\lim_{t\to t_f^-} a(t) = 0$;
\item[(v)] the following limiting behaviour applies:
\be \lim_{a\to 0^+} aH = +\infty, \qquad 
(aH)^2 \sim -K + \frac{\Lambda}{3}a^2,\quad a\to+\infty. \label{ah-lims} 
\ee
\end{itemize}
\end{definition}

We record the following: 
\begin{lemma}\label{lem:dec} The energy density and pressure of the FLRW background satisfy 
\begin{eqnarray} 8\pi\mu_{FLRW} + \Lambda &=& 3(H^2+Ka^{-2}), \label{flrw-density} \\
8\pi P_{FLRW} -\Lambda &=& -2H'-3H^2-Ka^{-2}. \label{flrw-pressure} 
\end{eqnarray}
Consequently, the dominant energy condition holds if and only if 
\be H^2+Ka^{-2} -\frac{\Lambda}{3} >0 \quad  \hbox{ and } \quad H'-Ka^{-2} < 0; \label{DEC} \ee
the strong energy condition holds if and only if 
\be H'+H^2-\frac{\Lambda}{3} < 0 \quad  \hbox{ and } \quad H'-Ka^{-2} < 0. \label{SEC} \ee
\hfill$\blacksquare$
\end{lemma}

In Definition \ref{def2}, the first term is technical, giving the level of differentiability required for the analysis below. It also guarantees a positive scale factor. The second item is the Big Bang condition. The second and third conditions enforce an `initially expanding' scenario (see the following corollary), and the fourth implies that recollapse leads to a Big Crunch rather than any alternative. We address the final condition below. 

\begin{corollary}\label{cor:init-exp}
If $K\leq 0$, then the universe is initially expanding. That is, there exists $t_\epsilon>0$ such that $H(t)>0$ on $(0,t_\epsilon)$.  
\end{corollary}

\noindent\textbf{Proof:} For $K\leq 0$, part (iii) of Definition \ref{def2} yields $H'(t)<0$ for all $t\in(0,t_f)$ and so $H(t_1)>H(t_2)$ for all $0<t_1<t_2<t_f$. For any $t_3\in(0,t_f)$, we have $a(t_3)>0$, and the mean value theorem then yields $t_\epsilon\in(0,t_3)$ with $H(t_\epsilon)>0$. Then $H(t)>H(t_\epsilon)$ for all $t\in(0,t_\epsilon)$ as required. \hfill$\blacksquare$

The conditions of Definition \ref{def2} severely limit the future evolution of flat and negatively curved isotropic spacetimes:

\begin{corollary}\label{cor:a-limits}
If $K\leq 0$, then either the universe is expanding for all $t>0$, or it reaches a point of maximum expansion and collapses thereafter. That is, either
\begin{itemize}
\item[(a)] $t_f=+\infty$ and $H(t)>0$ for all $t>0$, or
\item[(b)] $t_f<+\infty$ and there exists $t_{\rm{max}}>0$ such that $H(t)>0$ for $t\in(0,t_{\rm{max}})$ and $H(t)<0$ for $t\in(t_{\rm{max}},t_f)$. In this case, we must have $\Lambda<0$. 
\end{itemize}
\end{corollary}

\noindent\textbf{Proof:} By Corollary \ref{cor:init-exp}, $H=a'/a$ is initially positive. Either $a'$ remains positive (and (a) holds; note that a finite value of $t_f$ necessitates $a'$ becoming negative), or the function $a$ has a stationary point: there exists $t_0>0$ such that $a'(t_0)=0$. The weak energy condition (item (iii) of Definition \ref{def2}), with $K\leq 0$, then shows that this must be a local maximum of $a$ - and that there cannot be a local minimum. Thus $t=t_0$ corresponds to a global maximum of $a$, and option (b) holds. The condition $a''(t)<0$, which follows from the weak energy condition, and the existence of a collapsing phase, forces $a$ to decrease to $0$ in finite time. Thus $t_f<+\infty$.  Finally, we note that since $K\leq 0$, (\ref{DEC}) gives $H^2>\frac{\Lambda}{3}$, and so case (b) cannot arise if $\Lambda\geq 0$. \hfill$\blacksquare$

Within the class of spacetimes described by Definition \ref{def2}, we focus briefly on those backgrounds which are of particular importance in relation to standard cosmological models. We consider perfect fluids satisfying a linear equation of state $P=\omega \mu$ with $\omega\in(-\frac13,1]$, and perfect fluids in which the energy density is a non-interacting combination of dust and radiation. The latter are of particular importance in the $\Lambda CDM$ model. In these cases, the Friedmann equation (\ref{flrw-density}) may be written as, respectively, 
\begin{eqnarray} 
H^2 &=& -Ka^{-2} +\frac{\Lambda}{3} +\frac{C_\omega}{a^{3+3\omega}}, \label{hub-bar} \\
H^2 &=& -Ka^{-2} +\frac{\Lambda}{3} +\frac{C_m}{a^3} +\frac{C_r}{a^4}, \label{hub-mix} 
\end{eqnarray}
where $C_\omega,C_m,C_r$ are non-negative constants. In both cases, the limiting behaviour of (\ref{ah-lims}) applies. Our assumption that the strong energy condition holds rules out the accelerated expansion of the inflationary period (at least when $\Lambda=0$). Furthermore, other behaviours of the scale factor are both possible and important in cosmological modelling. However, in order to (i) maintain contact with important classes of models and (ii) limit the number of cases that are to be considered, we restrict to such cases as display the limiting behaviour of (\ref{ah-lims}). Hence this condition is included in our definition.

\begin{definition}\label{def3}
A \textbf{McVittie spacetime with an eternally expanding background} is a spacetime satisfying the conditions of Definition \ref{def2}, with $t_f=+\infty$ and $H(t)>0$ for all $t>0$. A \textbf{McVittie spacetime with a recollapsing background} is a spacetime satisfying the conditions of Definition \ref{def2} and with  $t_f<+\infty$.
\end{definition}

%As we have seen in the previous section, the $K>0$ McVittie spacetimes contain problematic singularities. The global structure in the case $K=0$ is well-understood. \textit{Thus in the remainder of this paper, we restrict to the case $K<0$}.
%
%\noindent\textbf{Notation:}As we have restricted to the case $K<0$ (so without loss of generality, $K=-1$), we adopt the notation $r_1=r_{1,(-1)}$ for the remained of the paper.

%%%%%%%%%%%%%%%%%%%%%%%%%%%%%%
%%%%%%%%%%%%%%%%%%%%%%%%%%%%%%
%%%%%%%%%%%%%%%%%%%%%%%%%%%%%%

\section{The boundary of the allowed region}\label{allowed boundary}

In this and the following sections, we consider the existence or otherwise of singular boundaries in the non-flat McVittie spacetimes. We compare and contrast with the corresponding situation in the flat McVittie case. We recall that in the flat case the surface $r=2M$ forms a singular boundary in two different ways: in every expanding $K=0$ McVittie spacetime with a Big Bang background (see Definition 4.1 of \cite{nolan2014particle}), every future directed causal geodesic of the spacetime reaches $r=2M$ at a finite (affine/proper) time in the past. In the case of a vanishing cosmological constant, there are families of ingoing radial null geodesics which meet $r=2M$ at a finite affine time in the future. We consider the corresponding boundaries in the non-flat cases. In the case $K<0$, we find that there is no corresponding divergence of scalar curvature terms along the boundary of the allowed region. See Proposition \ref{prop:k-neg-allowed-bdy} below. In the case $K>0$, there is a spacelike hypersurface that threads through the allowed region, along which the pressure diverges. See Proposition \ref{prop:k-pos-allowed-bdy}. 

We recall that the spacetime manifold is $\mm=\Omega\times \mathbb{S}^2$ where for $K<0$, 
\be \Omega = \Omega_{(-1)}
= \{(t,r):  r_{1(-1)}(t)<r, 0<t<t_f\}, \label{omega-def-neg} \ee
and for $K>0$,
\be \Omega = \Omega_{(+1)}
= \{(t,r):  r_{1(+1)}(t)<r<r_{2(+1)}(t), 0<t<t_f\}. \label{omega-def-pos} \ee
The boundary of $\Omega$ has the following disjoint decomposition:
\be \partial\Omega = {\cal{O}} \cup \partial\Omega_{\{t=0\}} \cup \partial\Omega_{\{\kappa=0\}}, \label{omega-bdy} \ee
where 
\begin{eqnarray}
{\cal{O}} &=& \{(t,r):t=r=0\},\label{bdy-O}\\
\partial\Omega_{\{t=0\}} &=& \{(t,r):t=0, r>0\}, \label{bdy-t0} \\
\partial\Omega_{\{\kappa=0\}} &=& \{(t,r): \kappa(t,r)=0, 0<t<t_f\}. \label{bdy-kap} 
\end{eqnarray}
(In fact we will find that ${\cal{O}}$ and $\partial\Omega_{\{t=0\}}$ are empty in the case $K>0$.) We will refer to $\omk$ as \textbf{the boundary of the allowed region} (or \textit{allowed boundary} in some instances). In the case $K<0$, this is the set
\be \partial\Omega_{\{\kappa=0\}} = \{(t,r): r=r_{1,(-1)}(t), t>0\}, \label{bdy-neg}\ee
and in the case $K>0$, this set has two components:
\begin{eqnarray} \partial\Omega_{\{\kappa=0\}}
&=& \{(t,r): r=r_{1,(+1)}(t)\} \cup \{(t,r): r=r_{2,(+1)}(t)\}. \label{bdy-pos}
\end{eqnarray}

In the case $K=0$, it is immediate from (\ref{P-def}) and (\ref{sig-k-zero}) that the pressure diverges at the boundary of the allowed region (i.e.\ at $r=2M$). Comparing (\ref{sig-k-neg}) and (\ref{sig-k-pos}), it is not clear that this carries over to the non-flat cases: in fact it does not. Two different situations emerge in the cases of $K<0$ and $K>0$. We consider $K<0$ first. To evaluate the pressure at $\sigma=0$, we require this lemma which will also be crucial for the analysis of the radial null geodesics of the spacetime.  %This is the first indication that these boundaries do not correspond to singularities of the spacetime. 

\begin{lemma}\label{sig-at-bdy}
Let $K<0$ and define
\be x=\frac{a}{r},\quad \alpha=\frac{a}{r_{1,(-1)}},\label{x-alpha-def} \ee
so that the allowed region corresponds to $0<x<\alpha$, and the boundary of the allowed region corresponds to $x=\alpha$. 
Then the function $\sig(t,r)$ defined by (\ref{sig-k-neg}) satisfies
\be \sig = \sig_0(t) + \sig_1(t)(\alpha-x)^{1/2} + \sig_2(t)(\alpha-x) + \sigr, \label{sig-decomp} \ee
where 
\be \sigr\in C^1(\Omega\cup\balr), \qquad \sigr=O((\alpha-x)^{3/2}),\quad x\to\alpha^-.\ee
Hence 
\be \sig \in C^0(\Omega\cup\balr)\cap C^1(\Omega). \label{sig-diff} \ee
The coefficients $\sig_0,\sig_1$ and $\sig_2$ are given by 
\begin{eqnarray}
\sigma_0 &=& \frac{2H}{\nu\alpha^3}(1+2\epsilon)^{-1} = 2H\frac{r_{1,(-1)}^2}{a^2+3r_{1,(-1)}^2}, \label{sig0-def} \\
\sigma_1 &=& \frac{H}{\nu\epsilon^{3/2}\alpha^{7/2}}(1+2\epsilon)^{1/2}\bar{K}(\epsilon), \label{sig1-def} \\
\sigma_2 &=& -\frac{6H}{\nu\alpha^4}(1+2\epsilon)^{-2},\label{sig2-def}
\end{eqnarray}
where
\be \nu = \frac{2M}{a},\quad \beta = \alpha - \nu^{-1},\quad \epsilon = \frac{\beta}{\alpha} \label{nu-beta-ep-def} \ee
and 
\be \bar{K}(\epsilon) = 3\epsilon^{3/2}J(\epsilon) -2, \label{kbar-def} \ee
with 
\be J(\epsilon) = \int_0^1 (1-\xi)^{-1/2}(2\xi+\epsilon)(\xi^2+\epsilon\xi+\epsilon)^{-5/2}d\xi. \label{j-def} \ee
\qedapp
\end{lemma}

Combining with (\ref{P-def}), this yields the following results:

\begin{proposition}
The pressure of a spacetime representing a point mass in an isotropic universe with $K<0$ satisfies 
\be (8\pi P-\Lambda)|_{\partial\Omega_{\{\kappa=0\}}} = -\frac{a^2+3r_{1,(-1)}^2}{r_{1,(-1)}^2}(H'-Ka^{-2})-3(H^2+Ka^{-2}), \label{P-finite-neg}\ee
which is finite away from singularities of the FLRW background. 
\qed
\end{proposition}

%\noindent\textbf{Proof:} This follows from the previous result and from the expression (\ref{P-def}) for the pressure. \hfill$\blacksquare$

%We can then write down the following result:

\begin{proposition}\label{prop:k-neg-allowed-bdy}
In a  McVittie spacetime with $K<0$, the energy density and pressure are finite at all points of $\{(t,r): t\in I, r\geq r_{1,(-1)}\}$, where $I$ is the maximal interval on which the energy density and pressure of the FLRW background are finite. \hfill{$\blacksquare$}
\end{proposition}

In the case $K>0$, we have a similar result relating to the limiting values of $\sig$ at the boundaries of the allowed region. However, the conclusion regarding the pressure is different from both the $K=0$ and $K<0$ cases. 

\begin{lemma}\label{lem:sig-zero} In a McVittie spacetime with $K>0$ and background scale factor $a(t)$, defined on a maximal interval $I$, let $I_0$ be the subset of $I$ for which $a(t)>3\sqrt{3}M$ and $H(t)\neq0$. Then for each $t\in I_0$, there exists a unique value $r_0(t)$ of $r$ with $r_0(t)\in(r_{1,(+)},r_{2,(+)})$ such that 
\be \sig(t,r_0(t)) = 0. \label{sig-is-zero} \ee
\qedapp
\end{lemma}

\begin{proposition}\label{prop:k-pos-allowed-bdy}
In a McVittie spacetime with $K>0$, there exists a spacelike hypersurface $\Sigma_0=\{(t,r): t\in I_0, r=r_0(t)\}$, which is a subset of the allowed region, along which the pressure is infinite. The weak energy condition is violated throughout one of the regions ${\cal{R}}_1$ or ${\cal{R}}_2$, where 
\begin{eqnarray} {\cal{R}}_1 &=& \{(t,r): t\in I_0, r_{1,(+)}<r<r_0\},\\
{\cal{R}}_2 &=& \{(t,r): t\in I_0, r_0<r<r_{2,(+)}\}.
\end{eqnarray}
\end{proposition}

\noindent\textbf{Proof:}  The existence of $\Sigma_0$ follows from the previous lemma, and its spacelike nature then follows immediately from (\ref{mcv-lel}). Since the weak energy condition is satisfied in the background, the vanishing of $\sig$ in (\ref{P-def}) cannot be compensated by vanishing of $H'-Ka^{-2}$, and so the pressure diverges on $\Sigma_0$. From (\ref{mu-def}) and (\ref{P-def}), we see that 
\begin{eqnarray} 8\pi(\mu+P) &=& -2H\sig^{-1}(H'-Ka^{-2}). \label{DEC-fail}
\end{eqnarray}
The background energy condition yields $H'-Ka^{-2}<0$ (see (\ref{DEC})), and at each time $t\in I_0$, the coefficient of this term has the same sign in each of ${\cal{R}}_{1}$ and ${\cal{R}}_{2}$, but must be positive in one and negative in the other (this follows from Lemma \ref{lem:sig-zero}). The conclusion follows. \hfill{$\blacksquare$}

%%%%%%%%%%%%%%%%%%%%%%%%%%%%%%
%%%%%%%%%%%%%%%%%%%%%%%%%%%%%%
%%%%%%%%%%%%%%%%%%%%%%%%%%%%%%

%\begin{comments}
%With the definition 
%\be \kappa := h^{\alpha\beta}\nabla_\alpha r\nabla_\beta r = 1-\frac{2M}{r}-K\frac{r^2}{a^2}, \ee
%the boundary of the allowed region $\omk=\{(t,r):\kappa(t,r)=0\}$ generalises to the case $K=0$ the surface $\{r=2M\}$ of the spatially flat McVittie model. However, we see that it plays a very different role in the non-flat cases. In particular, there is no scalar curvature singularity along $\omk$. As seen in Proposition \ref{prop:k-pos-allowed-bdy}, in the case $K>0$, there is instead a  scalar curvature singularity along the hypersurface $\{(t,r):\sigma(t,r)=0\}$. 

Thus we see that the problems analogous to those associated with the existence of the singularity along $\{r=2M\}$ in the flat McVittie model persist in the $K>0$ model, but are absent in the $K<0$ model. For this reason, we will focus for the remainder of this paper on the global structure of the $K<0$ model. \textbf{That being the case, we simplify the notation from this point onwards by taking $\Omega\equiv\Omega_{(-1)}$ (see (\ref{omega-def-neg})) and $r_1(t)\equiv r_{1,(-1)}(t)$ (see (\ref{r2-plus}))}.  We will briefly review some of the key features of the global structure of $K>0$ McVittie spacetimes at the end of the paper. 
%\end{comments}
%%%%%%%%%%%%%%%%%%%%%%%%%%%%%%
%%%%%%%%%%%%%%%%%%%%%%%%%%%%%%
%%%%%%%%%%%%%%%%%%%%%%%%%%%%%%

\section{Radial null geodesics and the horizon}\label{rngs-horizon}

In this brief section, we collect some key definitions and observations that play a role in the analysis below. We begin by noting that $t$ is a global time coordinate in any McVittie spacetime as can be seen from the line element (\ref{mcv-lel}). Thus $t$ can be used as a parameter along any causal geodesic. In particular, we can consider the radial null geodesics (RNGs) as curves in the $t-r$ plane and so several of the key features of RNGs can be determined by analysing the equations
\be \frac{dr}{dt}= \sigma(r\pm\kappa^{1/2}H^{-1}) \label{rng} \ee
which are derived by setting $ds=d\theta=d\phi=0$ in (\ref{mcv-lel}). (We will use the same term RNG to refer to either the relevant parametrised curves $s\mapsto (t(s),r(s))\in\Omega$ and to the solution curves of (\ref{rng}) in the $t-r$ plane.) There are two families of RNGs corresponding to the two signs: following the usual designation, we define \textit{outgoing} RNGs (ORNGs) to be radial null geodesics $r=r_o(t)$ satisfying
\be \frac{dr}{dt}= \sigma(r+\kappa^{1/2}H^{-1})=: F_o(t,r), \label{orng} \ee 
and we define \textit{ingoing} RNGs (IRNGs) to be radial null geodesics $r=r_i(t)$ satisfying
\be \frac{dr}{dt}= \sigma(r-\kappa^{1/2}H^{-1})=: F_i(t,r), \label{irng} \ee 

We study the geodesics on the domain $\Omega$ and we note that 
\be F_o(t,r), F_i(t,r) \in C^1(\Omega,\mathbb{R}). \label{F-is-C1} \ee
%The boundary of $\Omega$ has the following disjoint decomposition:
%\be \partial\Omega = {\cal{O}} \cup \partial\Omega_{\{t=0\}} \cup \partial\Omega_{\{\kappa=0\}}, \label{omega-bdy} \ee
%where ${\cal{O}}=\{(t,r):t=r=0\}$, $\partial\Omega_{\{t=0\}} =\{(t,r):t=0, r>0\}$ and $\partial\Omega_{\{\kappa=0\}}$ is defined in (\ref{bdy-neg}). 
The \textit{horizon} $\hor$ is defined to be the subset of $\Omega$ along which $\chi$ vanishes:
\be \hor=\{(t,r)\in\Omega:\chi(t,r)=0\}. \label{hor-def} \ee
Using (C3) of Definition \ref{def1}, (\ref{chi-new-def}) and (\ref{kap-new-def}), we can write
\be \chi = \kappa - r^2H^2. \label{eq:chi-kappa} \ee
We define the \textit{regular} region $\Omega_R$, the \textit{trapped} region $\Omega_T$ and the \textit{anti-trapped} region $\Omega_A$ as follows:
\begin{eqnarray}
\Omega_R &=& \{(t,r)\in\Omega: \chi>0\}, \label{def:reg-region}\\
\Omega_T &=& \{(t,r)\in\Omega: \chi<0 \wedge H<0\}, \label{def:trap-region}\\
\Omega_A &=& \{(t,r)\in\Omega: \chi<0 \wedge H>0\}. \label{def:anti-region}
\end{eqnarray}
We then have the (disjoint) decomposition
\be \Omega = \Omega_R\cup\Omega_A\cup\Omega_T\cup\hor. \label{omega-regions} \ee

Note that $\Omega_A$ (respectively $\Omega_T$), if non-empty, is a subset of the expanding (respectively collapsing) region of the spacetime. In $\Omega_R$, the proper radius increases (respectively decreases) with $t$ along outgoing (respectively ingoing) RNGs. In $\Omega_A$, $r$ increases along both families of RNGs and in $\Omega_T$, $r$ decreases along both families of RNGs. 

The structure of the horizon plays an important role in the following sections, and so we discuss this now (the horizons of spatially flat McVittie spacetimes have been studied in \cite{nolan1999point, kaloper2010mcvittie, lake2011more, faraoni2012making, afshordi2014horndeski}). As $K<0$ (and without loss of generality, equals $-1$), the horizon is defined by the equation
\be 1-\frac{2M}{r}-(H^2-a^{-2})r^2 =0.\label{ho-neg}\ee 
%Using this equation, it is straightforward to show that there is a horizon present at time $t$ if and only if
%\be M\sqrt{H^2(t)-a^{-2}(t)} <\frac{1}{3\sqrt{3}}. \label{hor-con} \ee
%Note that (\ref{sum-pos}) ensures existence of the root on the left hand side. 
Locally, the horizon can be represented in the form $r=r_h(t)$, and implicit differentiation yields
\be \left(1-\frac{3M}{r_h}\right)\frac{dr_h}{dt} = -Hr_h^3(H'+a^{-2}). \label{rh-diff} \ee
We know from the energy condition (\ref{DEC}) that $H'+a^{-2}<0$, and so (\ref{rh-diff}) tells us the sign of the derivative $r'_h(t)$ in terms of the sign of $H$ and of $r_h-3M$. Thus it is clear that the structure of the horizon depends on whether the universe expands for all time, or if recollapse occurs, so we discuss these cases separately. 

%%%%%%%%%%%%%%%%%%%%%%%%%
\subsection{The horizon in an eternally expanding background}
%%%%%%%%%%%%%%%%%%%%%%%%%

In this case, we have the following. 

\begin{lemma}\label{lem:hor-a}
Let $t_f=+\infty$ so that $H(t)>0$ for all $t>0$. Then there exists $H_0\geq 0$ such that
\be \lim_{t\to +\infty} (a(t),H(t)) = (+\infty,H_0). \label{a-h-lim}\ee
\hfill$\blacksquare$
\end{lemma}

Here, the structure of the horizon is essentially the same as in the $K=0$ case as described in \cite{nolan2014particle}. We summarise the key features: see \cite{nolan2014particle} for the relevant derivations. The energy conditions (\ref{DEC}) with $\Lambda\geq 0$ yield positivity of $H^2-a^{-2}$, implying that the horizon can be described by a positive, monotone function $v:(0,+\infty)\to \mathbb{R}_+:t\mapsto v(t)=\sqrt{H^2(t)-a^{-2}(t)}$ in the sense that we can write
\be (t,r)\in\hor \Leftrightarrow v(t) =\frac{1}{r}\sqrt{1-\frac{2M}{r}}. \label{w-hor} \ee
Since $t\mapsto v(t)$ is monotone, we can invert and write
\be (t,r)\in\hor \Leftrightarrow t=t_h(r) = v^{-1}(\frac{1}{r}\sqrt{1-\frac{2M}{r}}). \label{w-inv-hor} \ee
From (\ref{rh-diff}), we can deduce that $t\mapsto t_h(r)$ is decreasing on $(2M,3M)$, is increasing for $r>3M$ and so has a global minimum at $r=3M$ which we denote $t_{h,min}=t_h(3M)$. The horizon thus has two branches described by functions 
\be r_h^+:[t_{h,min},+\infty)\to [3M,r_+) \label{rh-plus}\ee
and
\be r_h^-:[t_{h,min},+\infty) \to (r_-,3M]. \label{rh-minus} \ee
Each function $t\mapsto r_h(t) = r_h^{\pm}(t)$ satisfies (\ref{rh-diff}) and we have 
\be \lim_{t\to +\infty} (r_h^+(t),r_h^-(t)) = (r_-,r_+), \label{rh-limits}\ee 
where for $H_0=0$, $r_+=+\infty$ and $r_-=2M$, while for $H_0>0$, $r_+$ and $r_-$ are respectively the larger and smaller positive roots of 
\be 1-\frac{2M}{r}-r^2H_0^2 = 0.\label{root-eq} \ee
There is a simple necessary and sufficient condition for the existence of these roots, and hence for the existence of a horizon in the case that $H_0>0$. This condition is 
\be MH_0<\frac{1}{3\sqrt{3}}, \label{hor-con} \ee
and we assume henceforth that this holds. In the case $H_0=0$, the horizon forms at time $t=t_{h,min}$, which satisfies $v(t_{h,min}) = \frac{1}{3\sqrt{3}M}$. Since $v(t)\to 0 $ as $t\to +\infty$, a horizon always forms in this case: there will always be values of $t$ for which $v(t)$ drops below the maximum value of $r^{-1}\sqrt{1-\frac{2M}{r}}$. The limiting value $H_0$ of the Hubble function is related to the cosmological constant by $\Lambda=3H_0^2$.

As $H>0$ throughout $\Omega$ in this case, $\Omega_T$ is empty and $\left.\frac{dr_i}{dt}\right|_P=0 \Leftrightarrow P\in\hor$. Furthermore, as $H'+a^{-2}<0$, we see from (\ref{rh-diff}) that $\frac{dr_h}{dt}$ is non-zero everywhere on $\hor$, and so IRNGs are nowhere tangent to $\hor$. Thus the horizon acts as a one-way membrane for IRNGs, which can only cross from $\Omega_A$ into $\Omega_R$ as $t$ increases. 
  
%%%%%%%%%%%%%%%%%%%%%%%%%%%%
\subsection{The horizon in a recollapsing background}
%%%%%%%%%%%%%%%%%%%%%%%%%%%%

We define $w(t)=H^2-a^{-2}$. From the assumption (\ref{ah-lims}), the energy conditions (\ref{DEC}) and Corollary \ref{cor:a-limits}, we see that $w$ decreases on $(0,t_{\rm{max}})$ from $+\infty$ to a negative minimum $w(t_{\rm{max}})<0$, and increases thereafter (i.e. on $(t_{\rm{max}},t_f)$), again approaching $+\infty$ as $t\to t_f^-$. Hence there exist $t_{w_1},t_{w_2} \in (0,t_f)$ with $t_{w_1}<t_{\rm{max}}<t_{w_2}$ such that 
\be w(t) \left\{ \begin{array}{ll} >0,& t\in(0,t_{w_1})\cup(t_{w_2},t_f); \\ <0, & t\in(t_{w_1},t_{w_2}). \end{array} \right. \label{w-profile} \ee 
Note also that the horizon $\chi=0$ and the allowed boundary $\kappa=0$ coincide at $t=t_{\rm{max}}$, whereat $H(t_{\rm{max}})=0$. 

For each fixed $t_0\in(t_{w_1},t_{w_2})$, the horizon structure is analogous to that of Schwarzschild-anti de Sitter spacetime: there is a unique value $r_{h_0}$ of $r$, with $r_{h_0}<2M$, such that 
\be \chi(t_0,r) \left\{ \begin{array}{ll} <0, & r_1<r<r_{h_0}; \\ =0, & r=r_{h_0}; \\ >0, & r>r_{h_0}. \end{array} \right. \label{hor-recoll-1} \ee   
(Recall that $r_1$ marks the boundary of the allowed region for $K<0$.) Since $w(t)$ diverges to $+\infty$ in the limit as $t\to0^+$ and in the limit as $t\to t_f^-$ (see (\ref{ah-lims})), there exist $t_{w_3}\in(0,t_{w_1})$ and $t_{w_4}\in(t_{w_2},t_f)$ such that $w(t)>0$ on $(0,t_{w_1})\cup(t_{w_2},t_f)$ and
\be w(t) \left\{ \begin{array}{ll} < \frac{1}{27M^2}, & t\in(t_{w_3},t_{w_1})\cup(t_{w_2},t_{w_4}); \\ >\frac{1}{27M^2},& t\in(0,t_{w_3})\cup(t_{w_4},t_f). \end{array}\right. \label{hor-recoll-2} \ee
Then for each fixed $t_0\in(t_{w_3},t_{w_1})\cup(t_{w_2},t_{w_4})$, the horizon structure is analogous to that of a Schwarzschild-de Sitter spacetime in which the horizon existence condition (\ref{hor-con}) is satisfied: there exist $r_{h_1}(t_0)\leq 3M \leq r_{h_2}(t_0)$ such that 
\be \chi(t_0,r) \left\{ \begin{array}{ll} <0, & r\in(r_1,r_{h_1})\cup(r_{h_2},+\infty); \\ =0, & r\in\{r_{h_1},r_{h_2}\}; \\ >0, & r\in(r_{h_1},r_{h_2}). \end{array} \right. \label{hor-recoll-3} \ee   The roots $r_{h_{1,2}}$ yield $C^1$ functions of $t$ with
\be r_{h_{1,2}}: [t_{w_3},t_{w_1})\cup(t_{w_2},t_{w_4}]\to (2M,+\infty),  \ee with $r_{h_1}(t_{w_3})=r_{h_2}(t_{w_3})=r_{h_1}(t_{w_4})=r_{h_2}(t_{w_4})=3M$ and 
\begin{eqnarray} 
\lim_{t\to t_{w_1}^-} r_{h_1} = \lim_{t\to t_{w_2}^+} r_{h_1} &=& 2M, \\
\lim_{t\to t_{w_1}^-} r_{h_2} = \lim_{t\to t_{w_2}^+} r_{h_2} &=& +\infty.
\end{eqnarray}
Furthermore, $r_{h_1}(t)$ (the inner branch of the horizon) is decreasing on $(t_{w_3},t_{w_1})$ and increasing on $(t_{w_2},t_{w_4})$, while the converse holds for $r_{h_2}(t)$ (the outer branch of the horizon). Note that in this case, the anti-trapped and trapped regions are, respectively, 
\begin{eqnarray} \Omega_A&=&\{(t,r):\chi(t,r)<0, t\in(0,t_{\rm{max}})\},\label{OA-lam-neg}\\ 
\Omega_T&=&\{(t,r):\chi(t,r)<0, t\in(t_{\rm{max}},t_f)\}. \label{OT-lam-neg}
\end{eqnarray}
Finally, we note that there is no horizon at time $t$ if $w(t)>\frac{1}{27M^2}$, i.e.\ for $t\in(0,t_{w_3})\cup(t_{w_4},t_f)$. 

Figure 1 illustrates the structure of the horizon and the allowed region of representative examples of eternally expanding and of recollapsing McVittie spacetimes with $K<0$. 

\begin{figure}\label{horizons}
	%\centering
{\includegraphics[scale=0.5]{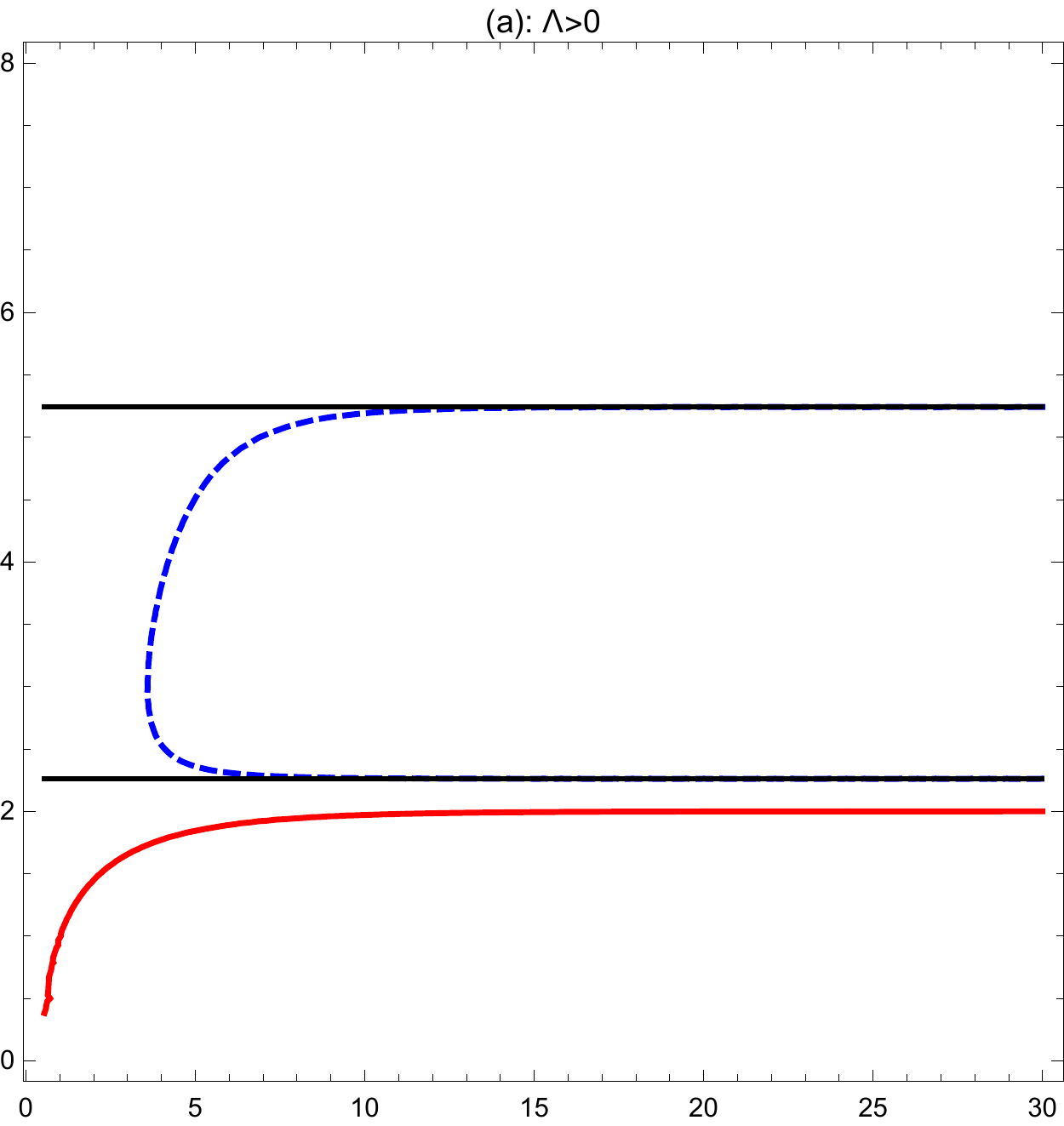} \includegraphics[scale=0.5]{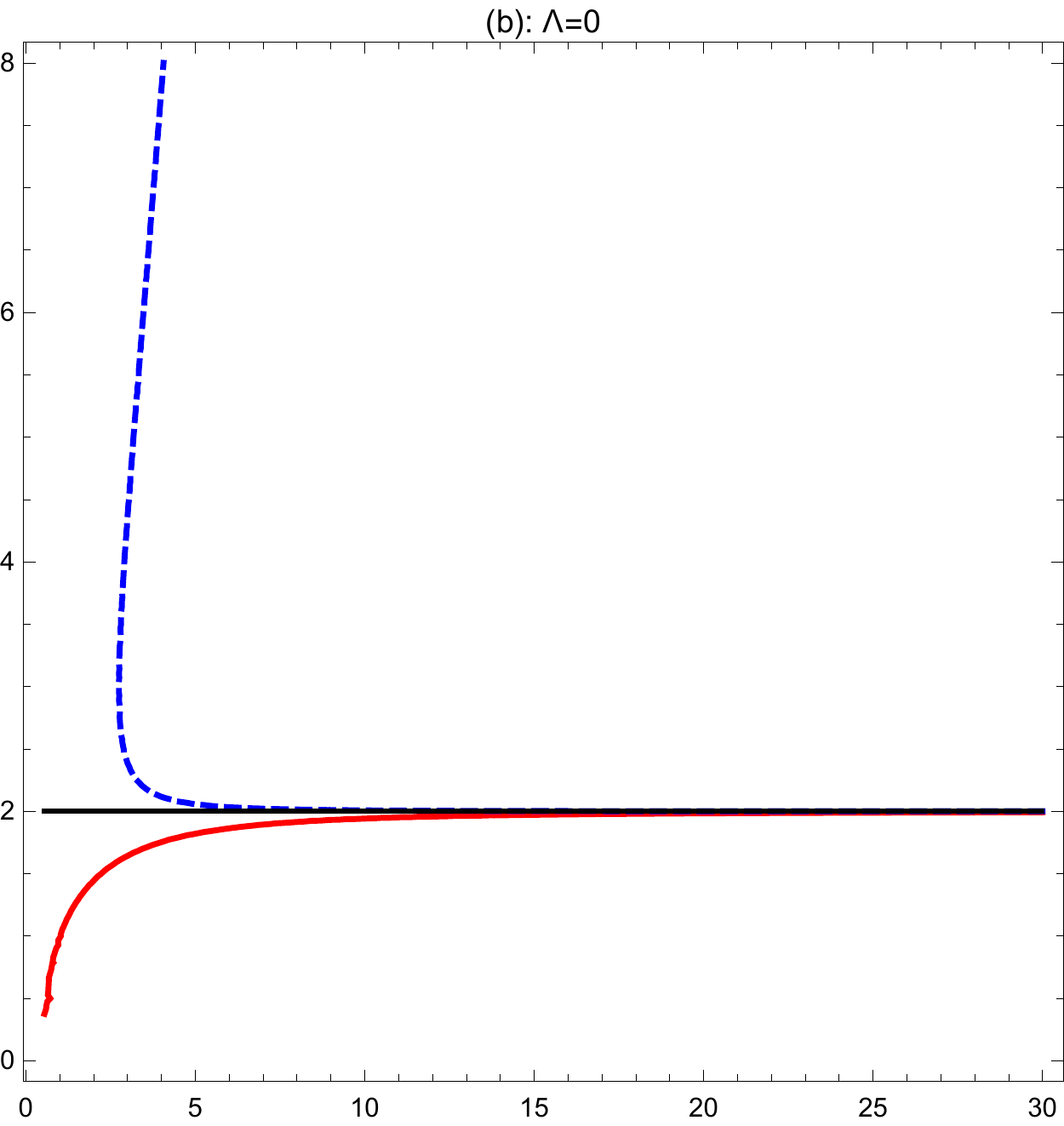} \includegraphics[scale=0.5]{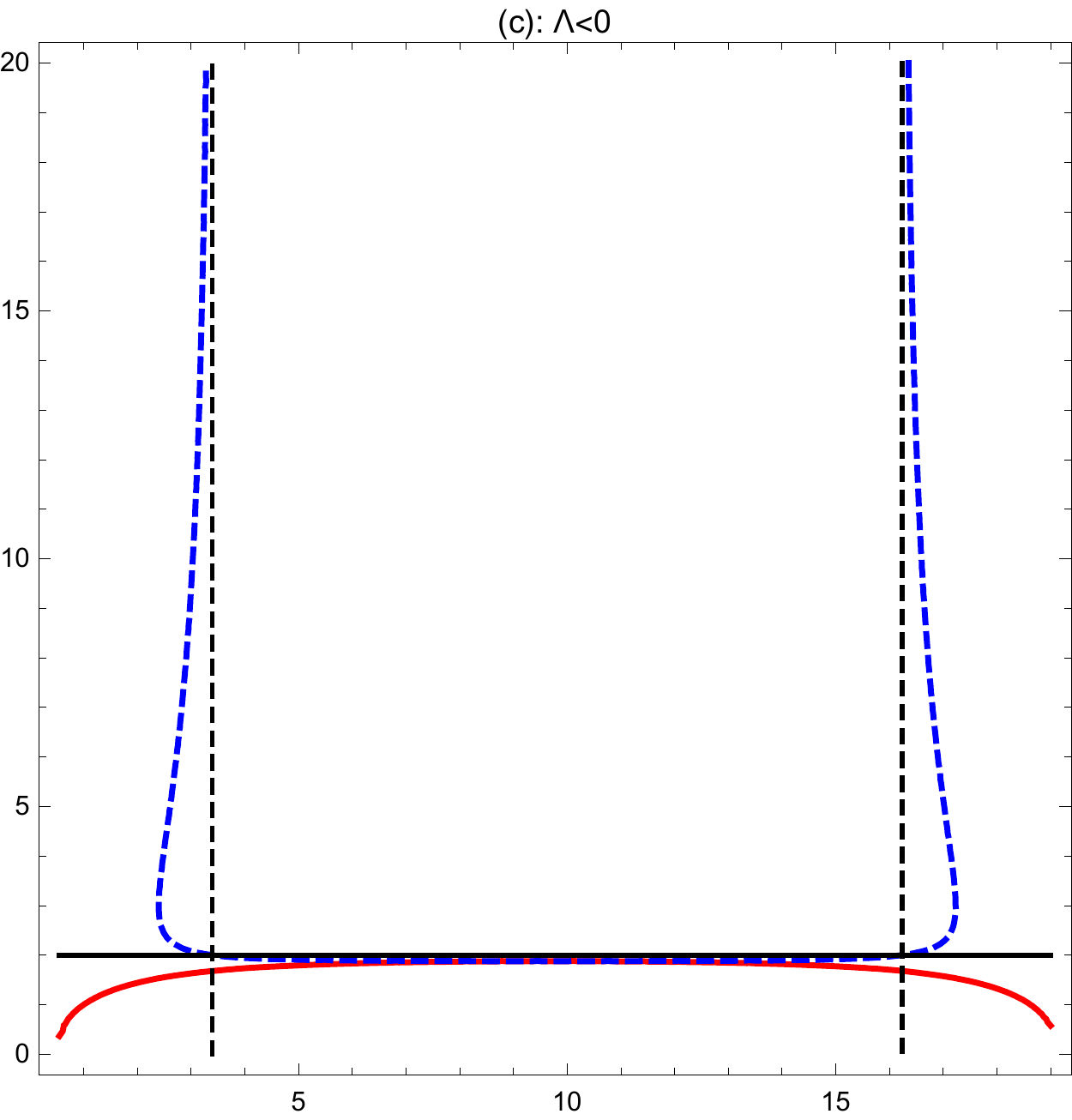}}
\caption{The horizon structure for dust-filled, $K<0$ McVittie spacetimes with positive, zero and negative cosmological constant and with $M=1$. In each case, the horizontal axis represents $t$ and the vertical axis represents $r$. The horizon $\hor$ is the bold-face dashed curved, and the solid curve is the boundary of the allowed region $r=r_1(t)$. The allowed region is $r>r_1(t)$. The regular region $\Omega_R$ is the region bounded by the horizon and bounded away from the origin. In cases (a) and (b), the complement of $\Omega_R\cup\hor$ in the allowed region is the anti-trapped region $\Omega_A$, and there is no trapped region. In case (c), the complement of $\Omega_R\cup\hor$ in the allowed region with $t<t_{\rm{max}}$ is the anti-trapped region $\Omega_A$ and the complement of $\Omega_R\cup\hor$ in the allowed region with $t>t_{\rm{max}}$ is the trapped region $\Omega_T$. In case (a), $\Lambda>0$ and the horizontal lines represent $r=r_{\pm}$ (see (\ref{rh-limits})). These are asymptotes (for $t\to+\infty$) of the inner and outer branches of the horizon. In case (b), $\Lambda=0$ and the horizontal line represents $r=2M$. This is an asymptote both of (the inner branch of) the horizon and of the boundary of the allowed region. Note that $\lim_{t\to+\infty}r_1(t)=2M$ (see (\ref{r2-plus})) in cases (a) and (b). In case (c), $\Lambda<0$ and again the horizontal line represents $r=2M$. The vertical dashed lines correspond to the times at which $w(t)=H^2-a^{-2}$ equals zero $(t=t_{w_3},t_{w_4})$: these are vertical asymptotes of the horizon. The horizon and the boundary of the allowed region meet at $t_{\rm{max}}$, whereat $H=0$. In this case, $t_{\rm{max}}\simeq 9.82$.}
\end{figure}

%%%%%%%%%%%%%%%%%%%%%%%%%%%%%%
%%%%%%%%%%%%%%%%%%%%%%%%%%%%%%
%%%%%%%%%%%%%%%%%%%%%%%%%%%%%%

\section{Past evolution of RNGs in $K<0$ McVittie spacetimes}\label{section:past}

In this section, we establish a key feature of negatively curved McVittie spacetimes: all radial null geodesics extend back either to the point $\co=\{(t,r)=(0,0)\}$ or to the boundary of the allowed region (\ref{bdy-neg}), reaching these destinations in finite affine time in the past. As $\co$ corresponds to the Big Bang of the initially expanding FLRW background, we see that this singularity is accessible to the spacetime, unlike the case for $K=0$ (and $K>0$ as we will see later). A further difference to the case $K=0$ stems from the fact that the boundary of the allowed region does not correspond to a scalar curvature singularity. Radial null geodesics originate at this surface, which must therefore be some milder form of singularity \cite{ellis1979classification}. The behaviour of RNGs which meet $\balr$ in the past is non-trivial, and depends strongly on details of the background. As we will see, $\balr$ comprises space-like regions and may also contain time-like regions. 
%At each point $P$ of a spacelike region, there is a unique future-pointing IRNG and a unique future-pointing ORNG of $\Omega$ that meet $P$ in the past. At each point $Q$ of a timelike region of $\balr$, there is unique, future-pointing ingoing RNG of $\Omega$ which meets $Q$ in the future, a unique future-pointing outgoing RNG of $\Omega$ which meets $Q$ in the past. 

Our first result establishes the fact that RNGs extend back to the boundary $\partial\Omega$ of $\Omega$ (recall the decomposition (\ref{omega-bdy}) of this boundary). To simplify the overall statement of our results, we also establish that this happens in finite affine time, although this requires the use of subsequent (independently proven) propositions.

\begin{proposition}\label{RNG-past-1} Let $({\mm},g)$ be an initially expanding McVittie spacetime with a Big Bang background for which $K<0$ and let $(t_0,r_0)\in\Omega$ with $r_0>r_1(t_0)$. Let $s$ be an affine parameter along an RNG $\gamma$ with $(t,r)|_{s=0}=(t_0,r_0)$. Then there exists $s_*\in(-\infty,0)$ such that $(t(s),r_{RNG}(s))\in\Omega$ for all $s\in(s_*,0)$ and $\lim_{s\to s_*^+}(t(s),r_{RNG}(s))\in\partial\Omega$. Here, $\gamma:(s_*,0]\to \Omega:s\mapsto (t(s),r_{RNG}(s))$ is the RNG.
\end{proposition}

\startproof Let $(t_0,r_0)$ be as in the statement of the theorem and let $(t_\alpha,t_0]$ be the left-maximal interval of existence of the terminal value problem 
\be \frac{dr}{dt}= \sigma(r\pm\kappa^{1/2}H^{-1}), \quad r(t_0)=r_0. \label{ivp} \ee
 We use the same notation for this interval for either choice of sign $\pm$. We write the solution as $r_{RNG}:(t_\alpha,t_0]\to\mathbb{R},t\mapsto r_{RNG}(t)$. Note that we must have $t_\alpha\geq0$ in order that $(t,r_{RNG}(t))\in\Omega$ for all $t\in(t_\alpha,t_0]$. First, we establish that the RNG meets $\partial\Omega$ in the limit as $t\to t_\alpha^+$. To see this, we note the following. The functions $F_{o,i}$ defined in (\ref{orng}) and (\ref{irng}) are $C^1$ on $\Omega$. As $t_{\alpha}$ is finite, it follows that the solution $(t,r_{RNG}(t))$ must exit any compact subset of $(t_\alpha,t_0]\times\Omega$ (see e.g.\ Theorem 2 of Section 2.4 of \cite{perkodifferential}). Thus either $r_{RNG}(t)$ is unbounded on $(t_\alpha,t_0]$, or the solution meets the boundary at a finite value of $t$ in the past (which must, but its definition, be $t=t_\alpha$). We rule out the former possibility by monotonicity arguments - there are different cases to consider. 

\begin{itemize}

\item[(a)] \textit{Eternally expanding universe with $\Lambda>0$}. In this case, $\frac{dr_o}{dt}>0$ throughout $\Omega$, and $\frac{dr_i}{dt}>0$ for all points $P\in\Omega$ with $r|_P>r_+$ (see (\ref{rh-limits}) \textit{et seq.} and panel (a) of Figure 1). It follows immediately that $r_{RNG}(t)$ cannot increase without bound as $t$ decreases, and so the geodesic meets $\partial\Omega$ at finite time in the past. 

\item[(b)] \textit{Eternally expanding universe with $\Lambda=0$}. Again, $\frac{dr_o}{dt}>0$ throughout $\Omega$. As noted in Section \ref{rngs-horizon} above, $\frac{dr_i}{dt}<0$ in $\Omega_R$, $\frac{dr_i}{dt}>0$ in $\Omega_A$ and $\frac{dr_i}{dt}=0$ along $\hor$, and the horizon is a one-way membrane for IRNGs which can only cross from $\Omega_A$ into $\Omega_R$ as $t$ increases. It follows that if $(t_0,r_0)\in\Omega_A\cup\hor$, then $(t,r_{RNG}(t))\in\Omega_A$ for all $t\in(t_\alpha,t_0)$. Consider then the alternative, that $(t_0,r_0)\in\Omega_R$. The structure of the horizon in this case indicates that the set $K=\{(t,r)\in \Omega_R\cup\hor:t\leq t_0\}$ is compact. Then Theorem 2 of Section 2.4 of \cite{perkodifferential} indicates that the geodesic $\gamma:(t_\alpha,t_0]$ exits $K$ as $t$ decreases. Thus the geodesic must enter $\Omega_A$ at some point in the past. So for all RNGs in this case, there exists $t_\beta\in(t_\alpha,t_0)$ such that $r_{RNG}'(t)>0$ for all $t\in(t_\alpha,t_\beta)$. This rules out the possibility that $r_{RNG}(t)$ is unbounded on $(t_\alpha, t_0)$, and so RNGs in this case must meet $\partial \Omega$ at finite time in the past. 

\item[(c)] \textit{Recollapsing universe with $\Lambda<0$}.
We define the early and late branches of the horizon by, respectively,
\begin{eqnarray}
\horm&=&\{(t,r)\in\hor: t<t_{\rm{max}}\}, \label{hor-minus} \\
\horp&=&\{(t,r)\in\hor: t>t_{\rm{max}}\}. \label{hor-plus} 
\end{eqnarray}
 Then $\horm$ is a one-way membrane for IRNGs, which are injected from $\Omega_A$ into $\Omega_R$ as $t$ increases, while $\horp$ is a one-way membrane for ORNGs, which are injected from $\Omega_R$ into $\Omega_T$ as $t$ increases. Recalling the notation of Section 5.2, and repeating the argument of case (b) above, we immediately see that all ORNGs with initial point in $\Omega_A\cup\Omega_R\cup\hor$ and all IRNGs with initial point $(t_0,r_0)$ where $t_0\leq t_{w_3}$ meet $\partial\Omega$ at a finite time in the past. So completing the proof in this case requires that we show that ORNGs in $\Omega_T$ meet $\horp$ at a finite time in the past, and that IRNGs with initial point $(t_0,r_0)$ with $t_0>t_{w_3}$ enter $\{(t,r):t\leq t_{w_3}\}$ at a finite time in the past. We outline the relevant arguments. Recall that IRNGs satisfy 
\be \frac{dr}{dt} = \frac{\sigma}{H}(rH-\kappa^{1/2}). \label{IRNG-rem}\ee
From (\ref{int-limit}) and (\ref{sig-k-neg}), we have, for each fixed $t\in(0,t_f)$,                                         
\be \sig(t,r) \sim H(t),\quad r\to+\infty. \label{sig-lim3}\ee
For IRNGs in $\Omega_R\cap\{t<t_{\rm{max}}\}$, so that $H>0$ and $\chi>0$, we have 
\be \frac{dr}{dt}>-\frac{\sigma}{H}\kappa^{1/2} \sim -\frac{r}{a},\quad r\to+\infty. \ee
This inequality is preserved along the geodesic as $t$ decreases. Integrating shows that $r$ cannot diverge to $+\infty$ as $t$ decreases, and so $r_{RNG}(t)$ must cross $t=t_{w_3}$ at finite $r$ as required. When the geodesic in initially in the region where $H<0$ and $\chi>0$, we can repeat the argument noting that 
\be \frac{dr}{dt}>-2\frac{\sigma}{H}\kappa^{1/2}, \ee
and when the geodesic is initially in the trapped region $\Omega_T$ (i.e.\ when $H<0$ and $\chi<0$), we can repeat the argument noting that 
\be \frac{dr}{dt} >2r\sigma \sim 2rH = -2r|H|,\quad r\to +\infty. \ee
ORNGs satsify 
\be \frac{dr}{dt} = \frac{\sigma}{H}(rH+\kappa^{1/2}). \label{ORNG-rem}\ee
For ORNGs in the trapped region $\Omega_T$, wherein $H<0$, $\chi<0$, we have 
\be \frac{dr}{dt} > r\sigma \sim rH = -r|H|,\quad r\to +\infty. \ee
As before, integrating shows that along these geodesics, $r_{RNG}(t)$ cannot diverge to $+\infty$ as $t$ decreases, and so the geodesics must meet $\horp$ as required, or must meet the boundary of the allowed region coming from the trapped region. We rule out the latter possibility in the next step of the proof. 
\end{itemize}

This establishes the fact that all RNGs meet $\partial\Omega$ at a finite value of $t$ in the past. In Proposition \ref{RNG-past-2}, we prove that such geodesics cannot extend to $\partial\Omega_{\{t=0\}}$, and so must extend back to a point $(t_*,r_{RNG}(t_*))\in\co\cup\balr$ where $t_*\geq0$. Furthermore, it follows from the results of Sections 5.1 and 5.2 that the geodesics meet the past boundary coming from the anti-trapped region: for each RNG, there exists $\delta>0$ such that 
\be (t,r_{RNG}(t))\in\Omega_A \quad\hbox{for all } t\in(t_*,t_*+\delta). \label{bdy-from-anti-trapped}\ee
Now let $s_*\in[-\infty,0)$ be the value of the affine parameter $s$ corresponding to $t_*$. Then there exists $\epsilon>0$ such that 
\be (t(s),r_{RNG}(s))\in\Omega_A \quad\hbox{for all } s\in(s_*,s_*+\epsilon), \label{bdy-from-anti-trapped-aff}\ee
from which it follows that 
\be \dot{r}(s)>0 \quad \hbox{for all } s\in(s_*,s_*+\epsilon). \label{rdot-near-bdy} \ee
The geodesic equations for RNGs yield
\be \ddot{r} = \frac{r\sigma}{H}(H'+a^{-2})\dot{t}^2, \label{affine}\ee
where the overdot(s) represent derivatives with respect to an affine parameter $s$ along the geodesic. The form of $\sigma$ in the $K<0$ is given by (\ref{sig-k-neg}). Combining this with the energy condition (\ref{DEC}) shows that $\ddot{r}<0$ everywhere along the geodesic. Since $\dot{r}>0$ in the approach to the boundary, this proves that the geodesic meets the boundary at a finite value of the affine parameter $s$, completing the proof of the proposition. \hfill$\blacksquare$

%We complete the proof in the following five steps. (i) We establish that every RNG extends back to the boundary $\partial\Omega$ of $\Omega$ defined in (\ref{omega-bdy}). (ii) We rule out the possibility that the RNG meets $\partial\Omega_{\{t=0\}}$ in the past. (iii) We establish the existence of RNGs that emerge from $\partial\Omega_{\{\kappa=0\}}$ and extend into the allowed region. This is technically challenging, and takes up a significant portion of the overall proof. (iv) We show that all other RNGs originate at $\co$ - and that at least some such geodesics exist. In step (v), we complete the proof by showing that the geodesics meet $\cal{O}\cup\partial\Omega_{\{\kappa=0\}}$ at finite affine time in the past. 

The next result shows that RNGs cannot extend back to $\partial\Omega_{\{t=0\}}=\{(t,r):t=0,r>0\}$: 

\begin{proposition}\label{RNG-past-2} Let $({\mm},g)$ be an initially expanding McVittie spacetime with a Big Bang background for which $K<0$ and let $(t_0,r_0)\in\Omega$ with $r_0>r_1(t_0)$. Let $s$ be an affine parameter along an RNG $\gamma$ with $(t,r)|_{s=0}=(t_0,r_0)$. Then there exists $s_*\in(-\infty,0)$ such that $(t(s),r_{RNG}(s))\in\Omega$ for all $s\in(s_*,0)$ and $\lim_{s\to s_*^+}(t(s),r_{RNG}(s))\in\co\cup\balr$.
\end{proposition}

\startproof We treat ORNGs and IRNGs separately. So suppose that there exists an ORNG for which there exists $r_i>0$ such that $\lim_{t\to 0}r_{RNG}(t) = r_i$. (This corresponds to the case $(t(s_*),r_{RNG}(s_*))\in\partial\Omega_{\{t=0\}}$.) We show that this leads to a contradiction, and so (in conjunction with Proposition \ref{RNG-past-1}), the conclusion follows.  In both cases of eternally expanding and recollapsing universes, there exists $t_1>0$ such that $H(t)>0$ and $r_{RNG}'(t)>0$ for all $t\in(0,t_1)$. Since $\sigma/H>0$, we have, along an ORNG, 
\be \frac{dr}{dt}>r\sigma > \frac{ra}{\sqrt{a^2+r^2}}\kappa^{1/2}H, \label{rpt1}\ee
and (using the monotone increasing function $a(t)$ as parameter along the ORNG)
\be \frac{dr}{da} >\frac{r\kappa^{1/2}}{\sqrt{a^2+r^2}}, \label{rpt2} \ee
where we have used (\ref{Sig-lr}) in the second inequality of (\ref{rpt1}). 
Noting that $r/\sqrt{a^2+r^2}<1$ and $\lim_{a\to 0^+}r/\sqrt{a^2+r^2}=1$, there exists $a_2>0$ such that $r/\sqrt{a^2+r^2}>1/2$ for all $a<a_2$. So 
\be \frac{dr}{da}>\frac12\kappa^{1/2} \quad \hbox{ for all } a<a_3=\min\{a(t_1),a_2\}. \ee
In the case $K<0$, $\kappa$ is an increasing function of $r$, and for a given $r_i>0$, $\kappa|_{r=r_i}$ diverges to $+\infty$ in the limit as $a\to 0^+$. So for all $r>r_i$ (which holds along the geodesic) and for all sufficiently small $a$, we have
\begin{eqnarray} \frac{dr}{da}  &>& \frac12 \kappa(a,r)^{1/2} \nonumber \\
&>& \frac12 \kappa(a,r_i)^{1/2} =\frac 12 \left(1-\frac{2M}{r_i}+\frac{r_i^2}{a^2} \right)^{1/2} \nonumber \\
&>& \frac{r_i}{2\sqrt{2}a}. \label{ineq} 
\end{eqnarray}
Integrating this inequality over $[a,a_4]$, with $a_4$ chosen sufficiently small, and letting $a\to 0^+$, shows that $r_{RNG}(t)\to 0$ before $a$ reaches zero, in contradiction of our assumption that the geodesic reaches $\partial\Omega_{\{t=0\}}$. Thus ORNGs cannot extend back to this boundary. 

For IRNGs, we again suppose that there is a geodesic that meets $\partial\Omega_{\{t=0\}}$ in the past. Noting that by part (v) of Definition 4.1, $aH>2$ for all sufficiently small $a$, and so 
\be \kappa=1-\frac{2M}{r}+\frac{r^2}{a^2} < 1-\frac{2M}{r}+\frac14r^2H^2. \label{kp-ineq1}\ee
Since $r$ is increasing along the IRNG for sufficiently small values of $a$ (IRNGs are confined to $\Omega_A$ for sufficiently small values of $a$), and since $H\to+\infty$ as $a\to 0^+$, we have
\be \kappa < \frac13r^2H^2 \label{kp-ineq2} \ee
along the IRNG for all sufficiently small values of $a$. Then the equation for IRNGs yields 
\be \frac{dr}{dt}>(1-\frac{1}{\sqrt{3}})r\sigma. \label{irng-ineq} \ee
We can now repeat the argument above for ORNGs from (\ref{rpt1}) onwards, changing an irrelevant factor in this inequality. 
\hfill$\blacksquare$

Next, we highlight a key difference between the spatially flat and non-flat McVittie spacetimes with $K<0$. In the latter, the Big Bang singularity $\co$ forms part of the past boundary of the spacetime. In the flat case, this is cut off by the singularity at $r=2M$. 

\begin{proposition}\label{RNG-past-3} Let $({\mm},g)$ be an initially expanding McVittie spacetime with a Big Bang background for which $K<0$. There exists $\epsilon>0$ such that every outgoing and ingoing RNG with initial point in the open set 
\be \co_\epsilon = \{(t,r):t>0,0<r<\epsilon,r^3>2Ma^2(t)\} \label{o-ep-def} \ee 
originates at $\co$ at finite affine time in the past. 
\end{proposition}

\startproof The proof involves comparing the slope of RNGs with the slope of the level set $\kappa(t,r)=1$ at a point of intersection. We note that 
\be \kappa = 1 \Leftrightarrow r^3 = 2Ma^2 \Leftrightarrow r=r_3(a), \label{kap-1}\ee
with an obvious definition of $r_3$. Then the slope of this curve satisfies
\be \frac{dr_3}{da} = \frac23\frac{r_3}{a}.\label{r3-slope}\ee
Since the surface $r=r_3(a)$ corresponds to $\kappa=1$, this surface lies within the allowed region $\kappa>0$ and so $r_3(a)>r_1(a)$ for all $a>0$. It follows that $\co_\epsilon\subset \Omega$. 

We work in a neighbourhood of $t=0$, and so we can use $a$ as a parameter along RNGs. Then the equation (\ref{rng}) of the RNGs yields
\be \left.\frac{dr}{da}\right|_{\kappa=1} = \frac{rH\pm1}{rH}\Sigma,\label{ORNG-kap-1}\ee
where evaluation along $r^3=2Ma^2$ should also be understood on the right hand side. The coefficient $\Sigma$ is given by 
\be \Sigma = \int_0^1(1+\frac{a^2}{r_3^2}\tau^2-\tau^3)^{-3/2}d\tau, \label{Sig-def}\ee
which we obtain from (\ref{sig-k-neg}) by evaluating along $\kappa=1$ using the change of variable $\tau=r/\bar{r}$. 
Define $x=a^2/r_3^2 =(2M)^{-2/3}a^{2/3}$. Then $x\to 0$ as $a\to 0$. In this limit, $\Sigma$ becomes singular: we need to determine its rate of divergence. Towards this end, we define
\be v_x(\tau)=1+x\tau^2-\tau^3,\quad x>0. \label{v-x-def}\ee
It is straightforward to show that $v_x$ has a unique root $\tau_0$, which is positive, and which can be written as $\tau_0=1+\delta=1+x/3+O(x^2)$. We then write $v_x(\tau)=(1+\delta-\tau)(\tau^2+\alpha\tau+\beta)$, where $\alpha = 1+\delta-x=1-2x/3+O(x^2)$, $\beta=(1+\delta)\alpha=1-x/3+O(x^2)$. Integrating by parts then yields 
\be \Sigma = \frac23x^{-1/2}-\lambda + O(x^{1/2}), \label{Sig-val}\ee
where
\be \lambda = 2 - \frac32\int_0^1(1-\tau)^{-1/2}(2\tau+1)(\tau^2+\tau+1)^{-5/2}d\tau \simeq 0.77. \label{lam-val}\ee  
From (\ref{ORNG-kap-1}), we then have
\begin{eqnarray*} \left.\frac{dr}{da}\right|_{\kappa=1} &=& \left(1\pm\frac{1}{r_3H}\right)\left(\frac23\frac{r_3}{a}-\lambda + O(\frac{a}{r_3})\right),\quad a\to 0.\nonumber \\
&=& \frac23\frac{r_3}{a}-\lambda + o(1),\quad a\to 0, \label{ORNG-kap-1a}\end{eqnarray*}
where we have used the condition (\ref{ah-lims}) of Definition \ref{def2}. Comparing with (\ref{r3-slope}), we see that sufficiently close to $\co$ (i.e.\ for sufficiently small $\epsilon$), an RNG of either family (ingoing or outgoing) with an initial point in $\co_\epsilon$ can only cross the hypersurface $r=r_3(a)$ from above in the $a-r$ plane, and so remains above $r=r_3$, and hence above $r=r_1$, as $a$ decreases. By Proposition \ref{RNG-past-2}, these geodesics extend back to $\co\cup \balr$ and so the result follows.\qed

As noted above, resolving the behaviour of RNGs at and near the allowed boundary $\balr$ is non-trivial. We begin with a couple of technical lemmas.  

\begin{lemma}\label{lem:r2prime} 
The function $r_1:(0,t_f)\to\real_+$ describing the boundary of the allowed region satisfies
\be \frac{dr_1}{dt} = \sigma_0(t) r_1,\label{r2-de} \ee
where $\sigma_0$ is defined in (\ref{sig0-def}). It follows that $r=r_1(t)$ formally satisfies both the ingoing and outgoing RNG equations. \qedapp
\end{lemma}

%\noindent\textbf{Proof:}

\begin{lemma}\label{lem-RNG-bdy}
Let $\rh=r-r_1(t)$. Then the RNG equations may be written
\be \frac{d\rh}{dt} = P_\pm(t)\rh^{1/2}+Q_\pm(t,\rh)\rh, \label{new-rng} \ee
where the upper and lower sign correspond to outgoing and ingoing RNGs respectively, the coefficients $P_\pm$ are given by
\be P_\pm = \alpha^{1/2}r_1^{-1/2}(r_1\sigma_1\pm H^{-1}\nu^{1/2}(1+2\epsilon)^{1/2}\sigma_0), \label{Ppm-def} \ee
and the coefficients $Q_\pm$ are $C^1$ on $\Omega\cup\balr =\{(t,\rh):t>0, \rh\geq 0\}$.  \qedapp
%and satisfy 
%\be \left.Q_\pm\right|_{\balr} = \sigma_0\pm\alpha r_1^{-1}H^{-1}\sigh\nu^{1/2}(1+2\epsilon)^{1/2}. \label{Qpm-lim} \ee
\end{lemma}

%\noindent\textbf{Proof:} 

We note that 
\be \rh=0 \Leftrightarrow x=\alpha \Leftrightarrow r=r_1 \ee
and
\be \rh>0 \Leftrightarrow x<\alpha \Leftrightarrow r>r_1, \ee
so that there exists an RNG of the spacetime that meets $\balr$ at time $t_0>0$ if and only if there exists a \textit{positive} solution $\rh$ of (\ref{new-rng}) with $\rh(t_0)=0$. The trivial solution $\rh\equiv0$ of (both cases of) (\ref{new-rng}) corresponds to the boundary of the allowed region: this is not a geodesic of the spacetime. Defining $u=\hat{r}^{1/2}$ then yields the following result. 
A crucial division by $u$ is permitted as we have restricted to positive $u$. Uniqueness follows by a standard theorem: the right-hand side of the ODE in (\ref{u-ivp}) is Lipshitz in $u$.

\begin{lemma}\label{u-sols}
Let $t_0\in(0,t_f)$, let $I$ be an interval containing $t_0$ and let $I_0$ be the interval $I$ punctured at $t_0$. There exists a solution $\rh:I\to[0,+\infty)$ of the initial value problem 
\be \frac{d\rh}{dt} = P_\pm(t)\rh^{1/2} + Q_\pm(t,\rh) \rh,\quad \rh(t_0)=0 \label{rh-ivp} \ee
with $\rh(t)>0$ for all $t\in I_0$ 
if and only if there exists a solution $u:I\to[0,+\infty)$ of the initial value problem
\be \frac{du}{dt} =\frac12P_\pm(t) + \frac12\tilde{Q}_\pm(t,u)u,\quad u(t_0)=0 \label{u-ivp} \ee
with $u(t)>0$ for all $t\in I_0$, where $\tilde{Q}_\pm(t,u)=Q_\pm(t,u^2)$. Furthermore, when such a solution exists, it is unique.
\qed
\end{lemma} 

%\noindent\textbf{Proof:} This is immediate on defining $u=\rh^{1/2}$.  \qed

It is evident from (\ref{u-ivp}) that the sign of $P_\pm(t_0)$ is crucial for the question of whether RNGs may originate or terminate at the point $(t_0,r_1(t_0))\in\balr$. The following lemma captures the key technical information regarding this, for application in the subsequent proposition. 

\begin{lemma}\label{u-pos-neg}
Let $t_0\in\real$, let $I$ be an interval containing $t_0$, let $p\in C^1(I)$ and let $q\in C^1(I\times (-u_0,u_0))$ for some $u_0>0$. Let $u:I\to\mathbb{R}$  be the unique solution of the initial value problem
\be \frac{du}{dt} = p(t) + q(t,u)u,\quad u(t_0)=0. \label{u-ivp1} \ee
\begin{enumerate}
\item If $p(t_0)>0$, then there exists $t_1<t_0$ and $t_2>t_0$ with $(t_1,t_2)\subset I$ such that $u(t)<0$ for all $t\in(t_1,t_0)$ and $u(t)>0$ for all $t\in(t_0,t_2)$. 
\item If $p(t_0)<0$, then there exists $t_1<t_0$ and $t_2>t_0$ with $(t_1,t_2)\subset I$ such that $u(t)>0$ for all $t\in(t_1,t_0)$ and $u(t)<0$ for all $t\in(t_0,t_2)$.
\item If $p(t_0)=0$ and $p'(t_0)>0$ (respectively $p'(t_0)<0$), then there exists $t_1<t_0$ and $t_2>t_0$ with $(t_1,t_2)\subset I$ such that $u(t)>0$ (respectively $u(t)<0$) for all $t\in(t_1,t_0)\cup(t_0,t_2)$.
\end{enumerate}
\qedapp
\end{lemma}

\begin{proposition}\label{prop:structure-of-all-bdy}
Define 
\be X_\pm(t) = aH\bar{K}(\epsilon) \pm\frac{2\epsilon}{1+2\epsilon},\quad t\in(t,t_f), \label{Xpm-def} \ee
where $\bar{K}$ and $\epsilon$ are defined in (\ref{kbar-def}) and (\ref{nu-beta-ep-def}) respectively.
Then $X_\pm\in C^1((0,t_f),\real)$ with $X_+(t)> X_-(t)$ for all $t\in(0,t_f)$. Assume that each of $X_\pm$ vanishes at only a finite number of points, and that the non-degeneracy condition $X_\pm'\neq0$ holds at zeros of $X_\pm$. Then the allowed boundary $\partial\Omega_{\{\kappa=0\}}$ decomposes into a union of disjoint intervals of the form
\begin{eqnarray}
\partial\Omega_{\{\kappa=0\}}^{(+,+)} &=& \{(t,r)\in\partial\Omega_{\{\kappa=0\}}: X_+(t)>0, X_-(t)>0\}; \label{bpp} \\
\partial\Omega_{\{\kappa=0\}}^{(+,-)} &=& \{(t,r)\in\partial\Omega_{\{\kappa=0\}}: X_+(t)>0, X_-(t)<0\}; \label{bpm} \\
%\partial\Omega_{\{\kappa=0\}}^{(-,+)} &=& \{(t,r)\in\partial\Omega_{\{\kappa=0\}}: X_+(t)<0, X_-(t)>0\}; \label{bmp} \\
\partial\Omega_{\{\kappa=0\}}^{(-,-)} &=& \{(t,r)\in\partial\Omega_{\{\kappa=0\}}: X_+(t)<0, X_-(t)<0\}, \label{bmm} 
\end{eqnarray}
and sets of a finite number of points of the form
\begin{eqnarray}
\partial\Omega_{\{\kappa=0\}}^{(+,0\uparrow)} &=& \{(t,r)\in\partial\Omega_{\{\kappa=0\}}: X_+(t)>0, X_-(t)=0,X_-'(t)>0\}; \label{bpz-plus} \\
\partial\Omega_{\{\kappa=0\}}^{(+,0\downarrow)} &=& \{(t,r)\in\partial\Omega_{\{\kappa=0\}}: X_+(t)>0, X_-(t)=0,X_-'(t)<0\}; \label{bpz-minus} \\
\partial\Omega_{\{\kappa=0\}}^{(0\uparrow,-)} &=& \{(t,r)\in\partial\Omega_{\{\kappa=0\}}: X_+(t)=0,X_+'(t)>0, X_-(t)<0\}; \label{bzm-plus}\\
\partial\Omega_{\{\kappa=0\}}^{(0\downarrow,-)} &=& \{(t,r)\in\partial\Omega_{\{\kappa=0\}}: X_+(t)=0,X_+'(t)<0, X_-(t)<0\}. \label{bzm-minus} 
\end{eqnarray}
Furthermore:
\begin{enumerate}
\item For each $P\in\partial\Omega_{\{\kappa=0\}}^{(+,+)}$, there is a unique ORNG and a unique IRNG of $\Omega$ that originate at $P$; 
\item For each $P\in\partial\Omega_{\{\kappa=0\}}^{(+,-)}$, there is a unique ORNG $\Omega$ that originates at $P$ and a unique IRNG that terminates at this point;
\item For each $P\in\partial\Omega_{\{\kappa=0\}}^{(-,-)}$, there is a unique ORNG and a unique IRNG of $\Omega$ that terminate at $P$.
\item If $P=(t_0,r_1(t_0))\in\partial\Omega_{\{\kappa=0\}}^{(+,0\uparrow)}$, then there is a unique ORNG that originates at $P$ and a unique IRNG of the spacetime that satisfies $r_{IRNG}(t_0)=r_1(t_0)$ and $r_{IRNG}(t)>r_1(t), t\neq t_0$. 
\item If $P=(t_0,r_1(t_0))\in\partial\Omega_{\{\kappa=0\}}^{(+,0\downarrow)}$, then there is a unique ORNG that originates at $P$ and no IRNG of the spacetime meets this boundary point. 
\item If $P=(t_0,r_1(t_0))\in\partial\Omega_{\{\kappa=0\}}^{(0\uparrow,-)}$, then there is a unique IRNG that terminates at $P$ and a unique ORNG of the spacetime that satisfies $r_{ORNG}(t_0)=r_1(t_0)$ and $r_{ORNG}(t)>r_1(t), t\neq t_0$. 
\item If $P=(t_0,r_1(t_0))\in\partial\Omega_{\{\kappa=0\}}^{(0\downarrow,-)}$, then there is a unique IRNG that terminates at $P$ and no ORNG of the spacetime meets this boundary point.
\end{enumerate}
Thus $\partial\Omega_{\{\kappa=0\}}^{(+,+)}$ is a past spacelike portion of $\partial\Omega_{\{\kappa=0\}}$; $\partial\Omega_{\{\kappa=0\}}^{(+,-)}$ is a timelike portion of $\partial\Omega_{\{\kappa=0\}}$ and $\partial\Omega_{\{\kappa=0\}}^{(-,-)}$ is a future spacelike portion of $\partial\Omega_{\{\kappa=0\}}$.
\end{proposition}

\noindent\textbf{Proof:} This is an application of Lemma \ref{u-pos-neg} to the IVP (\ref{u-ivp}) of Lemma \ref{u-sols}. We note that 
\be P_\pm = r_1^{-1/2}\nu^{-1/2}\alpha^{-5/2}\epsilon^{-1}(1+2\epsilon)^{1/2}X_\pm, \label{p-x} \ee
which is $C^1$ on $(0,t_f)$ allowing application of Lemma \ref{u-pos-neg}. For items (iv)-(vii), we note that \textit{positive} solutions of (\ref{u-ivp}) correspond to geodesics of the spacetime. So for item (v), part (iii) of Lemma \ref{u-pos-neg} applies and yields solutions of (\ref{u-ivp}) for putative IRNGs that satisfy $u(t)<0$ on an open interval punctured at $t_0$. These solutions are not geodesics of the spacetime: they occupy the region $\kappa<0$ except at $t=t_0$. \qed

%\begin{comments} 
The non-degeneracy condition on $X_\pm$ is generic in the sense that if we consider the function $X_-(t)$ that arises from an FLRW background for which the Hubble function satisfies (for example) (\ref{hub-bar}) and which has a zero $t_0$ for which $X_-'(t_0)=0$, a generic small perturbation of the parameters of (\ref{hub-bar}) or of the mass parameter $M$ will yield a function $X_-$ in which the degeneracy condition does not hold. In other words, the non-degeneracy hypothesis is not a strong restriction on the class of spacetimes we are considering.
%\end{comments}

We recap at this point: we have shown that all RNGs must extend back to the boundary of $\Omega$, and have ruled out the possibility that they extend back to $\Omega_{\{t=0\}}$. Thus all RNGs extend back either to the Big Bang, ${\cal{O}}=\{(t,r):t=r=0\}$, or to $\balr$, the boundary of the allowed region. The geodesics meet these past boundaries at finite affine time in the past. There always exist geodesics which extend back to $\co$, and Proposition \ref{prop:structure-of-all-bdy} provides a `handbook' for determining whether or not outgoing and ingoing RNGS originate at $\balr$, or if such RNGS `graze' the allowed boundary (e.g.\ the IRNGs of item (iv) of the proposition). We conclude this section by applying this handbook to determine the causal nature of $\balr$ in (classes of) actual McVittie spacetimes with $K<0$. In one sense, this is straightforward: all that is required is the calculation of $X_\pm$. However, as we will see, these quantities are sensitive to fine details of the spacetimes, and so it is difficult to obtain a complete classification for all cases defined by Definition \ref{def2}.  Some general conclusions are possible, and some statements will rely on numerical evidence. We begin by recording bounds and asymptotic behaviour for some key quantities; these are a straightforward consequence of the relevant definitions.

\begin{lemma}\label{r2-ep-lims}
With $r_1$ defined in (\ref{r2-plus}) and $\epsilon$ defined in (\ref{nu-beta-ep-def}), we have
\be \lim_{a\to+\infty} r_1 = 2M, \quad r_1 \sim (2Ma^2)^{1/3},\quad a\to 0, \label{r2-limits} \ee
\be \epsilon = 1-\frac{r_1}{2M} = \frac{r_1^2}{a^2+r_1^2}\in (0,1), \label{ep-bounds} \ee
\be \lim_{a\to+\infty} \ep = 0, \quad \lim_{a\to 0} \ep = 1. \label{ep-limits} \ee
\qed
\end{lemma}

\begin{lemma}\label{J-ep-integrals}
The integral $J(\epsilon)$ defined in (\ref{j-def}) can be written as
\be J(\epsilon) = -2\ep^{-3/2}(1-\frac{\ep}{4})^{-3/2}(1+2\ep)^{-1/4}\int_{\tau_0}^{\tau_1}\frac{\sin\tau\cos^{5/2}\tau}{\sqrt{\sin(\tau-\tau_0)}}d\tau,\label{j-ep-int}
\ee
where
\begin{eqnarray}
\tau_0 &=& -\arctan\left(\frac{2+\ep}{\sqrt{\ep(4-\ep)}}\right)\in(-\frac{\pi}{2},-\frac{\pi}{3}),\label{tau0-def}\\
\tau_1 &=& -\arctan\left(\frac{\ep}{\sqrt{\ep(4-\ep)}}\right)\in(-\frac{\pi}{6},0).\label{tau1-def}
\end{eqnarray}
\qedapp
\end{lemma}

%\startproof 

There is one crucial advantage of the representation (\ref{j-ep-int}) over (\ref{j-def}): the former captures all possible singular behaviour in $\ep$. This is important for determining the contribution of $\bar{K}$ to $X_\pm$. This representation allows us to determine some features of $\bar{K}$ analytically, and allows us to determine other features in a numerically robust manner.  These features are given in the following Observation, and \textbf{we will assume thereafter that they are true}. The integral of (\ref{j-ep-int}) can be written as a linear combination of seven different but standard elliptic integrals: this has not enabled us to determine a proof of all of the statements below.

%%%%%%%%%
%%% FROM HERE
%%%%%%%%%

\begin{observation} The quantity $\bar{K}$ of (\ref{kbar-def}) satisfies the following relations:
\begin{eqnarray}
\bar{K}(0)&=&0, \label{bk0} \\
\bar{K}(1)&\simeq& 0.47, \label{bk1} \\
\bark'(\ep) &\sim& \frac{1}{12}\ep^{-1/2},\quad \bar{K}(\epsilon)\sim\frac16\ep^{1/2},\quad \ep\to 0, \label{bkprime} \\
\bar{K}(\ep) &>& 0 \quad \hbox{for all } \ep\in(0,1), \label{bk-pos} 
%2\bar{K}(\ep) &>& \frac{2\ep}{1+2\ep}\quad \hbox{for all } \ep\in(0,1), \label{bk-bnd2} 
\end{eqnarray}
and there exists $\ep_*\in(0,1)$ such that
\be \bar{K}(\ep) \left\{ \begin{array}{cc} > \frac{2\ep}{1+2\ep}, & \ep\in(0,\ep_*);\\
< \frac{2\ep}{1+2\ep}, & \ep\in(\ep_*,1). \end{array} \right. \label{bk-bnd1} 
\ee
\end{observation}

\begin{figure}\label{keyplots}
\includegraphics[scale=1]{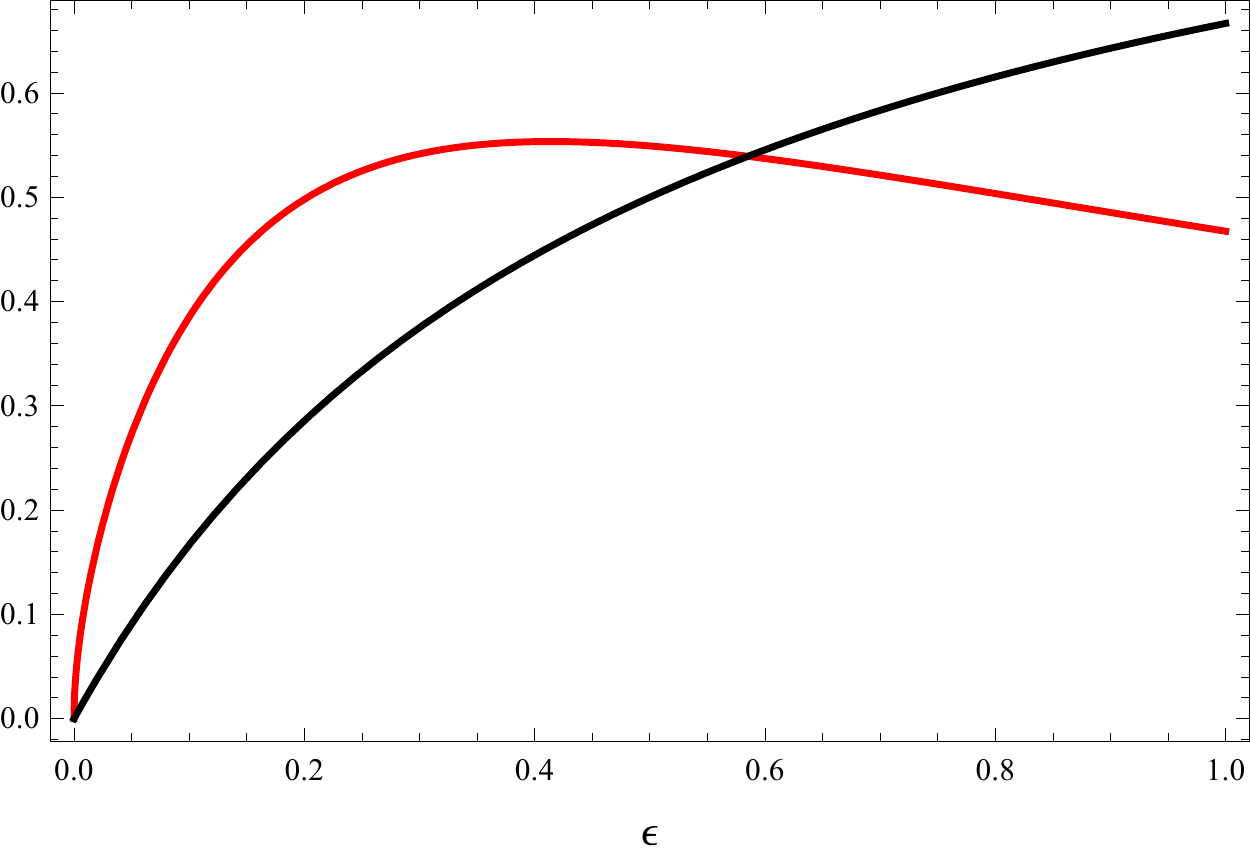}
\caption{Numerical plots of the quantities $\bark(\ep)$ and $\frac{2\ep}{1+2\ep}$ for $\ep\in(0,1)$. These provide numerical evidence for the validity of (\ref{bk-pos})-(\ref{bk-bnd1}). The quantities plotted here are either rational functions, or functions defined by non-oscillatory definite integrals.} 
\end{figure}

Of these, we have been able to prove (\ref{bk0}) and (\ref{bkprime}). The other relations rely on numerical evidence. To see that (\ref{bk0}) holds, we define $L(\epsilon)=\ep^{3/2}J(\ep)$ (so that $\bar{K}(\ep)=3L(\ep)-2$). Then from Lemma \ref{J-ep-integrals}, we can write down
\be L(0) = -2\int_{-\frac{\pi}{2}}^0 \frac{\sin\tau\cos^{5/2}\tau}{\sqrt{\sin(\tau+\frac{\pi}{2})}} d\tau. \label{l-zero} \ee
This can be evaluated exactly and yields $L(0) = -2/3$, and so $\bar{K}(0)=0$. The first part of (\ref{bkprime}) follows by a direct but somewhat lengthy calculation that relies on the fact that the integrals encountered can be evaluated exactly, as with (\ref{bk0}). Integrating and using (\ref{bk0}) yields the second equation of (\ref{bkprime}). Note that (\ref{bkprime}) proves the inequality (\ref{bk-bnd1}) in a neighbourhood of $\ep=0$. We have
\be L(1) = J(1) = -\int_{-\frac{\pi}{3}}^{-\frac{\pi}{6}} \frac{\sin\tau\cos^{5/2}\tau}{\sqrt{\sin(\tau+\frac{\pi}{3})}} d\tau, \label{l1-int} \ee
which we evaluate numerically (using Mathematica) to give (\ref{bk1}). For the inequalities (\ref{bk-pos}) and (\ref{bk-bnd1}), we have the numerical evidence of Figure \ref{keyplots}. We note that (\ref{bk-bnd1}) is not required in the proofs below, but this inequality aids in understanding the relative contributions of $\bar{K}$ and $2\epsilon/(1+2\epsilon)$ to $X_{\pm}$.%\textit{We emphasise that we will assume henceforth the validity of (\ref{bk0})-(\ref{bk-bnd1}).}

\subsection{The past boundary of eternally expanding $K<0$ McVittie spacetimes.}

In this case, we have $\Lambda\geq0$ and $H(t)>0$ for all $t>0$. We can immediately write down the following results:

\begin{proposition}\label{Xpm-eternal}
In an eternally expanding $K<0$ McVittie spacetime, we have
\begin{enumerate}
\item $X_+>0$ for all $t>0$;
\item there exists $t_1>0$ such that $X_->0$ for all $t\in(0,t_1)$;
\item there exists $t_2>0$ such that $X_->0$ for all $t>t_2$. 
\end{enumerate}
\end{proposition}

\startproof Part (i) is immediate from the definition (\ref{Xpm-def}) and positivity of $a, H, \bar{K}$ and $\epsilon$. From Lemma \ref{r2-ep-lims}, we have $\ep\to 1$ as $t\to 0$ (equivalently $a\to 0$). Thus
\be X_- \sim aH\bar{K}(1) - \frac23,\quad a\to 0^+, \label{xm-lim-azero} \ee
and so $X_-\to +\infty$ as $a\to 0^+$ using (\ref{ah-lims}) and (\ref{bk1}). Thus part (ii) follows. As $a\to+\infty$ (equivalently $t\to+\infty$), we have $aH\sim(1+\Lambda a^2/3)^{1/2}$. This limit corresponds to $\epsilon\to 0$, so that $\bar{K}(\epsilon)\sim\frac16\epsilon^{1/2}$, and $2\epsilon/(1+2\epsilon)\sim 2\epsilon$. It follows that $X_-$ is dominated by $aH\bar{K}(\epsilon)$ in this limit and so the conclusion (iii) follows. \qed

 With the smoothness property $X_-\in C^1((0,+\infty),\real)$ established in Proposition \ref{prop:structure-of-all-bdy} and the asymptotic positivity properties at $a=0$ and $a\to+\infty$ as established above, we see that every change in sign of $X_-$ from positive to negative must be followed by a change in sign from negative to positive. The total number of such sign changes is determined by the behaviour of $H$ as a function of $a$ (and the implicit function theorem tells us that $H$ is always locally a function of $a$). But with the freedom allowed by Definition \ref{def2}, it is possible that the total number of sign changes may be arbitrarily high. Likewise, it is possible that there are no sign changes. Suffice to say that each pair of sign changes gives rise to a characteristic structure in the conformal diagram of the spacetime. Successive pairs of sign changes give rise to chains of these structures. Figure 3 shows plots of the key function $X_-$ for examples which have respectively no sign changes (so that $X_-$ remains positive for all values of $a$) and exactly two sign changes.

\begin{figure}\label{example}
\includegraphics[scale=1]{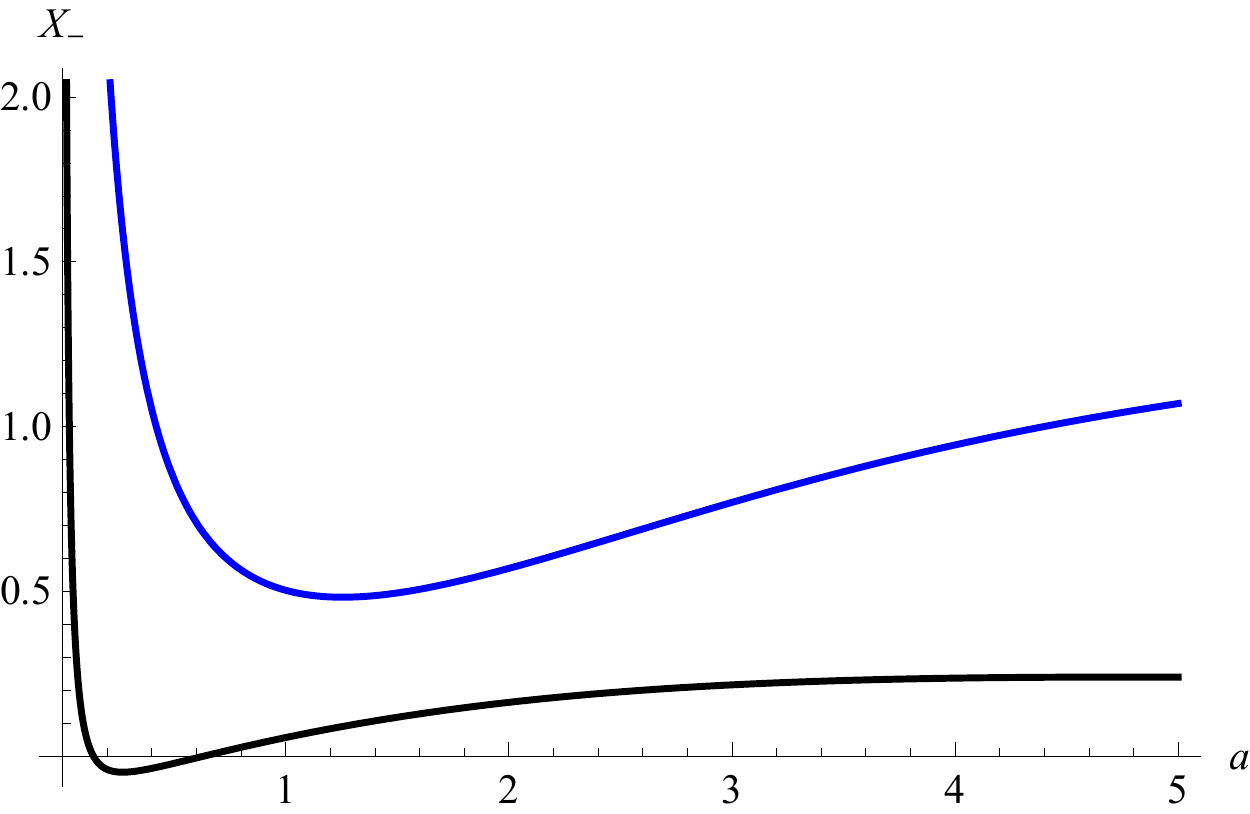}
\caption{Numerical plots of the quantity $X_-$ defined in (\ref{Xpm-def}). In an expanding universe, this function plays a key role in determining the causal structure of the allowed boundary $\balr$ as described in Proposition \ref{prop:structure-of-all-bdy}. The examples show plots of $X_-$ for eternally expanding $K=-1$ McVittie spacetimes satisfying (\ref{hub-mix}), and thus satsifying Definition \ref{def2}. The upper plot corresponds to parameter values $M=\Lambda=\Omega_m=\Omega_r=1$, and the lower plot corresponds to parameter values $M=1,\Lambda=\Omega_m=\Omega_r=0.01$. In the former case, $\balr$ is spacelike everywhere; in the latter case, $\balr$ has a timelike portion corresponding to values of $a$ for which $X_-$ is negative.} 
\end{figure}

\subsection{The past and future boundaries of recollapsing $K<0$ McVittie spacetimes.} 

In a McVittie spacetime with a recollapsing background (\textit{cf.} Definition \ref{def3}), the recollapsing phase is essentially the time reversal of the expanding phase. This assertion can be made rigorous by defining a spacetime subject to the conditions of \ref{def1}, and with scale factor $\hat{a}:(0,t_f)\to \mathbb{R}_+, \hat{a}(t)=a(t_f-t)$ where $a$ is the scale factor of a McVittie spacetime with a recollapsing background. This new spacetime is also a McVittie spacetime with a recollapsing background. RNGs of the new spacetime are the time reversal of RNGs of the original spacetime. From this we can deduce a generalistion of Propositions \ref{RNG-past-1}, \ref{RNG-past-2} and \ref{RNG-past-3}:

\begin{proposition}\label{RNGs-recollapse}
Let $({\mm},g)$ be an initially expanding, recollapsing McVittie spacetime with a Big Bang background for which $K<0$ and let $(t_0,r_0)\in\Omega$ with $r_0>r_1(t_0)$. Let $s$ be an affine parameter along an RNG $\gamma$ with $(t,r)|_{s=0}=(t_0,r_0)$. Then there exists $s_\alpha\in(-\infty,0)$ and $s_\omega\in(0,+\infty)$ such that $(t(s),r_{RNG}(s))\in\Omega$ for all $s\in(s_\alpha,s_\omega)$ and $\lim_{s\to{s_\alpha,s_\omega}}(t(s),r_{RNG}(s))\in\co\cup\balr$. Here, $\gamma:(s_\alpha,s_\omega)\to \Omega:s\mapsto (t(s),r_{RNG}(s))$ is the RNG. There exist open subsets $\Omega_{1,\alpha},\Omega_{2,\alpha}\subset\Omega$ for which geodesics with initial points in $\Omega_{1,\alpha}$ and $\Omega_{2,\alpha}$ originate at $\co$ and $\balr$ respectively. Likewise, there exist open subsets $\Omega_{3,\omega},\Omega_{4,\omega}\subset\Omega$ for which geodesics with initial points in $\Omega_{3,\omega}$ and $\Omega_{4,\omega}$ terminate at $\co$ and $\balr$ respectively.
\qed
\end{proposition}

Proposition \ref{prop:structure-of-all-bdy} also holds in recollapsing McVittie spacetimes, and corresponding to Proposition \ref{Xpm-eternal} we have the following (recall that $t_{\rm{max}}$ is the point of maximum expansion; $H(t_{\rm{max}})=0$ - \textit{cf.} Corollary \ref{cor:a-limits}):

\begin{proposition}\label{Xpm-recollapse}
In a recollapsing $K<0$ McVittie spacetime, 
\begin{enumerate}
\item there exists $t_1\in(0,t_f)$ such that $X_+(t)>0$ and $X_-(t)>0$ for all $t\in(0,t_1)$;
\item there exists $t_2\in(0,t_f)$ such that $X_+(t)<0$ and $X_-(t)<0$ for all $t\in(t_2,t_f)$;
\item there exist $t_3,t_4\in(t_1,t_2)$ such that $t_{\rm{max}}\in(t_3,t_4)$, $X_+(t)>0$ and $X_-(t)<0$ for all $t\in(t_3,t_4)$.
\end{enumerate}
\end{proposition}

\startproof Parts (i) and (ii) follow in the same way as parts (i) and (ii) of Proposition \ref{Xpm-eternal}. Part (iii) follows from parts (i) and (ii), the differentiability of $X_\pm$ and the fact that $X_\pm(t_{\rm{max}}) = \pm 2\epsilon/(1+2\epsilon)$. \qed

 It follows from this proposition and the `handbook' Proposition \ref{prop:structure-of-all-bdy} that in a recollapsing $K<0$ McVittie spacetime, the allowed boundary $\balr$ is initially past-spacelike, has a timelike portion, and is future-spacelike as the big crunch singularity $t=t_f$ is approached. There may be additional timelike portions within the time intervals $(t_1,t_3)$ and $(t_4,t_2)$.

%%%%%%%%%%%%%%%%%%%%%%%%%%%%%%
%%%%%%%%%%%%%%%%%%%%%%%%%%%%%%
%%%%%%%%%%%%%%%%%%%%%%%%%%%%%%

\section{The future boundary for $K<0$ and $\Lambda\geq 0$: black holes galore}\label{section:future}

It has been shown that there are IRNGs in $K=0$ McVittie spacetimes that reach the radius $r_-$ (see (\ref{rh-limits})) in finite affine parameter time \cite{kaloper2010mcvittie}, \cite{nolan2014particle}. The implications for the interpretation of McVittie spacetimes as containing black holes were laid out \cite{kaloper2010mcvittie}. The following results show that this feature also holds in eternally expanding $K<0$ McVittie spacetimes. The black hole interpretation requires a full understanding of the future evolution of RNGs, both ingoing and outgoing, which we establish in the following results.

\begin{proposition}\label{ORNGs}
Let $\Omega$ be the allowed region of an eternally expanding $K<0$ McVittie spacetime and let $(t_0,r_0)\in\Omega$. Then the unique ORNG $s\mapsto (t(s),r_{{\sc{ORNG}}}(s))$ with $(t(0),r_{{\sc{ORNG}}}(0))=(t_0,r_0)$ is future-complete and satisfies $\lim_{s\to+\infty}(t(s),r_{{\sc{ORNG}}}(s))=(+\infty,+\infty)$. 
\end{proposition}

\startproof 
First, we prove that all ORNGs with an initial value $r_0\leq2M$ extend into the region $r>2M$ (part (a) below). We then show that ORNGs with initial value $r_0>2M$ are future complete with the asymptotic behaviour stated (part (b)). We recall first that ORNGs satisfy 
\be \frac{dr}{dt}  = \sigma(r+\kappa^{1/2}H^{-1}) >0, \label{orng-rem} \ee
the inequality holding throughout $\Omega$. 

\begin{itemize}

\item[(a)] If $r_0=2M$, it is immediate from (\ref{orng-rem}) that $r_{{\sc{ORNG}}}(t)>2M$ for $t>t_0$. So suppose that there exists an ORNG $\gamma$ with $r_0<2M$ and $r_{{\sc{ORNG}}}(t)<2M$ for all $t\in[t_0,t_\omega)$, the right-maximal interval of existence. It follows from part (i) of Proposition \ref{Xpm-eternal} and from Proposition \ref{prop:structure-of-all-bdy} that the geodesic cannot meet $\balr$ at finite value of $t>t_0$. Thus 
\be r_1(t) < r_{{\sc{ORNG}}}(t) <2M \hbox{ for all } t\in[t_0,t_\omega). \label{r-max-bnds}\ee
It follows from the existence of these bounds that the right hand side of (\ref{orng-rem}) is bounded (\textit{cf.} also Lemma \ref{sig-at-bdy}) and so the right-maximal interval of existence is $[t_0,+\infty)$. Then the function $t\mapsto r_{{\sc{ORNG}}}(t)$ is increasing and bounded above by $2M$ and so $r_\infty:=\lim_{t\to+\infty} r_{{\sc{ORNG}}}(t)$ exists and, using (\ref{r-max-bnds}) and (\ref{r2-limits}), satisfies $r_\infty= 2M$. 

We introduce the 1-parameter family of curves of the $a-r$ plane
\be {\cal{K}}_\lambda = \{(a,r): \kappa(a,r)=(1-\lambda)\frac{r^2}{a^2}\},\quad \lambda\in[0,1]. \label{kap-lam} \ee
Then 
\be (a,r)\in{\cal{K}}_\lambda \Leftrightarrow 1-\frac{2M}{r}+\lambda\frac{r^2}{a^2}=0 \Leftrightarrow r=r_\lambda(a) \label{kap-la-props} \ee
where $\rl:[0,\infty)\to [0,2M)$ is a smooth, increasing function. Implicit differentiation yields
\be \rl'(a) = 2\lambda\frac{\rl^3}{a^3}\left(1+3\lambda\frac{\rl^2}{a^2}\right)^{-1}. \label{rlp}\ee
The allowed boundary $\balr$ is $ {\cal{K}}_1$, and the surface $\{r=2M\}$ is $ {\cal{K}}_0$. The family ${\cal{K}}_\lambda, \lambda\in[0,1]$ foliates the region bounded by $\{r=2M\}$, $\balr$ and $\{a=0\}$.

Using (\ref{sig-form1a}), we can calculate the slope of the ORNG $\gamma$ at its point of intersection with $\kl$:
\be \left.\frac{dr}{da}\right|_{r=\rl(a)}=\sqrt{1-\lambda}\frac{\rl^2}{a^2}(1+\frac{\sqrt{1-\lambda}}{aH})\int_0^{a/\rl}(1+y^2-\frac{2M}{a}y^3)^{-3/2}dy. \label{orng-on-rl}\ee  

Since $\rl(a)<2M$ for all $\lambda\in(0,1]$ and $a>0$, the integral here satisfies
\begin{eqnarray} \int_0^{a/\rl}(1+y^2-\frac{2M}{a}y^3)^{-3/2}dy &>& \int_0^{a/2M}(1+y^2-\frac{2M}{a}y^3)^{-3/2}dy \nonumber \\ &\sim& 1, \quad a\to\infty. \label{int-bound}\end{eqnarray}
This provides a lower bound for the slope of the RNG as it crosses $\kl$. With the lower bound in hand, and comparing (\ref{rlp}) and (\ref{orng-on-rl}), we see that we can choose $\lambda\in(0,1)$ and a value $a_0$ of $a$ such that
\be r_{ORNG}(a_0)=\rl(a_0),\quad r'_{ORNG}|_{r=\rl(a_0)}>\rl'(a_0)  \label{key-bound1} \ee
and furthermore 
\be r'_{ORNG}|_{r=\rl(a)}>\rl'(a),  \quad a>a_0. \label{key-bound2} \ee
This follows from the fact that the right hand sides of (\ref{rlp}) and (\ref{orng-on-rl}) are $O(a^{-3})$ and $O(a^{-2})$ respectively in the limit as $a\to+\infty$. Then for this value of $\lambda$,  $r_{ORNG}(a)>\rl(a)$ for all $a>a_0$. The definition of $\kl$ then yields $\kappa>\sqrt{1-\lambda}\frac{r^2}{a^2}$ along the geodesic, and so we obtain the bound
\begin{eqnarray} \frac{dr}{da} &>& \sqrt{1-\lambda}\frac{r^2}{a^2}(1+\frac{\sqrt{1-\lambda}}{aH})\int_0^{a/2M}(1+y^2-\frac{2M}{a}y^3)^{-3/2}dy \nonumber \\
& \sim & \sqrt{1-\lambda}\frac{r^2}{a^2},\quad a\to+\infty. \label{rng-lower-bnd} 
\end{eqnarray}
along the ORNG $\gamma$.  Let $r_0=r_{ORNG}(a_0)$. Integrating (using the leading order estimate for the right hand side) shows that $r_{ORNG}(a)$ exceeds $2M$ - yielding a contradiction to the existence of $\gamma$  - if  
\be \frac{\sqrt{1-\lambda}}{a_0}-\frac{1}{r_0}+\frac{1}{2M}>0. \label{ineq1} \ee
The validity of this inequality follows from the condition that $\gamma$ crosses $\kl$ from below, as expressed by (\ref{key-bound1}). To see this, we note that using the crossing condition $r_0=r_\lambda(a_0)$ shows that (\ref{ineq1}) is equivalent to 
\be \frac{1-\lambda}{\lambda} >\frac{r_0^2}{4M^2}\left(\frac{2M}{r_0}-1\right). \label{ineq2}\ee
Examining leading order terms, the condition that the geodesic crosses from below yields the inequality 
\be \frac{1-\lambda}{\lambda} >4\left(\frac{2M}{r_0}-1\right). \label{ineq3}\ee
Then the validity of (\ref{ineq2}), and hence of (\ref{ineq1}) follows from the fact that $r_0<2M$. 

\item[(b)] We apply positivity of $H$ and the inequality (\ref{Sig-lr}) to the equation governing ORNGs to obtain
\begin{eqnarray}
\frac{dr}{dt} &=& \sigma(r+\kappa^{1/2}H^{-1}) \nonumber\\
&>& \sigma r \nonumber\\
&>& \frac{arH\kappa^{1/2}}{\sqrt{a^2+r^2}} \label{ORNG-ineq1}
\end{eqnarray}
or equivalently
\be 
\frac{dr}{da}>\frac{r\kappa^{1/2}}{\sqrt{a^2+r^2}}. \label{ORNG-ineq2}
\ee
Along the ORNG, $\frac{dr}{da}>0$, and so $\kappa>\epsilon+\frac{r^2}{a^2}$ where $0<\epsilon=1-2M/r_0<1$. Then it is straightforward to verify that 
\be
\frac{dr}{da}>\epsilon^{1/2}\frac{r}{a}.\label{ORNG-ineq3} 
\ee
Integrating proves that $r\to+\infty$ as $a\to+\infty$ and hence as $t\to+\infty$. Since $\ddot{r}<0$ (\textit{cf.} (\ref{affine})), this cannot happen in finite affine parameter time. 
\end{itemize}
\qed

\begin{proposition}\label{BH galore}
Let $\Omega_R$ be the regular region of an eternally expanding, $K<0$ McVittie spacetime and let $(t_0,r_0)\in\Omega_R$. Then there exists $s_\omega<+\infty$ such that $\lim_{s\to s_\omega^-} (t(s),r_{IRNG}(s)) = (+\infty, r_-)$ where $s\mapsto (t(s),r_{IRNG}(s))$ is the unique IRNG with initial point $(t(0),r_{IRNG}(0))=(t_0,r_0)$ and $r_-$ is given in (\ref{rh-limits}). 
\end{proposition}

\startproof We recall that the governing equation for IRNGs is 
\be \frac{dr}{dt} = \sigma(r-\kappa^{1/2}H^{-1}) \in C^1(\Omega), \label{irng-rem} \ee
and that the regular region (in an eternally expanding universe) is characterised by $r<\kappa^{1/2}H^{-1}$ (see (\ref{eq:chi-kappa}) and (\ref{def:reg-region}). The boundary of the regular region is the horizon $\chi=0$, which (as described in Section 5 above) can be described by functions of the form $t\mapsto r_h(t)$ satisfying (\ref{rh-diff}). Noting that IRNGs satisfy $\left.\frac{dr}{dt}\right|_{\chi=0} = 0$, we see that IRNGs cannot exit the regular region $\Omega_R$ at a finite value of $t$. By standard ODE results (see e.g.\ Section 2.4 of \cite{perkodifferential}), we see that IRNGs $t\mapsto r_{IRNG}(t), r(t_0)=r_0$ with $(t_0,r_0)\in\Omega_R$ exist globally in $t$. A straightforward argument shows that these must satisfy $r_{IRNG}'(t)\to 0$ as $t\to+\infty$, and so $r_{IRNG}(t)\to r_-$ as $t\to+\infty$. It remains to prove that these geodesics are incomplete. This follows immediately from the fact that $\dot{r}(s)<0$ for all $s>0$ and from (\ref{affine}) which yields $\ddot{r}(s)<0$ using the dominant energy condition. \qed

Our final propositions provide details of the behaviour of ingoing RNGs that originate outside the horizon - that is, in the anti-trapped region of the spacetime. When $\Lambda>0$, these IRNGs escape to infinity, a possibility that arises due to the eternal expansion of the universe (Proposition \ref{IRNGs-future-1}). Our expectation is that as with the case of spatially flat McVittie spacetimes, the case $\Lambda=0$ is more difficult to deal with in full generality (see e.g.\ Proposition 5.5 of \cite{nolan2014particle}). However, in the spatially flat case, the dominant asymptotic (late time) behaviour of $H$ is sensitive to the equation of state (Lemma 5.2 of \cite{nolan2014particle}). This is not so in the present case - \textit{cf.} (\ref{ah-lims}). Thus with a slightly weaker hypothesis than applies in the spatially flat case we can in fact determine the general behaviour of IRNGs when $\Lambda=0$: all such geodesics must enter the regular region, and are then subject to Proposition \ref{BH galore}. This is the content of Proposition \ref{IRNGs-future-2} below. We note that the hypothesis of Equation (\ref{H-asymp-irng}) includes the cases of (\ref{hub-bar}) and (\ref{hub-mix}) discussed in Section 4 above.

\begin{proposition}\label{IRNGs-future-1}
Let $\Omega$ be the allowed region of an eternally expanding $K<0$ McVittie spacetime with $\Lambda>0$ and let $(t_0,r_0)\in\Omega$ with $r_0>r_+$. Then the unique IRNG $s\mapsto (t(s),r_{IRNG}(s))$ with $(t(0),r_{IRNG}(0))=(t_0,r_0)$ is future-complete and satisfies $\lim_{s\to+\infty}(t(s),r_{IRNG}(s))\to(+\infty,+\infty)$. 
\end{proposition}

\startproof The analysis of Section 5 shows that $\frac{dr}{da}>0$ along such IRNGs throughout the right-maximal interval of existence. Suppose that $r$ is bounded above. Then (\ref{irng-rem}) remains finite along the geodesic, which thus exists globally. Then $\lim_{a\to+\infty} r_{IRNG}(a)$ exists and is finite. Call this limiting value $r_\infty$. As $a\mapsto r_{IRNG}(a)$ is monotone and bounded above by this limiting value, we can find
\begin{eqnarray}
\sigma(a,r)  &=& H\kappa^{1/2}\int_0^{a/r}(1+y^2-\frac{2M}{a}y^3)^{-3/2}dy \nonumber \\
&\sim& H_0\left(1-\frac{2M}{r_\infty}\right)^{1/2}\int_0^\infty(1+y^2)^{-3/2}dy,\quad a\to+\infty. 
\label{sig-lim-irng}
\end{eqnarray}
where we have used the representation (\ref{sig-form1a}) for $\sigma$. Note that the integral evaluates to 1.
Likewise, we find
\be r-\kappa^{1/2}H \sim r_\infty-\left(1-\frac{2M}{r_\infty}\right)^{1/2}H_0>0,\quad a\to+\infty.\label{chi-lim-irng}\ee
Using (\ref{irng-rem}), these asymptotic relations yield
\be \frac{dr}{da} \sim \left(1-\frac{2M}{r_\infty}\right)^{1/2}\left(r_\infty-\left(1-\frac{2M}{r_\infty}\right)^{1/2}H_0\right)a^{-1}, \label{rp-lim-irng}\ee
which, on integrating, contradicts the assumption that $r_{IRNG}(a)$ is bounded above. Thus $r\to+\infty$ along the IRNG, and has infinite affine length by the usual argument. \qed

\begin{proposition}\label{IRNGs-future-2}
Let $\Omega$ be the allowed region of an eternally expanding $K<0$ McVittie spacetime with $\Lambda=0$ and with 
\be H^2 \sim a^{-2} + \frac{C}{a^\lambda},\quad a\to+\infty \label{H-asymp-irng} \ee
for some $\lambda>2$. Let $(t_0,r_0)\in\Omega_A$. Then the unique IRNG $s\mapsto (t(s),r_{IRNG}(s))$ with $(t(0),r_{IRNG}(0))=(t_0,r_0)$ enters the regular region $\Omega_R$ and terminates in finite affine time at $t=+\infty, r=r_-$.
\end{proposition}
%For $K<0$ and $\Lambda\geq0$, the regular region is always non-empty, and all IRNGs in the regular region extend to a 'horizon' at finite affine time in the future. 
%

\startproof Suppose there exists an IRNG which does not satisfy the stated conclusion. Since the horizon is a one-way membrane for IRNGs, this IRNG must remain in the anti-trapped region $\chi<0$ for all $a>a(t_0)$. From (\ref{H-asymp-irng}), along such an IRNG we have
\be \chi \sim 1-\frac{2M}{r} - \left(\frac{r^2}{a^2}\right)\frac{C}{a^{\lambda-2}},\quad a\to+\infty. \label{chi-neg} \ee
If $r/a$ is bounded along the geodesic, this quantity is eventually positive - i.e.\ the geodesics enters the regular region. Thus $r/a$ must be unbounded as $a\to+\infty$. This allows us to estimate as follows:
\be \kappa \sim \frac{r^2}{a^2},\quad a\to+\infty, \label{kapp-est} \ee
\be \int_0^{a/r}(1+y^2-\frac{2M}{a}y^3)^{-3/2}dy \sim \int_0^{a/r}(1+y^2)^{-3/2}dy \sim \frac{a}{r},\quad a\to+\infty,\label{int-est}\ee
which, using (\ref{irng-rem}), lead  to the estimate 
\be \frac{dr}{da} \sim \frac{r}{a}\left(1-\frac{1}{aH}\right) \sim \frac{Cr}{2a^{\lambda-1}}. \label{dr-da-est}\ee
Integrating this last relation, and using $\lambda>2$, shows that $r$ is bounded in the limit as $a\to+\infty$, a contradiction. Thus every IRNG must enter the regular region, and the final conclusion follows from Proposition \ref{BH galore}. \qed

%%%%%%%%%%%%%%%%%%%%%%%%%%%%%%
%%%%%%%%%%%%%%%%%%%%%%%%%%%%%%
%%%%%%%%%%%%%%%%%%%%%%%%%%%%%%

\section{Conformal diagrams for $K<0$ McVittie spacetimes}\label{conformal diagrams}

%%%%%%%%%%%%%%%%%%%%%%%%%%%%%%
%%%%%%%%%%%%%%%%%%%%%%%%%%%%%%
%%%%%%%%%%%%%%%%%%%%%%%%%%%%%%
We can translate the results above into conformal diagrams by the usual process of straightening out outgoing and ingoing radial null geodesics to lines oriented at $\pm 45^\circ$ (respectively) to the horizontal. The diagrams below share the features that boundaries corresponding to curvature singularities (i.e.\ the Big Bang and Big Crunch surfaces) are shown serrated, the boundary of the allowed region $\Omega_{\kappa=0}$ is shown as a double line, and the black hole horizon is shown dashed. 

\subsection{Eternal expansion with $\Lambda =0$.}\label{conf:lam-zero}

The key features of the conformal diagram in this case are as follows. The past boundary is formed by the union of the Big Bang $\co=\{(t,r):t=r=0\}$ and the boundary of the allowed region. In the example shown, this is formed by the ``staircase'' structure $P_1-P_2-P_3-P_4$. The portions $P_1-P_2$ and $P_3-P_4$ are of the type $\partial\Omega_{\{\kappa=0\}}^{(+,+)}$ of Proposition \ref{prop:structure-of-all-bdy} and the portion $P_2-P_3$ is of the type $\partial\Omega_{\{\kappa=0\}}^{(+,-)}$. The points $P_2$ and $P_3$ are of the types $\partial\Omega_{\{\kappa=0\}}^{(+,0\downarrow)}$ and $\partial\Omega_{\{\kappa=0\}}^{(+,0\uparrow)}$ respectively. This conformal diagram corresponds to an eternally expanding McVittie spacetime with $\Lambda=0$ in which the key variable $X_-$ defined in (\ref{Xpm-def}) changes sign twice (as with the example shown in the lower curve of Figure 3). The sign changes occur at values of $t$ corresponding to $P_2$ and $P_3$. As noted in Comment 8, it is possible that there are no sign changes, or an even number of them. In the former case, the boundary of the allowed region would be a horizontal (or at least spacelike) curve from $P_1$ to $P_4$. Additional sign changes would give rise to further ``risers'' and ``steps'' in the staircase of Figure 4,   in addition to the single riser $P_2-P_3$ and step $P_3-P_4$. Future null infinity, ${\cal{I}}^+$, is null in this case: this follows from Proposition \ref{IRNGs-future-2}. Both $t$ and $r$ extend to positive infinity along this surface. The black hole horizon is at $r=2M$. The time coordinate $t$ extends to positive infinity along the IRNGs reaching the horizon, but only a finite amount of affine parameter time elapses along each of the geodesics. 

\begin{figure}\label{conformal1} 
\vskip -1.5in
\includegraphics[scale=1]{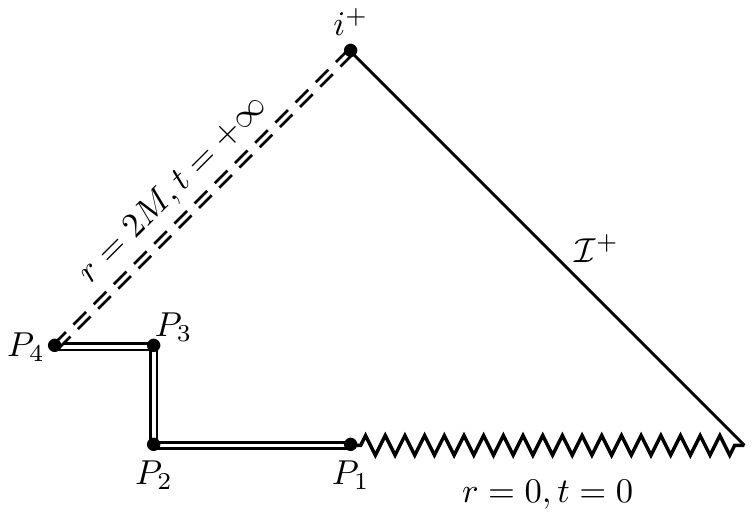}
\vskip-8in
\caption{Conformal diagram for McVittie spacetimes with $K<0$ and $\Lambda=0$.}
\end{figure}

\subsection{Eternal expansion with $\Lambda >0$.}\label{conf:lam-pos}

The only difference between the $\Lambda>0$ and the $\Lambda=0$ cases is in relation to the causal nature of future null infinity. As proven in Proposition \ref{IRNGs-future-1}, IRNGs with an initial point satisfying $r(t_0)>r_+$ extend to infinity, and have infinite affine length. On the other hand, IRNGs that enter the regular region $\Omega_R$ run into the black hole horizon in finite affine time (as described in Proposition \ref{BH galore}). This yields a spacelike ${\cal{I}}^+$ as indicated in Figure 5. In all other details, the boundary of the spacetime has the same features as that of Figure 4.

\begin{figure}\label{conformal2} 
\vskip -1.5in
\includegraphics[scale=1]{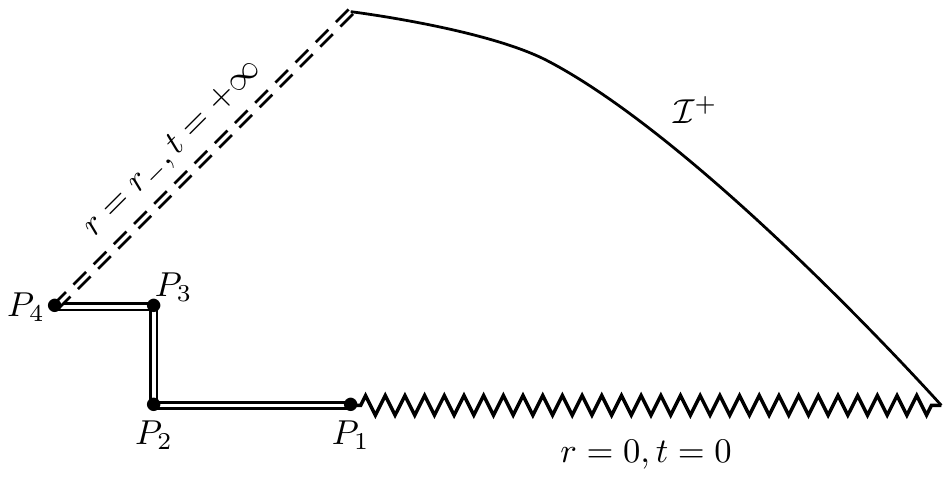}
\vskip-8in
\caption{Conformal diagram for McVittie spacetimes with $K<0$ and $\Lambda>0$.}
\end{figure}

\subsection{Recollapse: $\Lambda<0$.}\label{conf:lam-neg}

\begin{figure}\label{conformal3} 
\vskip -1.5in
\includegraphics[scale=1]{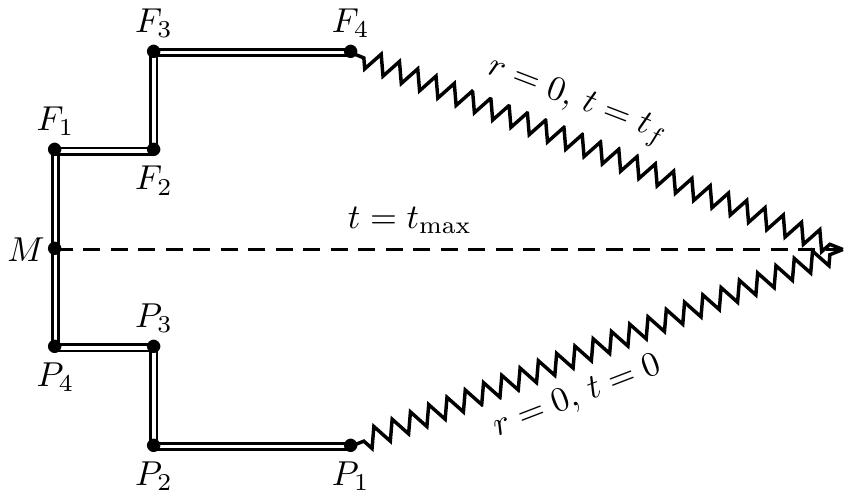}
\vskip-8in
\caption{Conformal diagram for McVittie spacetimes with $K<0$ and $\Lambda<0$.}
\end{figure}

In the recollapsing case, the conformal diagram is symmetric in time about $t=t_{\rm{max}}$, the time of maximum expansion of the FLRW background. At this instant, $H(t)=0$. The past boundary is formed by the union of the Big Bang singularity $\co=\{(t,r):t=r=0\}$ and the portion $P_1-P_2-P_3-P_4$ of the boundary of the allowed region.  The future boundary is formed by the union of the Big Crunch singularity $\{(t,r):t=t_f,r=0\}$ and the portion $F_1-F_2-F_3-F_4$ of the boundary of the allowed region. The timelike portion $P_4-F_1$ (which is of the type $\partial\Omega_{\{\kappa=0\}}^{(+,-)}$ in the notation of Proposition \ref{prop:structure-of-all-bdy}) contains the point $M$ corresponding to the time of maximum expansion. The comments of Section \ref{conf:lam-zero} in relation to the staircase structure of the boundary of the allowed region and the numbers of steps and risers also apply to the present case. There may be additional risers and steps moving to the left between $P_3$ and $P_4$, and there may be additional risers and steps moving to the right between $F_1$ and $F_2$. See figure \ref{conformal3}.

\section{The positive curvature case.}\label{positive curvature}

For completeness, we summarise some key global features of $K>0$ McVittie spacetimes. As with the cases of $K\leq 0$, the global structure depends strongly on that of the FLRW background. In the background, eternal expansion is possible if the cosmological constant is positive and is sufficiently large relative to the density  (see e.g.\ Section 9.3 of \cite{ellis2012relativistic}): otherwise, recollapse occurs. So we will consider the same possibilities of eternal expansion and recollapse in the positive $K$ case as we considered above in the negative $K$ case. 

From Lemma \ref{lem-allowed} and \ref{lem:a-m-cond}, we know that at each time $t$, the allowed region has a minimum and a maximum radius, and moreover this region is empty for times $t$ with $a(t)<3\sqrt{3}M$. In the eternally expanding case, there exists a unique $t_i>0$ such that 
\be a(t) \left\{ 
\begin{array}{rl} < 3\sqrt{3}M, & 0<t<t_i;\\
= 3\sqrt{3}M, & t=t_i; \\
> 3\sqrt{3}M, & t>t_i.
\end{array}
\right.
  \label{tmin-exp} \ee
and in the recollapsing case, there are unique values $t_i, t_j$ such that 
\be a(t) \left\{ 
\begin{array}{rl} < 3\sqrt{3}M, & 0<t<t_i, t_j<t<t_f;\\
= 3\sqrt{3}M, & t=t_i, t_j; \\
> 3\sqrt{3}M, & t_i<t<t_j.
\end{array}
\right.
  \label{tmin-recoll} \ee

By implicit differentiation of the equation
\be \kappa=1-\frac{2M}{r}-\frac{r^2}{a^2}=0 \label{kpos-bdy} \ee
defining the boundary of the allowed region for $K>0$, we see that for $H(t)>0$, the inner boundary $r=r_{1,(+)}(t)$ of the allowed region is a decreasing function of $t$ and the outer boundary $r=r_{2,(+)}(t)$ is an increasing function of $t$ (and vice versa when $H(t)<0$). In the eternally expanding case, $r_{1,(+)}$ decreases monotonicallly from $r=3M$ at $t=t_i$ and satisfies $\lim_{t\to+\infty} r_{1,(+)}=2M$. The outer boundary increases monotonically, without bound, from $r=3M$ at $t=t_i$. In the recollapsing case, $r_{1,(+)}$ decreases to a minimum value $r_{1,(+)}(t_{\rm{max}})$ and then increases again to $r=3M$ at $t=t_j$. The outer boundary increases and then decreases. Here, $t_{\rm{max}}$ is the time of maximum expansion. 

\begin{figure}\label{horizons-pos}
	%\centering
{\includegraphics[scale=0.5]{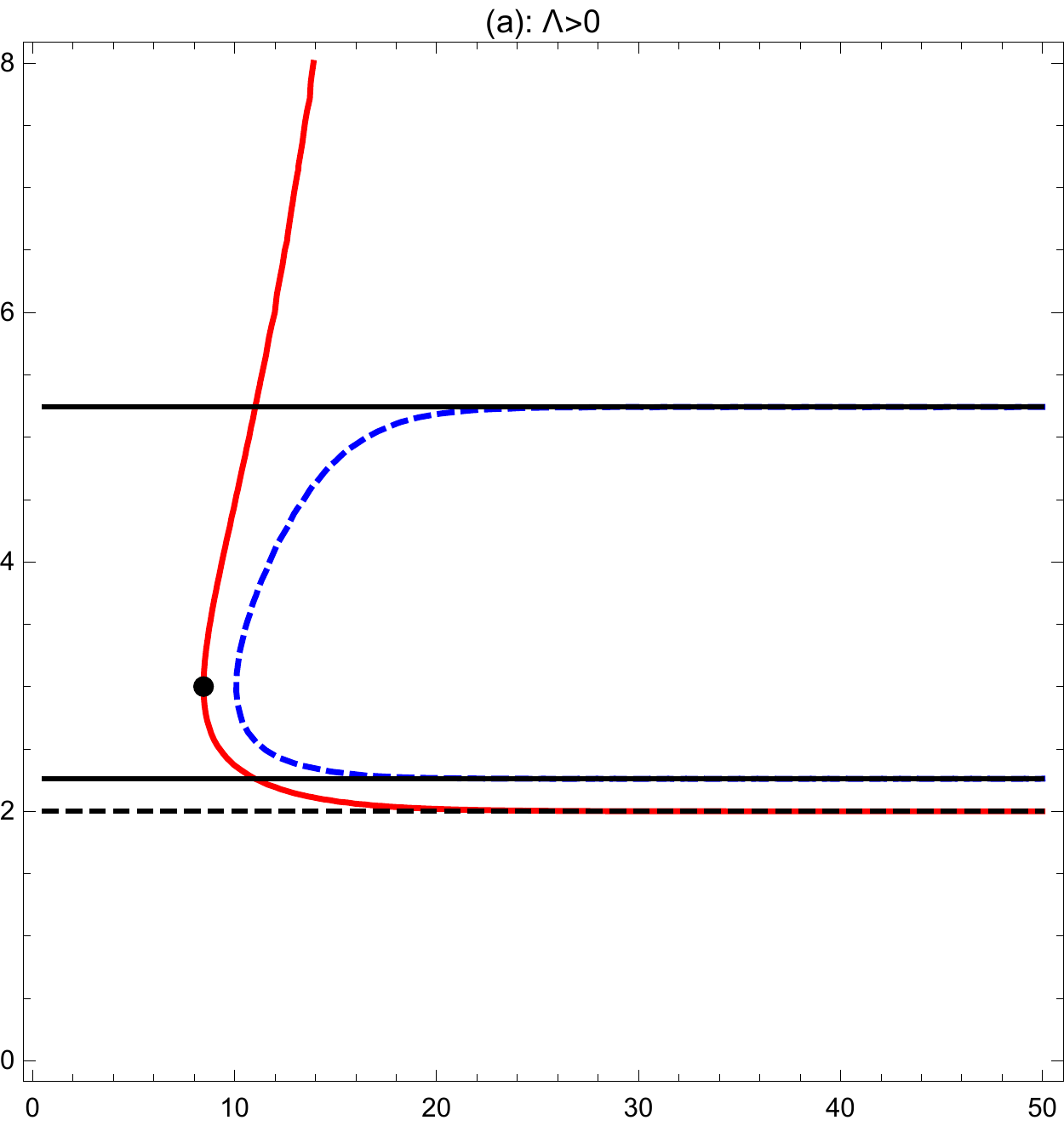}\hskip10pt \includegraphics[scale=0.5]{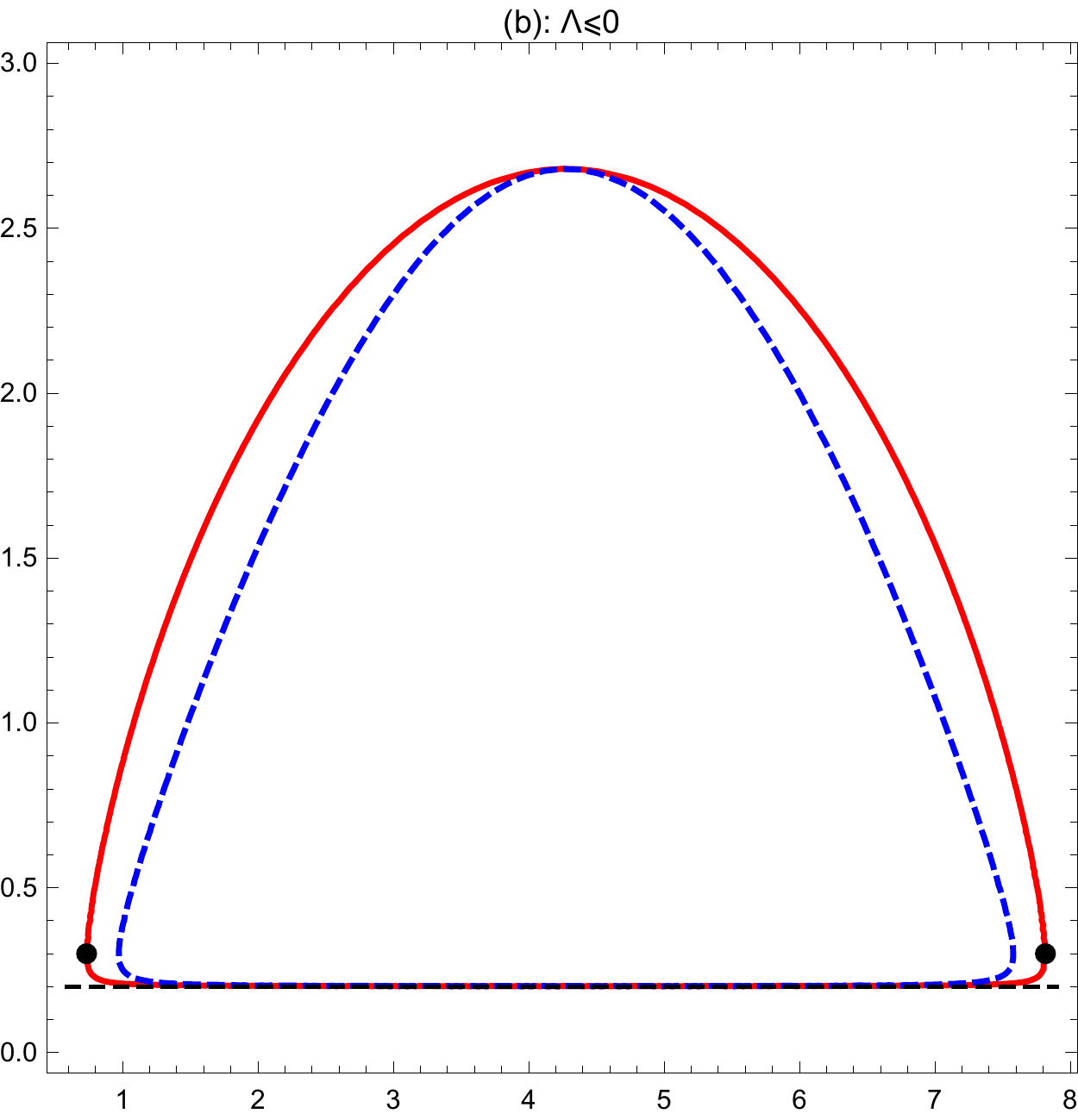}}
\caption{The allowed region (bounded by the solid curve) and the horizon (dashed curve, contained in the allowed region) for $K>0$ McVittie spacetimes with dust-filled FLRW backgrounds.  In each case, the horizontal axis represents $t$ and the vertical axis represents $r$.  In the left panel, $M=1$, the cosmological constant is positive and is sufficiently large to ensure eternal expansion. The horizonta solid lines mark the asymptotes of the horizon, and the horizontal dashed line is $r=2M$, which is an asymptote for the inner boundary of the allowed region. The boldface point marks the event $(t_i,3M)$ corresponding to the earliest point of the allowed boundary, whereat $a(t_i)=3\sqrt{3}M$ (see Lemma \ref{lem:a-m-cond}). In this case $t_i\simeq 8.47$. In the right panel, the cosmological constant is negative, but the same structure arises for zero cosmological constant. $M=0.1$ in this case, and the horizontal dashed line is $r=2M$. The horizon and the boundary of the allowed region meet at $t_{\rm{max}}$, whereat $H=0$. In this case, $t_{\rm{max}}\simeq 4.28$. The boldface points mark the events $(t_i,3M)$ and $(t_f,3M)$, the earliest and latest events on the allowed boundary. $a(t)=3\sqrt{3}M$ at both points, and $t_i\simeq 0.74, t_f\simeq 7.82$. In both cases, the allowed region is bounded away from $r=0$: the Big Bang singularity is cut-off as in the $K=0$ case.}
\end{figure}

The horizon is the zero set of 
\be \chi = \kappa - r^2H^2  = 1-\frac{2M}{r}-r^2(H^2+a^{-2}), \label{hor-pos}\ee
and so is a subset of the allowed region. In the eternally expanding case, the cosmological constant must be positive and we have $H\to H_0>0$. It follows that in this case, the structure of the horizon in this case is similar to the $K<0$ case with a positive cosmological constant. In the recollapsing case, the horizon has the same general structure as the boundary of the allowed region. The paragraphs above describing this boundary also describe the horizon, but with the scale factor $a(t)$ replaced by the function $a(t)(1+a^2(t)H^2(t))^{-1/2}$. See Figure \ref{horizons-pos}. 

The spacelike hypersurface $\Sigma_0$ of Proposition \ref{prop:k-pos-allowed-bdy}, along which the pressure of the spacetime diverges, is a subset of the allowed region, emanating from $(t,r)=(t_i,3M)$. In the recollapsing case, $\Sigma_0$ threads through the allowed region, connecting $(t_i,3M)$ and $(t_f,3M)$. This gives rise to two spacetime regions in each case: that bounded to the past by the portion of the allowed boundary with $r<3M$ and to the future by $\Sigma_0$, and that bounded to the future by the future by the portion of the allowed boundary with $r>3M$ and to the past by $\Sigma_0$. 

This brief analysis of the global structure of $K>0$ McVittie spacetimes is sufficient to establish that this family is, in a sense, more similar to the $K=0$ case than the $K<0$ studied above. With the focus of this paper being the $K<0$ case, we set aside the analysis of the RNGs in the $K>0$ case. But we conclude with one observation on this topic: the structure of the allowed boundary and the horizon in the eternally expanding case strongly suggests that the arguments of Section \ref{section:future} above will follow through and so establish that these closed universes contain black hole horizons. 

\section{Black holes in the case $K=0$?}\label{BH-flat}
In \cite{nolan1999point}, we established various results relating to the global structure of McVittie spacetimes with flat ($K=0$) FLRW backgrounds and with vanishing cosmological constant. In \cite{kaloper2010mcvittie}, the observation that the corresponding spacetimes with a postivie cosmological constant contain black hole horizons was made. As seen above for the case $K<0$, the horizon appears at radius $r=r_-$, the (inner) black hole horizon radius of the corresponding Schwarzschild-de Sitter spacetime. This marks the endpoint of ingoing radial null geodesics that do not escape to infinity. In the case where $\Lambda=0$, these geodesics terminate at $r=2M$. We claimed in \cite{nolan2014particle} that this rules out the black hole interpretation, as the geodesics terminate at a singularity. However, this claim is incorrect. As shown in \cite{nolan2014particle}, when $\Lambda=0$, the ingoing radial null geodesics that terminate (in finite affine time) at $r=2M$ (with $t\to+\infty$ in the limit) approach this boundary surface \textit{through the regular region of the spacetime} (see Section 5 of \cite{nolan2014particle}). That is, $\chi>0$ along the geodesics, where 
\be \chi =1-\frac{2M}{r}-r^2H^2. \label{chi-flat} \ee
In the class of spatially flat McVittie spacetimes studied in \cite{nolan2014particle}, we assumed that the FLRW background admits an equation of state $P_{FLRW}=g(\mu_{FLRW})$ subject to the technical condition that $\xi:=(3/2)(1+g'(0))>0$ (which allows a wide variety of standard fluid models). Applying the Einstein equations then yields (see Lemma 5.2 of \cite{nolan2014particle})
\be \lim_{t\to+\infty} -\frac{H'}{H^2}=\xi>0.\label{hplim}\ee
When $K=0$, we have
\be 8\pi P = -2H'(1-\frac{2M}{r})^{-1/2}-3H^2. \label{p-flat}\ee
Then in the regular region, using the energy condition $H'<0$, we have 
\be 8\pi P < -\frac{2H'}{rH}-3H^2, \ee
and the inequality $8\pi P>-3H^2$ applies throughout the spacetime. Taking the limit $t\to+\infty$ along an ingoing radial null geodesic $\gamma$ that terminates at $r=2M$, we obtain
\be \lim_{t\to+\infty} 8\pi P|_\gamma = 0, \ee
where we use $H\to 0$ and (\ref{hplim}). 
Thus the IRNGs \textit{do not} terminate at a curvature singularity as claimed in \cite{nolan2014particle}, and this obstruction to the interpretation of $r=2M$ as a black hole horizon is not present. Consequently, it appears that it is a case of black holes galore in  
the flat case as in the open case as established above. 

\section{Conclusions} Debate continues on the influence of inhomogeneities on large scale cosmological structures \cite{green2014well,buchert2015there}. It is clear, however, that there are strong arguments for considering the role that inhomogeneities may have on these structures, and on the overall expansion history of the universe \cite{bolejko2011inhomogeneous,ellis2011inhomogeneity}. It is therefore of use to have available \textit{exact} as well as perturbative models of such inhomogeneities. Exact models can act as a guide to understanding issues such as averaging in cosmology \cite{andersson2011inhomogeneous}. This is the source of the motivation for the study of McVittie spacetimes. At first glance, one sees in McVittie metrics a Schwarzschild spacetime embedded in an isotropic cosmological background. However, as the literature testifies \cite{nolan1999point,kaloper2010mcvittie,lake2011more,da2013expansion}, understanding the global structure of these spacetimes is a delicate issue. In order to make the link with the complete family of FLRW backgrounds, it is important that we understand McVittie spacetimes for which the background may lie in any one of the three families of isotropic universes: flat ($K=0$), open ($K<0$) and closed ($K>0$). In this paper, we have determined details of the global structure of these inhomogeneous spacetimes in the $K\neq 0$ cases, adding to the literature available on the $K=0$ case. This is a non-trivial task, owing primarily to the lack of the availability of  a metric which may be written in terms of elementary functions of the naturally occuring coordinates. Our results are summarised above, and so do not need to be repeated, but we draw attention to some key features. 

First, we note that the Big Bang (which arises in our coordinate representation as the point $\{(t,r):t=r=0\}$) forms a part of the past boundary of $K\neq 0$ McVittie spacetimes. This is in contrast to the $K=0$ case, in which there is a minimum radius $r=2M$ in the past. The remainder of the past boundary $\{(t,r): \kappa(t,r)=0, t>0\}$ of the spacetime is a non-scalar curvature singularity (the Weyl curvature (\ref{weyl-form1}), density and pressure are all finite along $\kappa=0$ - see (\ref{mu-def}), (\ref{P-def}) and Proposition 4.1 and Lemma 4.2). Furthermore, a heuristic inspection of the geodesic equations indicates that the components $u^\alpha = \frac{dx^\alpha}{d\lambda}$ of the tangent to a causal geodesic $\lambda\in I\mapsto x^\alpha(\lambda)$, in the local coordinates employed in the paper, remain finite in the approach to the allowed boundary. This (along with the behaviour of the curvature noted above) is sufficient to ensure that, considered as a singular boundary, the surface $\kappa=0$ is gravitationally weak \cite{tipler1977singularities,nolan1999strengths}. Extending this initial analysis to parallel propagated frames carried along the geodesic suggests that the singularity is of quasi-regular type \cite{ellis1979classification}, allowing for an extension of the spacetime through this boundary \cite{clarke1975singularities}.   However, extending through to negative values of $\kappa$ yields a space of Riemannian signature: the signature changes across $\kappa=0$ ($\kappa$ is forced by definition (\ref{def1}) to be positive). The Einstein equations are still satisfied, but the ``fluid flow" vector $u^\alpha$ is now spacelike. It is notable that this signature change occurs in the absence of any singularity along the boundary hypersurface. 

Second, we note that black holes are a universal feature of eternally expanding $K<0$ McVittie spacetimes. The black hole horizon is identified with the boundary $\{(t,r): t=+\infty, r=r_-\}$ formed by the endpoints of ingoing radial null geodesics which do not escape to infinity. These geodesics have finite affine length. The radius $r_-$ is defined in (\ref{rh-limits}), and corresponds to the inner horizon radius (i.e.\ the black hole horizon radius) of the Schwarzschild(-de Sitter) spacetime with mass parameter $M$ and cosmological constant $\Lambda = 3H_0^2$, where $0\leq H_0=\lim_{t\to+\infty} H(t)$. The density and pressure both drop to zero at the black hole horizon, suggesting that the black hole interior is vacuum, as in the extension through the horizon proposed in \cite{lake2011more}. 

In this paper, we have derived the global structure of a class of spacetimes that represent embedding of the Schwarzschild((anti) de Sitter) family into an FLRW background of arbitrary curvature index, with the focus on the cases $K\neq 0$. When $K<0$, there are simple conditions with clear physical and geoemtric content that give rise to a unique spacetime (once the background FLRW geometry and the mass parameter are specified). We have not been able to identify a corresponding uniqueness condition in the case $K>0$, and it would be interesting to find such a condition. Our analysis gives an understanding of how the localised inhomogeneity generated by the mass parameter $M$ (which corresponds to the renormalised Hawking mass of the spacetime \cite{hawking1968gravitational}) disrupts the local and global structure of the isotropic background. By addressing the cases $K\neq0$, we progress the programme begun in \cite{nolan1999point}. Clear differences emerge between the different values of $K$, the most prominent being the continued presence of the Big Bang singularity of the FLRW background in the $K<0$ case: this is cut off by a minimum radius condition that applies when $K\geq 0$. But the presence of a black hole horizon is common to both families - and appears likely also to apply to the closed family in the presence of a sufficiently large cosmological constant.

\ack I thank Daniel Guariento, Abe Harte, Ko Sanders and Peter Taylor for useful discussions. I am very grateful to an anonymous referee for their comments on an earlier version of this paper. 

\appendix

\section{Proofs}

\section*{Proof of Lemma \ref{sig-at-bdy}}

In (\ref{sig-k-neg}), we substitute $y=a/\bar{r}$ to obtain
\begin{eqnarray} \sig &=& H\kappa^{1/2} \int_0^x (1+y^2-\nu y^3)^{-3/2} dy, \label{sig-form1a} \\ 
&=& Hx^{-1}\nu^{-1}(x^2+\beta x+\gamma)^{1/2}(\alpha-x)^{1/2} \Sigma, \label{sig-form1} 
\end{eqnarray} 
where 
\be \Sigma =\int_0^x(\alpha-y)^{-3/2}(y^2+\beta y+\gamma)^{-3/2}dy, \label{Sig-int-def} \ee
and where terms are defined in (\ref{x-alpha-def}) and (\ref{nu-beta-ep-def}) above. This relies on the observation that 
\be \kappa(t,r) = x^{-2}q(x),\quad q(x) = 1 +x^2-\nu x^3,\label{kap-x-def} \ee
where the cubic $q(x)$ has a unique real root. This root is positive, and corresponds to the value $r=r_{1,(-1)}(t)$ and thus is given by $x=\alpha$.
This allows us to write $q(x) =\nu(\alpha-x)(x^2+\beta x + \gamma)$, where the terms $\beta, \gamma$ depend only on $\alpha$ and $\nu$, and we calculate  
\be \beta = \alpha-\nu^{-1},\quad \gamma=\alpha\beta. \label{beta-gamma-def}\ee 
The quadratic factor in $q$ is positive definite (and is bounded away from zero). 

Integrating by parts yields 
\be
\Sigma = 2(\alpha-x)^{-1/2}(x^2+\beta x+\gamma)^{-3/2}-2\alpha^{1/2}\gamma^{-3/2}+3\Sigma_1, \label{Sig-int1}\ee
where
\be 
\Sigma_1 = \int_0^x(\alpha-y)^{-1/2}g_1(y)dy,\quad g_1(y)=(2y+\beta)(y^2+\beta y+\gamma)^{-5/2}. \label{Sig1-int-def}\ee
Integrating by parts again yields
\be 
\Sigma_1 = -2(\alpha-x)^{1/2}g_1(x)+2\alpha^{1/2}\beta\gamma^{-5/2}-4\Sigma_2.\label{Sig1-int1}\ee
The term $\Sigma_2$ satisfies
\be \Sigma_2'(x) = -\frac12(\alpha-x)^{1/2}g_1'(x). \label{Sig2-prime} \ee
Since $g_1$ is analytic at $x=\alpha$, so too is its derivative. This allows us to write $g_1'$ as a power series in (\ref{Sig2-prime}) and to integrate term-by-term to obtain 
\be
\Sigma_2 = \Sigma_2(\alpha) + \sum_{k=0}^\infty h_k(\alpha-x)^{k+3/2}, \label{Sig2-sum}\ee
for some coefficients $h_k$, where the series converges uniformly on an interval of the form $0\leq \alpha-x < \delta$. It follows that 
\be
\Sigma_1(x) = \Sigma_1(\alpha)-2(2\alpha+\beta)(\alpha^2+\alpha\beta+\gamma)^{-5/2}(\alpha-x)^{1/2}+O((\alpha-x)^{3/2}),\label{Sig1-int2}\ee
where
\begin{eqnarray} 
\Sigma_1(\alpha)&=& \int_0^\alpha (\alpha-y)^{-1/2}(2y+\beta)(y^2+\beta y+\gamma)^{-5/2} dy \nonumber \\
&=& \alpha^{-7/2} J(\epsilon). \label{Sig1-int3} 
\end{eqnarray}
To obtain (\ref{Sig1-int3}), we have introduced the quantity 
\be \epsilon=\frac{\beta}{\alpha} = 1-\frac{r_{1,(-1)}}{2M} = \frac{r_{1,(-1)}^3}{2Ma^2}, \label{ep-def}\ee
and made the change of variable $y=\alpha\xi$ in the integral. The result of the lemma follows by collecting half-integer powers of $\alpha-x$ first in $\Sigma$ and then in $\sigma$, where a Taylor expansion of various coefficients around $x=\alpha$ is required to obtain the coefficients (\ref{sig0-def})-(\ref{sig1-def}) above. The smoothness properties stated in the lemma follow from the integral representations above and from the series representation (\ref{Sig2-sum}). \qed

\section*{Proof of Lemma \ref{lem:sig-zero}}

We may take $K=+1$ and so $\kap = 1-\frac{2M}{r}-\frac{r^2}{a^2}$. With $t\in I_0$, the allowed region is non-empty. Applying l'Hopital's rule to the integral in (\ref{sig-k-pos}) yields
\begin{eqnarray}
\lim_{r\downarrow r_{1,(+)}} \sig(t,r) &=& H\frac{2r_{1,(+1)}^2}{3r_{1,(+1)}^2-a^2}, \label{sig-lim1} \\
\lim_{r\uparrow r_{2,(+)}} \sig(t,r) &=& H\frac{2r_{2,(+1)}^2}{3r_{2,(+1)}^2-a^2}.\label{sig-lim2}
\end{eqnarray}
(The free function $s_M(t)$ makes no contribution, as $\kappa$ vanishes in the relevant limit.) From (\ref{r1-pos}) and (\ref{r2-pos}), we see that these terms have opposite signs whenever $H(t)\neq 0$. It follows that for each $t\in I_0$ for which $H(t)\neq0$, $\sig(t,r)$ changes sign (and so equals zero) at least once in the interval $(r_{1,(+1)},r_{2,(+1)})$. From the PDE (\ref{sig-eqn}), we can read-off 
\be \kap\partial_r\sig|_{\sig=0} = \frac{H}{a^2}r, \ee
which has constant sign at each value of $t$ ($\kap>0$ by definition). Thus $\sig(t,r)$ has a unique zero in the interval $(r_{1,(+1)},r_{2,(+1)})$ for each $t\in I_0$. \hfill{$\blacksquare$}

\section*{Proof of Lemma \ref{lem:r2prime}}
The boundary of the allowed region is $\partial\Omega_{\{\kappa=0\}}=\{(t,r)\in\bar{\Omega}:\kappa(t,r)=0\}$, and in the present case $(K<0)$, is described by the function $r_1:(0,t_f)\to\mathbb{R}, t\mapsto r_1(t)$ (see (\ref{r2-plus})). Then 
\be 1-\frac{2M}{r_1(t)}+\frac{r_1^2(t)}{a^2}=0\quad \hbox{ for all } t\in(0,t_f), \label{allowed-bdy} \ee 
and implicit differentiation yields 
\be r_1'(t) = 2\frac{Hr_1^3}{a^2}\left(1+3\frac{r_1^2}{a^2}\right)^{-1}.\ee 
Comparing with (\ref{sig0-def}) yields (\ref{r2-de}). Comparing with (\ref{rng}) proves the last statement of the lemma. 
\qed

\section*{Proof of Lemma \ref{lem-RNG-bdy}}
This is a (reasonably) straightforward calculation that relies on the defining equation (\ref{rng}) for RNGs, Lemma \ref{lem:r2prime} and Lemma \ref{sig-at-bdy}. The quantities $\nu$ and $\epsilon$ are defined in Lemma \ref{sig-at-bdy}. Note that 
\be \alpha-x = \alpha(r_1+\rh)^{-1}\rh, \label{alpha-rh} \ee
and recall that $\kappa(t,r) = \nu x^{-2}(\alpha-x)(x^2+\beta x+\gamma)$.
%, yielding $\kappa=\kappa_1(\alpha-x)$ with 
%\be \left.\kappa_1\right|_\balr = \nu\alpha^{_2}(\alpha^2+2\alpha\beta) = \nu(1+2\epsilon). \label{kap1} \ee
The precise form of $Q_\pm$ will not be required below: it is sufficient to know that these functions are $C^1$ on the region indicated. \qed

\section*{Proof of Lemma \ref{u-pos-neg}}
The hypotheses of the lemma indicate that $u\in C^2(I)$. Differentiating (\ref{u-ivp1}) yields
\be u''(t) = p'(t)+q(t,u)u'+(\partial_1q(t,u)+\partial_2q(t,u)u')u,\label{udd} \ee
and so 
\be u''(t_0)=p'(t_0) + q(t_0,0)p(t_0), \label{udd0} \ee
and we note that $u'(t_0)=p(t_0)$. Now apply Taylor's theorem: for sufficiently small $|h|$, 
\be u(t_0+h) = p(t_0)h + \frac12 u''(t_h)h^2,\label{taylor} \ee
where $t_h$ lies between $t_0$ and $t_0+h$. For $p(t_0)\neq 0$, parts (i) and (ii) follow by taking $h$ to be sufficiently small. For $p(t_0)=0$ and $p'(t_0)\neq 0$, part (iii) follows by observing that by continuity, $u''(t_h)$ and $p'(t_0)$ have the same sign for sufficiently small values of $h$. \qed

\section*{Proof of Lemma \ref{J-ep-integrals}}
In the integral (\ref{j-def}) defining $J$, we make the change of variable $\zeta=(1-\xi)^{1/2}$ and define $m=1+\frac{\ep}{2}$, $n^2=\frac{\ep}{4}(4-\ep)$ to write
\be J(\ep)=4\int_0^1(m-\zeta^2)\left[(\zeta^2-m)^2+n^2\right]^{-5/2} d\zeta. \label{j-zeta} \ee
The further change of variable $n\tan\tau=\zeta^2-m$ (and some algebraic manipulations) then yields (\ref{j-ep-int}). We note that in this integral, the lower limit $\tau_0$ is an increasing function of $\ep$, the upper limit is a decreasing function of $\ep$ and
\begin{eqnarray}
\lim_{\ep\to 0} (\tau_0,\tau_1) = (-\frac{\pi}{2},0),\\
\lim_{\ep\to 1} (\tau_0,\tau_1) = (-\frac{\pi}{3},-\frac{\pi}{6}).
\end{eqnarray}\qed

%\section*{Proof of Proposition \ref{IRNGs-future-1}} 

%\section*{Proof of Proposition \ref{IRNGs-future-2}}

\section*{References}
\bibliographystyle{unsrt}
\bibliography{mybib}

\end{document}